\documentclass[10pt]{iopart}
\RequirePackage[colorlinks,citecolor=blue,urlcolor=magenta,linkcolor=blue]{hyperref}
\usepackage{graphicx}
\usepackage{enumerate}
\usepackage{url}
\usepackage{amssymb}
\usepackage{amsfonts}
\usepackage[thinc]{esdiff}
\usepackage{setspace}
\usepackage{verbatim}
\usepackage{dutchcal}
\usepackage{tabularx}
\usepackage{scalerel}
\usepackage{braket}
\usepackage{tikz}
\usepackage{tikz-cd}
\usetikzlibrary{svg.path}

\usepackage{bm}
\usepackage{graphicx}
\usepackage{enumitem}
\usepackage{physics}
\usepackage{dsfont}
\usepackage[normalem]{ulem}
\usepackage{mathrsfs}
\usepackage{calrsfs}
\usepackage{comment}
\newcommand*{\red}{\textcolor{red}}

\usepackage[caption=false]{subfig}
\renewcommand{\vec}[1]{\mathbf{#1}}
\usepackage{varioref}
\usepackage[active]{srcltx}

\expandafter\ifx\csname package@font\endcsname\relax\else
\expandafter\expandafter
\expandafter\usepackage
\expandafter\expandafter
\expandafter{\csname package@font\endcsname}%
\fi
\hypersetup{
	colorlinks=true,       
	linkcolor=blue,          
	citecolor=blue,        
}

\usepackage[section]{placeins}

\usepackage{fancybox}
\usepackage{soul}

\usepackage{moreenum}
\usepackage[utf8]{inputenc}

\labelformat{equation}{Eq.~(#1)} 
\labelformat{figure}{Fig.~#1} 
\labelformat{subfigure}{Fig.~\thefigure#1} 
\labelformat{table}{Tab.~#1} 
\labelformat{appendix}{Appendix #1}

\definecolor{orcidlogocol}{HTML}{A6CE39}
\tikzset{
  orcidlogo/.pic={
    \fill[orcidlogocol] svg{M256,128c0,70.7-57.3,128-128,128C57.3,256,0,198.7,0,128C0,57.3,57.3,0,128,0C198.7,0,256,57.3,256,128z};
    \fill[white] svg{M86.3,186.2H70.9V79.1h15.4v48.4V186.2z}
                 svg{M108.9,79.1h41.6c39.6,0,57,28.3,57,53.6c0,27.5-21.5,53.6-56.8,53.6h-41.8V79.1z M124.3,172.4h24.5c34.9,0,42.9-26.5,42.9-39.7c0-21.5-13.7-39.7-43.7-39.7h-23.7V172.4z}
                 svg{M88.7,56.8c0,5.5-4.5,10.1-10.1,10.1c-5.6,0-10.1-4.6-10.1-10.1c0-5.6,4.5-10.1,10.1-10.1C84.2,46.7,88.7,51.3,88.7,56.8z};
  }
}

\newcommand\orcidicon[1]{\href{https://orcid.org/#1}{\mbox{\scalerel*{
\begin{tikzpicture}[yscale=-1,transform shape]
\pic{orcidlogo};
\end{tikzpicture}
}{|}}}}
\usepackage[square,numbers,compress]{natbib}

\usepackage{bm}
\usepackage{xparse}
\usepackage{dsfont}
\usepackage[mathscr]{euscript}
\usepackage{float}
\usepackage{etoolbox}
\DeclareUnicodeCharacter{2212}{-}
\begin{document}

\title[A non-local origin for massive gravity and late-time acceleration]{A non-local origin for massive gravity and late-time acceleration}

\author{Susobhan Mandal$^{1,2}$ and S.~Shankaranarayanan$^{1}$}
%
\address{$^1$ Department of Physics, Indian Institute of Technology Bombay, Mumbai - 400076, India}
\address{$^2$Department of Physics, Indian Institute of Science Education and Research Tirupati, Tirupati -  517507, India }
\ead{sm12ms085@gmail.com}
\ead{shanki@iitb.ac.in}

\begin{abstract}
The accelerated expansion of the universe poses a significant challenge to General Relativity. Non-local modifications to gravity have emerged as a compelling class of theories to address this dark energy puzzle. Building upon earlier proposals~\cite{Deser:2007jk, Deser:2013uya, Dodelson:2013sma, Maggiore:2014sia, Capozziello:2021krv, Modesto:2013jea, Nojiri:2019dio, Nojiri:2007uq, Elizalde:2018qbm, Jhingan:2008ym}, we investigate a specific non-local modified gravity action incorporating terms like $R\Box^{-2}R$, $R^{\mu\nu}\Box^{-2}R_{\mu\nu}$, $R^{\mu\nu\sigma\delta}\Box^{-2}R_{\mu\nu\sigma\delta}$ and demonstrate that it provides a dynamical origin for a massive graviton by reducing to the standard and extended Fierz-Pauli action at the linearized level. A fixed-point analysis of the background cosmology reveals a stable de Sitter attractor, ensuring the model naturally drives accelerated expansion. Crucially, we investigate the cosmological perturbations and show that the theory's six propagating degrees of freedom are free from ghost instabilities. We further demonstrate that all large-scale tensor modes are dynamically stable and decay on the accelerating background. 
This ghost-free massive gravity extension provides distinct predictions for gravitational wave polarizations and is theoretically consistent with $\mathbf{\Lambda CDM}$ at late times, positioning it as a unique alternative to scalar-tensor models like $f(R)$ and Galileons.
This robust stability at both the background and perturbative levels establishes our model as a consistent and compelling alternative to the standard $\Lambda$CDM paradigm. 
%
\end{abstract}

\section{Introduction}
%
The observed accelerated expansion of the Universe~\cite{DiValentino:2021izs,Kamionkowski:2022pkx} is a large-scale, infrared (IR) phenomenon, prompting numerous proposals for modifying General Relativity (GR) in this regime~\cite{2016-Joyce.etal-ARN,Shankaranarayanan:2022wbx,Mandal:2025xuc}. One way to conceptualize modifications to a force law at large distances is through the idea of \emph{screening}. In theories like electrodynamics, charge screening occurs due to the presence of opposite charges that diminish the effective field strength at large distances. Gravity, however, presents a unique challenge: it possesses only one type of \emph{charge} --- mass-energy --- which is universally attractive. Consequently, a screening mechanism analogous to that in electrodynamics, relying on the cancellation by opposite charges, is not viable for gravity~\cite{Boulware:1972yco}.

Instead, cosmic acceleration itself may arise from an effective screening (weakening) of gravity at cosmological or IR scales. If gravity's influence diminishes sufficiently at cosmological scales, it could lead to the observed acceleration without invoking dark energy~\cite{Dvali:2007kt,deRham:2007rw}. Such a modification, inherently tied to large distance scales, naturally suggests that the underlying theory describing this gravitational behavior might be non-local. Thus, non-local modifications to gravity have emerged as a compelling class of theories to address this dark energy puzzle ~\cite{Deser:2007jk, Deser:2013uya, Dodelson:2013sma, Maggiore:2014sia, Capozziello:2021krv, Modesto:2013jea, Nojiri:2019dio, Nojiri:2007uq, Elizalde:2018qbm, Jhingan:2008ym}.
Non-local field theories, characterized by terms that depend on spacetime integrals or inverse differential operators, are well-suited to describe interactions that are not confined to a single spacetime point and can manifest as long-range modifications~\cite{Jaccard:2013gla}. The inherent scale introduced by non-local operators can effectively alter the gravitational force law in the IR, providing a mechanism for this \emph{gravitational screening}~\cite{Nojiri:2010pw,Maggiore:2014sia}.

Non-local theories, characterized by operators like $\Box^{-1}$, naturally introduce scale-dependent effects, hence, modifying the gravitational potential at cosmological distances~\cite{Capozziello:2008gu,Gambuti:2021meo}. These terms arise in quantum effective actions when integrating out massless modes (e.g., gravitons) or in classical averaging of inhomogeneities \cite{Bautista:2017enk, Barvinsky:2023exr, Barvinsky:2002uf, Elias:2017wkr, Donoghue:2015nba, Barvinsky:2003kg, Barvinsky:1994cg}. Unlike phenomenological dark energy models, non-local terms (e.g., involving 
inverse powers of differential operators like the d'Alembertian ($\Box^{-1}$), can emerge naturally in effective field theories from integrating out short-wavelength modes or from quantum loop corrections of light/massless particles. These theories can dynamically screen the cosmological constant or mimic its effects, addressing fine-tuning.  However, constructing consistent non-local theories is highly non-trivial: the equations of motion are integro-differential, requiring careful handling of causality and gauge invariance, and their predictions must avoid ghosts and instabilities while recovering GR at solar-system scales~\cite{Jaccard:2013gla}. While non-local gravity has been widely explored as a viable alternative to explain dark energy~\cite{Deser:2007jk, Deser:2013uya, Dodelson:2013sma, Maggiore:2014sia, Capozziello:2021krv, Modesto:2013jea, Nojiri:2019dio, Nojiri:2007uq, Elizalde:2018qbm, Jhingan:2008ym}, much of the existing literature has focused on making it non-local using well-established auxiliary field method \cite{Cusin:2015rex}. In this work, we investigate a novel non-local gravitational action \ref{MT 1a}, which offers a previously unexplored connection between the non-local gravity and massive gravity theories at linear order.
  
In massive gravity, a graviton mass ($m_g$) introduces a Yukawa suppression $\sim e^{-m_g r}/r $ to the Newtonian potential at scales $ r \gtrsim m_g^{-1}$~\cite{Bernard:2014bfa,deRham:2014zqa,Mazuet_2018,Gumrukcuoglu:2021gua,Mandal:2024nhv}. In other words, these theories modify gravity at large distances --- precisely in the IR regime where cosmic acceleration is observed. This IR modification, leading to a weakening of gravity at cosmological scales, shares a characteristics with the effects produced by non-local theories~\cite{Jaccard:2013gla}. This leads us to the question: Is it plausible that some theories of massive gravity could emerge from, or be related to, a more fundamental non-local framework? We address this in the work and show that such non-local theories lead to a stable, late-time de Sitter attractor and show that the tensor perturbations are free from ghost instabilities.

This work is organized as follows. In Sec.~\ref{sec:Model}, we introduce the non-local gravity model and show that it can be mapped to Extended Fierz-Pauli action. In Sec.~\ref{sec:attractor} we show explicitly that the model leads to a de Sitter attractor.  Sec.~\ref{sec:Stability} provides a rigorous investigation into the stability of cosmological
perturbations and show that they are ghost-free. Sec.~\ref{sec:Obs} provides a overview of the observational implications of the non-local model w.r.t cosmological and gravitational wave observations. We conclude and compare this model with other modified gravity models in Sec.~\ref{Sec:Conclusions}. The six appendices contain detailed calculations presented in the main text. 

In this work, we use the the metric signature $(-,+,+,+)$ and natural units where $c=1, \kappa = \sqrt{8\pi G}$.

\section{Non-local gravity model}
\label{sec:Model}
As mentioned, here, we explore this connection more concretely by demonstrating the existence of a class of pure gravity theories, formulated as non-local field theories, which reduce to the Fierz-Pauli (FP) action in the linearized gravity limit~\cite{Fierz:1939ix}. More specifically, we show that the linearized limit of the non-local action around Minkowski space reproduces the FP massive gravity action. The class of non-local theory we consider is described by the following action:
\begin{equation}\label{MT 1}
S  = \frac{1}{2\kappa^{2}}\int d^{4}x\sqrt{-g} \ \left[ R +  \mathcal{L}_{\rm NL} \right] \, ,
\end{equation}
where  
\begin{equation}\label{MT 1a}
\!\!\!\! \mathcal{L}_{\rm NL}  = \mu_{1} R
\frac{1}{\Box^{2}}R + \mu_{2}R^{\mu\nu}\frac{1}{\Box^{2}}R_{\mu\nu} + \mu_{3}R^{\mu\nu\rho\sigma}\frac{1}{\Box^{2}}R_{\mu\nu\rho\sigma} 
\end{equation}
 $\mu_{1}, \mu_{2}, \mu_{3}$ are couplings with mass dimension $2$, $\Box = g^{\mu\nu}\nabla_{\mu}\nabla_{\nu}$, and inverse d’Alembertian 
$\Box^{-1}$ introduces a mass scale. 
We linearize the above gravity action around Minkowski spacetime, decomposing the metric as follows:
\begin{equation}\label{MT 2}
g_{\mu\nu} = \eta_{\mu\nu} + 2\kappa h_{\mu\nu}, \ g^{\mu\nu} = \eta^{\mu\nu} - 2\kappa h^{\mu\nu}
+ \mathcal{O}(\kappa^{2}),
\end{equation}
where $\eta_{\mu\nu} = \text{diag}(-1, 1,\ldots, 1)$ and $h_{\mu\nu}$ is the metric perturbation around Minkowski spacetime. Upon linearizing $R$, $R_{\mu\nu}$, and $R_{\mu\nu\rho\sigma}$ in terms of $h_{\mu\nu}$, we get:
\begin{eqnarray}
& & \hspace*{-2.4cm} 
\sqrt{-g}R/(2 \kappa^2)  \equiv \mathcal{L}_{m = 0}  = 
- \frac{1}{2}\partial_{\lambda}h_{\mu\nu}\partial^{\lambda}
h^{\mu\nu} + \partial_{\mu}h_{\nu\lambda}\partial^{\nu}h^{\mu\lambda} 
- \partial_{\mu} h^{\mu\nu}\partial_{\nu}h + \frac{1}{2}\partial_{\lambda}h\partial^{\lambda}h  \nonumber \\
& & \hspace*{-2.4cm}  \frac{1}{2\kappa^{2}}\sqrt{-g}R\frac{1}{\bar{\Box}^{2}}R  = 2(\alpha - 1)^{2}h^{2} + \mathcal{O}
(\kappa); \quad \frac{1}{2\kappa^{2}}\sqrt{-g}R^{\mu\nu}\frac{1}{\bar{\Box}^{2}}R_{\mu\nu}  = \frac{1}{2}h_{\mu
\nu}h^{\mu\nu} - \alpha h^{2} + \mathcal{O}(\kappa) \nonumber \\
\label{MT 4}
& & \frac{1}{2\kappa^{2}}\sqrt{-g}R^{\mu\nu\rho\sigma}\frac{1}{\bar{\Box}^{2}}R_{\mu\nu\rho\sigma}  
= 2(h_{\mu\nu}h^{\mu\nu} - \alpha^{2}h^{2}) + \mathcal{O}(\kappa), 
\end{eqnarray}
where $\bar{\Box} = \eta^{\mu\nu}\partial_{\mu}\partial_{\nu}$. In deriving the expressions, we have used the relation $\partial_{\mu} h^{\mu\nu} = \alpha \partial^{\nu} h$ where $\alpha$ is unknown parameter that will be determined using the physical conditions. This relation is not a gauge-fixing condition. Rather it is one of the equations that can be obtained from massive gravity theory with mass deformation parameter being $\alpha$ which is discussed in Ref.~\cite{Mandal:2024nhv}. 

To see this, let us first consider GR. In GR, the equations derived from the variation of the action is the Einstein tensor $G_{\mu\nu}$. Applying the Bianchi identity, $\nabla^\mu G_{\mu\nu} = 0$, leads to $\nabla^\mu T_{\mu\nu} = 0$ (energy-momentum conservation), which ensures that the theory is consistent without needing to satisfy $\partial^\nu h_{\mu\nu} - \partial_\mu h = 0$ (the divergence of the linearized Einstein tensor vanishes by identity).

In the massive gravity theory, the linearized action contains a mass term ($\mathcal{L}_{\text{mass}}$) that breaks the diffeomorphism invariance ($\delta h_{\mu\nu} \neq \nabla_{(\mu} \xi_{\nu)}$). Hence, it is not possible to freely choose a gauge to eliminate all non-physical DOFs. Instead, the field equations derived from the massive gravity action often contain a specific combination of terms that, when manipulated (by taking the divergence of the full field equation), lead to an auxiliary constraint on the physical fields. This constraint, $\partial^\nu h_{\mu\nu} - \partial_\mu h = 0$ (or the equivalent terms in non-local theories), is required to ensure consistency and eliminate the potentially dangerous scalar ghost (the Boulware-Deser ghost) that generically appears in massive gravity theories.
In summary, the key reason it is not a gauge condition is that the mass term has already broken the gauge symmetry it would normally be used to fix. Instead, this equation must be satisfied as a condition of motion derived from the action to ensure the theory propagates only the five or six healthy physical degrees of freedom expected of a massive spin-2 field, and not a ghost.

Using this, we can self-consistently show the massive gravity can be equivalent to non-local gravity in the linearised limit upto quadratic terms by choosing $\alpha = 1, 1/2$. Since we considered only the $\mathcal{O}(\kappa^{0})$ terms in the action, we can safely write $\frac{1}{\Box} = \frac{1}{\bar{\Box}} + \mathcal{O}(\kappa)$ where the operators with $\kappa$ factors come depending on whether $\Box$ acts on scalar or tensor quantities on its right.
Detailed calculations are provided in \ref{App:Linearised gravity}.

Now applying the inverse d'Alembertian operator $\Box^{-2}$ (which in Fourier space corresponds to multiplication by $1/k^4$), ${\cal L}_{NL}$ yields terms quadratic in $h_{\mu\nu}$. Specifically, at order $\mathcal{O}(\kappa^{0})$, these terms reduce to:
\begin{equation}\label{MT 5}
\begin{split}
\frac{\sqrt{-g} }{2\kappa^{2}} \mathcal{L}_{NL} & = h_{\mu\nu}h^{\mu\nu}\left[\frac{\mu_{2}}{2} 
+ 2\mu_{3}\right] + h^{2}\left(2\mu_{1}(\alpha 
- 1)^{2} - \mu_{2}\alpha - 2\mu_{3}\alpha^{2}\right) 
\end{split}
\end{equation}
The above action has the form similar to the generalized FP action~\cite{Gambuti:2021meo,Mandal:2024nhv}:
\begin{equation}\label{eq. 0.1}
S_{GFP} = \int d^{D}x \Big[ \mathcal{L}_{m = 0} - \frac{1}{2}m^{2}(h_{\mu\nu}h^{\mu\nu}
 - \alpha_{\rm GFP} \, h^{2})\Big] 
\end{equation}
where $m$ is the mass of spin-2 field, $\alpha_{\rm GFP}$ is referred to as the massive deformation parameter and $\mathcal{L}_{m = 0}$ is the Lagrangian density corresponding to the linearized  Einstein-Hilbert action. Typically $\alpha_{\rm GFP} = 1$ for the standard FP theory to propagate 5 degrees of freedom and avoid the Boulware-Deser ghost~\cite{Boulware:1972yco}. Therefore, to ensure that our non-local theory \ref{MT 1} reduces to a linearised massive gravity theory described by the FP action with a graviton mass $m$ and the deformation parameter $\alpha$ is related to $\alpha_{\rm GFP} $, the coefficients of $h_{\mu\nu}h^{\mu\nu}$ and $h^2$ from  \ref{MT 5} must satisfy specific conditions. Setting $\alpha  = \alpha_{\rm GFP}$, these conditions are:
\begin{equation}\label{MT 6}
\mu_{2} + 4\mu_{3}  = - m^{2}, ~ 
4\mu_{1}(\alpha - 1)^{2} - 2\mu_{2}\alpha - 4\mu_{3}\alpha^{2}  = \alpha \,  m^{2} .
\end{equation}
The viability of specific $\alpha$ values often relates to ensuring the absence of ghosts and potentially other theoretical consistencies. In quantum field theories, ghost typically appears from higher derivative quadratic-order terms such that the propagator in momentum space contain a term with negative residue at the positive-energy pole. Thus, the linearised analysis of quadratic order terms would be sufficient to establish the absence of ghosts. This is explicitly shown in \ref{App:Linearised gravity} and \ref{App: Anomaly}. With three unknown coupling constants ($\mu_1, \mu_2, \mu_3$) and two equations \ref{MT 6}, the system is generally underdetermined. However, by considering specific, physically motivated values for $\alpha$, we can find solutions: 
\begin{enumerate}[leftmargin=0.4cm]
\item {\bf Standard FP gravity ($\alpha = 1$):}  For this choice, the conditions in \ref{MT 6} simplify significantly. The first term in the second equation vanishes, $( \alpha-1)^2 = 0$. The relations become:
    \begin{equation}\label{MT 7}
    \begin{split}
    \mu_{2} + 4\mu_{3}  = - m^{2} \, , & \quad
    - 2\mu_{2} - 4\mu_{3}  = m^{2}
    \end{split}
    \end{equation}
The two equations yields $ \mu_2 = 0 $ and $\mu_3 = - m^{2}/4$. In this case, $\mu_1$ remains unconstrained by these conditions, meaning any value of $\mu_1$ (or $\mu_1=0$ for minimality) would still yield the FP mass term ($\alpha=1$). This implies that the $R \Box^{-2} R$ term is not necessary to recover the standard FP action.

\item {\bf  Extended FP theory ($\alpha = 1/2$):} 
Recently, we showed that the extended FP gravity ($\alpha = 1/2$) arises from spontaneous symmetry breaking in the matter sector~\cite{Mandal:2024nhv}. Hence, $\alpha = 1/2$ provides a more natural origin for the graviton mass parameters and could explain cosmic acceleration without dark energy~\cite{Mandal:2024nhv}. For this value,  \ref{MT 6} becomes:
    \begin{equation}\label{MT 8}
     \mu_{1} - \mu_{2} - \mu_{3} = m^{2}/2 \quad    
   \mu_{2} + 4\mu_{3} = - m^{2}  \, .
      \end{equation}
 This yields  
    \begin{equation}\label{MT 8a}
    \mu_1 = -m^2/2 - 3\mu_3,~~\mu_2 = -m^2 - 4\mu_3 \, .
    \end{equation}
 Imposing the additional simplifying assumption $\mu_3 = 0$, we get:
 $\mu_1 = -m^2/2, \mu_2 = -m^2$. 
\end{enumerate}
The couplings $\mu_1, \mu_2, \mu_3$ are introduced in the action as proportionality constants for the non-local terms. Their physical scales are determined by the requirement that the non-local terms only become dynamically relevant and dominate the energy budget at cosmological scales (the infrared limit): To drive the current accelerated expansion, the energy density associated with the non-local operator must be of the order of the observed dark 
energy density:
$$\rho_{\text{non-local}} \sim \rho_{\Lambda} \sim (10^{-3} \text{ eV})^4 \sim M_P^2 H_0^2$$
Since our non-local action is quadratic in curvature terms and involves the inverse Laplacian operator $\Box^{-2}$, the dimensional analysis requires the couplings to be related to the square of the current Hubble parameter, $H_0$:
    $${\mu_i \sim H_0^2}$$
Setting $\mu_i$ to this scale ensures the non-local effects are screened at laboratory and Solar System scales, only appearing at the largest cosmological distances (the degravitation effect). While the couplings are physically motivated to be of order $H_0^2$, this scale $\left(H_0 \sim 10^{-33} \text{ eV}\right)$ is still vastly smaller than the fundamental scale of gravity (the Planck mass, $M_P \sim 10^{18} \text{ GeV}$). Therefore, a fine-tuning of the fundamental parameters is still required to select a scale that matches the present-day acceleration. Our model does not solve the theoretical origin of the $H_0^2$ scale, but rather models the dynamics that result from such a scale choice. 
It is important to mention that the model is effectively an IR modification of gravity. The earliest epoch (CMB generation at $z \approx 1100$) is expected to be \emph{indistinguishable from GR/$\Lambda$CDM}, as the non-local mass scale $m \sim H_0$ should only become dynamically relevant at low redshift.

The non-local theory contains one massive spin-2 mode and one additional massive scalar mode present as the propagating degrees of freedom with masses $\mu_{1} = - m^{2}/2 - 3\mu_{3}, \ \mu_{2} = - m^{2} - 4\mu_{3}$. This corresponds to extended FP action in linearised limit as shown in the manuscript. This is shown also in \cite{Mandal:2024nhv}. The additional scalar mode is not a ghost which is shown explicitly in our earlier work in \cite{Mandal:2024nhv}.    
In quantum field theories, ghost typically appears from higher derivative quadratic-order terms such that the propagator in momentum space contain a term with negative residue at the positive-energy pole. For example, in the case of $(\Box - m_{1}^{2})(\Box - m_{2}^{2})$ where $m_{1} > m_{2}$, the momentum space propagator can be expressed as
\begin{eqnarray}
\hspace*{-2.5cm}\frac{1}{(k^{2} + m_{1}^{2})(k^{2} + m_{2}^{2})}  = \frac{1}{m_{1}^{2} - m_{2}^{2}}\left(\frac{1}{k^{2} 
+ m_{2}^{2}} - \frac{1}{k^{2} + m_{1}^{2}}\right) & & \\
 =\frac{1}{m_{1}^{2} - m_{2}^{2}}\Big[\frac{1}{[k^{0} - \omega_{1}(\vec{k})][k^{0} + \omega_{1}(\vec{k})]}
 - \frac{1}{[k^{0} - \omega_{2}(\vec{k})][k^{0} + \omega_{2}(\vec{k})]}\Big], & & 
 \nonumber 
\end{eqnarray}
where $\omega_{i}(\vec{k}) = \sqrt{\vec{k}^{2} + m_{i}^{2}}$. Thus, the linearised analysis of quadratic order terms would be sufficient to establish the absence of ghosts.  \ref{App:Linearised gravity} and \ref{App: Anomaly} contain the detailed calcualtions.

A critical feature of the non-local theory  \ref{MT 1} is the emergence of the non-commutativity of two limits at the level of interaction :
\begin{enumerate}
\item Massless limit ($m \to 0$) where $\mu_{i} \sim m^{2}$
\item Linearization limit ($\kappa \to 0$), where the theory is expanded around Minkowski space.
\end{enumerate}
The order in which these limits are taken leads to inequivalent theories. This is explicitly shown at the level of interaction terms in these theories in the \ref{App: Anomaly}. Specifically, if the massless limit is taken first (i.e., $\mu_i \to 0$ in the full nonlinear action), the non-local terms in \ref{MT 1} vanish identically. Subsequent linearization yields the standard linearized Einstein-Hilbert action, preserving full diffeomorphism invariance at \(\mathcal{O}(\kappa)\):  
$
h_{\mu\nu} \to h_{\mu\nu} + \partial_\mu \xi_\nu + \partial_\nu \xi_\mu.  
$  
However, if one first linearizes the gravity theory (as in \ref{MT 5}–\ref{MT 8}  and then takes the $m \rightarrow 0$ limit, the outcome is different. Besides the linearized Einstein-Hilbert action, residual non-zero interaction terms stemming from the non-local part of the action persist. This follows from the fact that $\bar{\Box}h_{\mu\nu} = m^{2}h_{\mu\nu}$ and $\frac{1}{\bar{\Box}}h_{\mu\nu} = \frac{1}{m^{2}}h_{\mu\nu}$ considering the on-shell relation $k^{2} = - m^{2}$ at the interaction level as we are doing perturbative expansion around the massive gravity theory.

The full diffeomorphism invariance (the gauge symmetry of GR) is not restored in the linearized non-local action as the mass parameter $m \rightarrow 0$. This occurs because the linearization procedure and the massless limit are non-commutative operations:
$$\lim_{m \rightarrow 0} \left( \text{Linearized Action} \right) \neq \text{Linearized} \left( \lim_{m \rightarrow 0} \text{Full Action} \right)$$
Taking $m \rightarrow 0$ after linearization leaves behind a residual vDVZ (van Dam-Veltman-Zakharov) discontinuity~\cite{vanDam:1970vg,Zakharov:1970cc}. This discontinuity implies that the linearized theory of the massless field (in this case, GR) does not match the massless limit of a massive field (the non-local gravity theory considered here). The residual effect in the linearized theory is the extra degree of freedom --- the scalar mode (scalaron) --- which remains coupled to matter even as $m \rightarrow 0$. If unsuppressed, this scalar mode mediates a fifth force that would violate Solar System tests. However, the key to passing solar system tests lies in the full, non-linear theory, not the linearized one. Since $m$ is set to be of the order of the current Hubble constant, $m \sim H_0 \approx 10^{-33} \text{ eV}$, its direct influence is negligible on solar system scales. Crucially, any remaining scalar coupling must be screened due to Vainshtein mechanism. In the non-linear theory, the strong gravitational fields near massive objects non-linearly suppress the kinetic term of the scalar degree of freedom, effectively decoupling the fifth force from the matter source. {The full action contains additional interaction terms, as detailed in \ref{App:Linearised gravity} and \ref{App: Anomaly}}, which are necessary to trigger this non-linear screening mechanism. A detailed analysis confirming the screening of the scalar mode for this class of theories can be found in Ref.~\cite{Mandal:2024nhv}.

The persistence of these terms in the second scenario apparently indicates the breakdown of diffeomorphism invariance ($h_{\mu\nu} \rightarrow h_{\mu\nu} + \partial_{\mu}\xi_{\nu} + \partial_{\nu}\xi_{\mu}$) at the $\mathcal{O}(\kappa)$ level, which is not present in standard GR or in the first limiting procedure. However, the diffeomorphism symmetry is shown to be preseved in \cite{Mandal:2024nhv} once we decompose the metric perturbation in terms of scalar, vector, and tensor modes following Stueckelberg mechanism. This non-commutativity reveals a structural anomaly akin to the vDVZ discontinuity in massive gravity~\cite{vanDam:1970vg}, where $m \to 0$ limit fails to restore GR due to the persistence of scalar modes that can lead to non-unitarity in the regular perturbation theory~\cite{Hell:2022wci, Banerjee:2025fph}.  (See \ref{App: Anomaly} for details.) 


%

\section{de Sitter attractor}
\label{sec:attractor}
A crucial test for any proposed modification to GR is its ability to generate a period of late-time cosmic acceleration, thereby explaining the current expansion of the Universe without resorting to a cosmological constant or an ad-hoc dark energy component. To demonstrate this capability for the non-local action \ref{MT 1}, we derive the modified Friedmann equations for a spatially flat Friedmann-Lemaître-Robertson-Walker (FLRW) background. To do so, we first localize the action through the well-established auxiliary field method~\cite{Maggiore:2014sia,Nojiri:2007uq}. This technique, previously employed in various contexts of non-local gravity, allows for a more tractable analysis of the field equations. While this technique is well-known, our application to this specific action reveals a stable de Sitter attractor. We introduce a set of auxiliary scalar and tensor fields ($\Phi, S, \Phi_{\mu\nu}, S_{\mu\nu}$) and Lagrange multipliers ($\mathcal{A}_{1,2}, \mathcal{A}^{\mu\nu}, \mathcal{B}^{\mu\nu}$), rewriting action \ref{MT 1} as:
\begin{eqnarray}
S & =& \frac{1}{2\kappa^{2}}\int d^{4}x \ \sqrt{-g}\Big[ R - \mu_{1}RS - \mu_{2}R_{\mu\nu}
S^{\mu\nu} 
- \mathcal{A}_{1}(\Box\Phi + R)  \nonumber  \\
\label{MT 10}
 & -& \mathcal{A}_{2}(\Box S + \Phi) -  \mathcal{A}^{\mu\nu}(\Box\Phi_{\mu\nu} + R_{\mu\nu}) - \mathcal{B}^{\mu\nu}(\Box S_{\mu\nu} 
 + \Phi_{\mu\nu})\Big],
 \end{eqnarray}
Varying this action with respect to the Lagrange multipliers and auxiliary fields yields the following non-local constraints and field equations
\begin{eqnarray}
\label{MT 11}
 \hspace*{-1.8cm} S = - \frac{1}{\Box}\Phi = \frac{1}{\Box^{2}}R, \ S_{\mu\nu} = - \frac{1}{\Box}\Phi_{\mu\nu} = 
\frac{1}{\Box^{2}}R_{\mu\nu}, 
\mathcal{A}_{2}  = \mu_{1}\Phi, \ \mathcal{A}_{1} = \mu_{1}\frac{1}{\Box^{2}}R = \mu_{1}S  & & \nonumber \\
\mathcal{B}^{\mu\nu} =  \mu_{2}\Phi^{\mu\nu}, \ \mathcal{A}^{\mu\nu} = \mu_{2}\frac{1}{\Box^{2}}
R^{\mu\nu} = \mu_{2}S^{\mu\nu}.  & & 
\label{MT 12}
\end{eqnarray}
Considering the metric variation of the non-local action $S_{NL}$, we obtain
the modified Einstein field equations:
\begin{equation}\label{MT 13}
G_{\mu\nu} - \mu_{1}\mathcal{G}_{\mu\nu}^{(1)} + \mu_{2}\mathcal{G}_{\mu\nu}^{(2)} = \kappa^{2}
T_{\mu\nu},
\end{equation}
where $T_{\mu\nu}$ is the energy-momentum tensor for matter and $\mathcal{G}_{\mu\nu}^{(1,2)}$ represent the contributions from the non-local terms (their explicit forms are given in \ref{App: covariant field equations}.
%
\begin{figure}
\includegraphics[height = 12cm, width = 15cm]{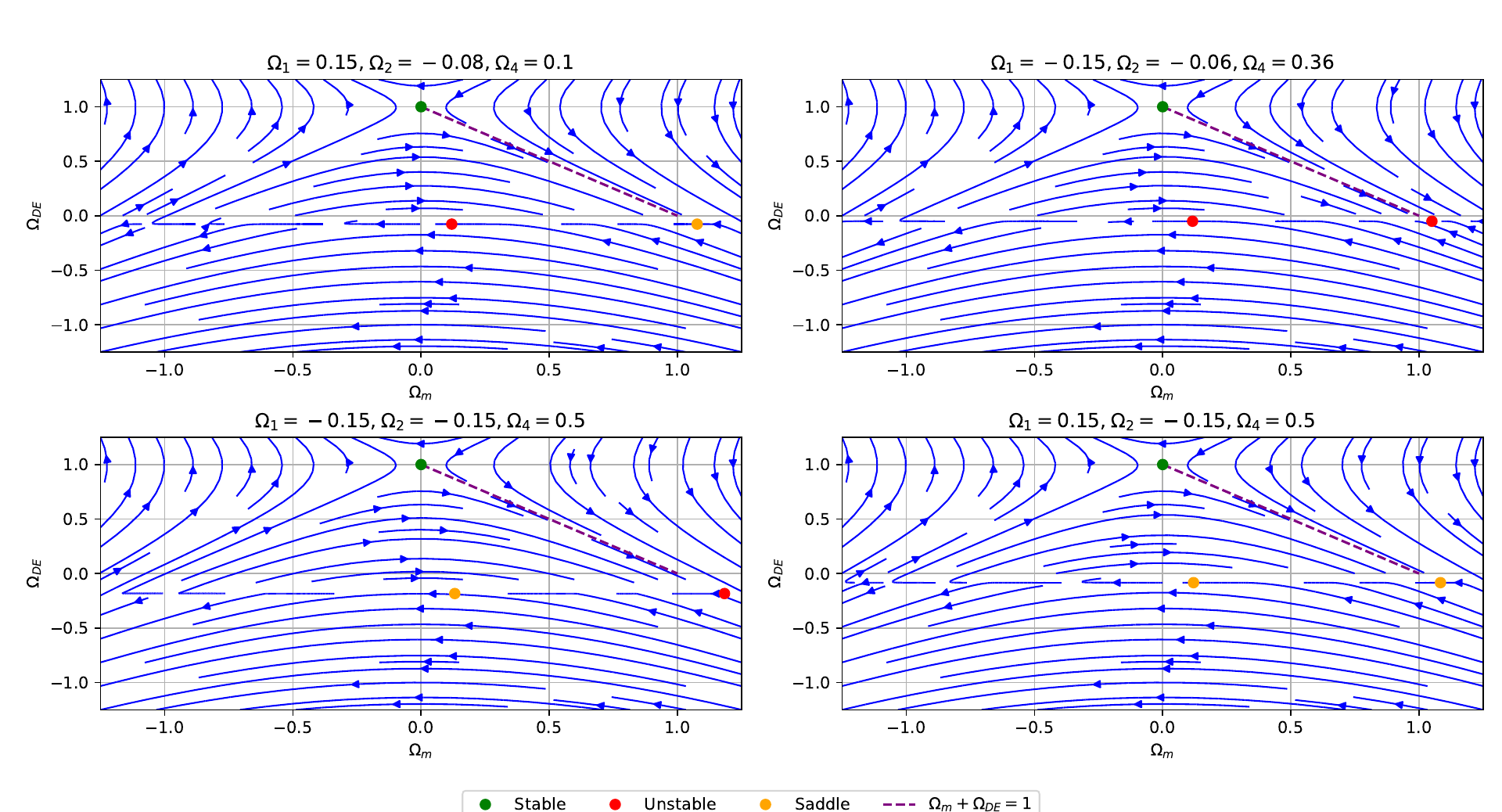}
\caption{Directional flow in the density parameter space with four initial configurations and their fixed point analysis. The above plots show universality of deSitter attractor in our non-local model with mass parameter $m$ being arbitrary and it is absorbed in the density parameters. Here directional flow is plotted in $\Omega_{m}-\Omega_{DE}$ plane and the physical configurations are located in $\Omega_{m} + \Omega_{DE} = 1$ line with $\Omega_{m}$ and $\Omega_{DE}$ lie between $0$ and $1$.}
\label{Figure 1}
\end{figure}
%
For a spatially flat FLRW background with pressureless matter and considering the simplified case where $\mu_{2} = 0$, we derive the modified Friedmann constraint equation. This equation can be expressed in terms of dimensionless density parameters:
\begin{equation}\label{MT 15a}
1 = \Omega_{m} + \Omega_{1} + \Omega_{2} + \Omega_{3} \equiv \Omega_{m} + \Omega_{DE},
\end{equation}
where $\Omega_{DE}$ represents the effective dark energy density arising from the non-local terms and are given by:
{\small
\begin{equation}
\hspace*{-0.4cm} 
\Omega_{m} = \frac{\kappa^{2}\rho}{3H^{2}(1 - 2\mu_{1}S)},~ 
\Omega_{1} = \frac{2 \mu_{1}\dot{S}}{H(1 - 2\mu_{1}S)}, \ \Omega_{2} =  \frac{\mu_{1}\Phi^{2}}{6H^{2}(1 - 2\mu_{1}S)}, \ 
\Omega_{3} = - \frac{\mu_{1}\dot{S}\dot{\Phi}}{3H^{2}(1 - 2\mu_{1}S)}
 \label{MT 15}
\end{equation}
}
where $H = \dot{a}(t)/a(t)$ is the Hubble parameter and $a(t)$ is the scale factor. By defining an additional variable, $\Omega_{4} = {12\Omega_{2}}/{\Phi}$, we obtain the following closed set of five autonomous differential equations with respect to the number of e-folds, $N = -\ln(1+z)$:
\begin{eqnarray}
{d\Omega_{i}}/{dN} = f_i(\Omega_j).
\label{MT 16}
\end{eqnarray}
(See \ref{app.Background field equations} and \ref{app.background field equations 2} for the complete expressions.) This system fully describes the cosmic dynamics of the model. To analyze the future evolution of the universe, we study the system for $N \to \infty$ (i.e., forward in cosmic time). The complete evolutionary history predicted by this model is best visualized through a \emph{phase space analysis}, with the results for the $\mu_2=0$ case depicted in \ref{Figure 1}. This figure plots the cosmic trajectories within a projection of the multi-dimensional phase space. Each streamline represents a possible evolutionary path for the universe, starting from different initial conditions in the early cosmos. As shown, regardless of their origin, all trajectories converge towards a single, stable critical point. This point is the \emph{global attractor} of the dynamical system, signifying the ultimate fate of the universe in this model. At this attractor, the matter density parameter vanishes ($\Omega_m \to 0$) while the effective dark energy density dominates completely ($\Omega_{DE} \to 1$). This state corresponds to a de Sitter universe, driving a period of eternal, accelerated expansion. The existence and stability of this fixed point are not assumed but are direct consequences of the theory's structure, proving that the model naturally culminates in a late-time accelerating phase.
\begin{center}
\begin{figure}
\includegraphics[height = 12cm, width = 15cm]{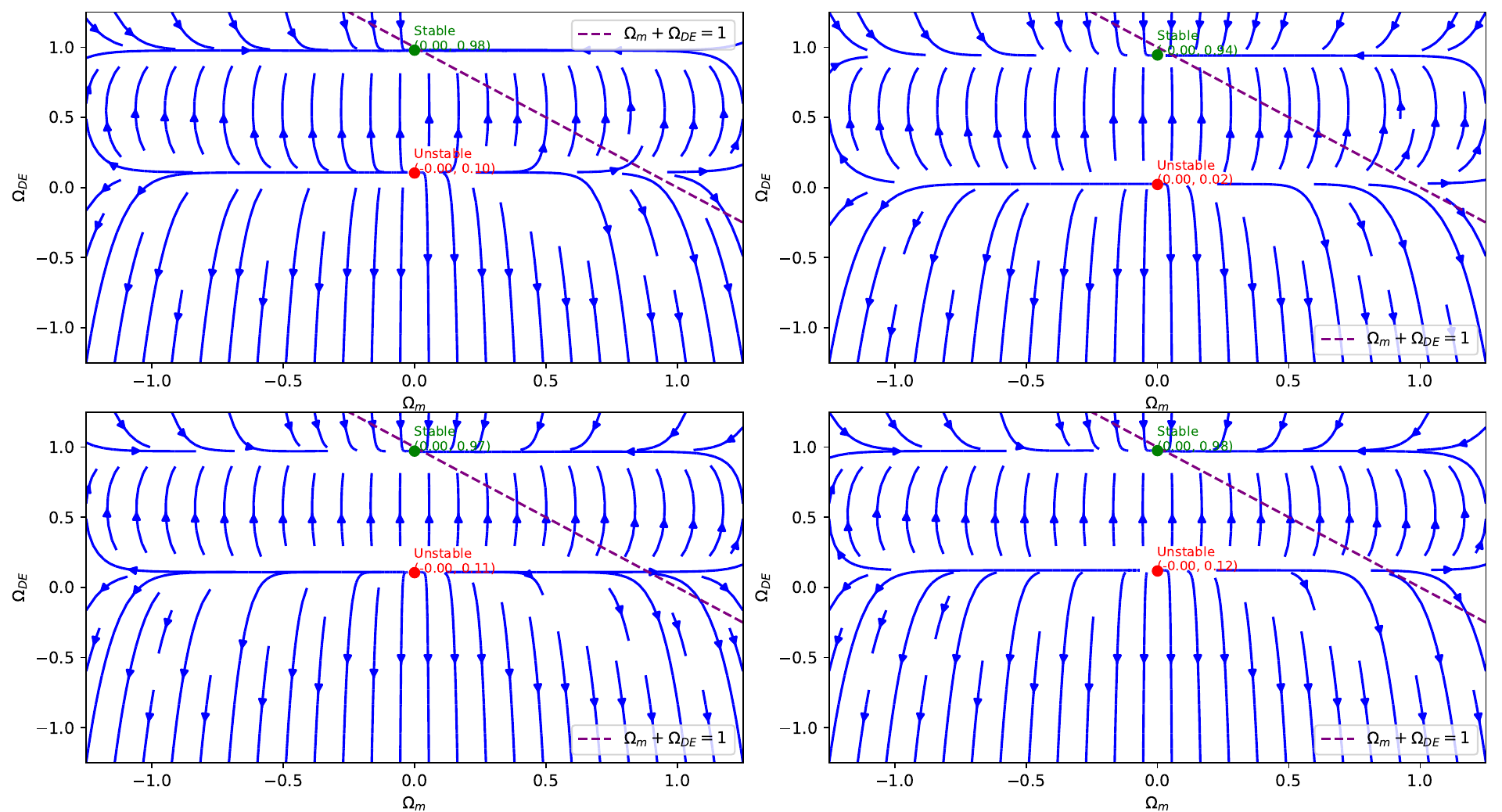}
\caption{Directional flow in the density parameter space when both $\mu_{1}, \mu_{2} \neq 0$, related to mass parameter $m$ and the fixed point analysis. Like earlier, here directional flow is also plotted in $\Omega_{m}-\Omega_{DE}$ plane and the physical configurations are located in $\Omega_{m} + \Omega_{DE} = 1$ line with $\Omega_{m}$ and $\Omega_{DE}$ lie between $0$ and $1$.}
\label{Figure 2}
\end{figure}
\end{center}
To ensure this result is not an artifact of our simplification, we extend our analysis to the more general case where $\mu_{2} \neq 0$. This introduces additional terms and variables, increasing the dimensionality of the autonomous system. The detailed derivation, presented in the \ref{app.background field equations 2}, again yields a closed set of coupled differential equations. We perform the same fixed-point analysis on this more complex system, with the corresponding phase space portrait shown in \ref{Figure 2}. Remarkably, the qualitative behavior remains the same. The figure clearly demonstrates that even with the inclusion of the $R_{\mu\nu} {\Box^{-2}} R^{\mu\nu}$ term, the system is once again dominated by a \emph{stable late-time attractor}. This fixed point is also a de Sitter solution, confirming that the mechanism driving cosmic acceleration is a \emph{robust and generic feature} of the non-local action in \ref{MT 1}.

Before proceeding, it is crucial to compare our results with the existing literature, particularly the analysis by Cusin et al~\cite{Cusin:2015rex}. They investigated a non-local action incorporating a Weyl-squared term ($C_{\mu\nu\rho\sigma}\Box^{-2}C^{\mu\nu\rho\sigma}$) and concluded that the term analogous to our $\mu_2$ term leads to cosmologically \emph{non-viable scenarios}. Non-local Weyl terms (as explored by Cusin et al. in \cite{Cusin:2015rex}) primarily affect the tensor sector due to conformal invariance and typically fail to introduce the scalar dynamics needed to drive homogeneous cosmic acceleration and sustain a stable de Sitter phase. 
Our use of a non-local Riemann term (which inherently contains the Ricci scalar $R$) successfully introduces a dynamical scalar degree of freedom.
In Ref.~\cite{Cusin:2015rex}, the existence of growing mode solution in the case of $R_{\mu\nu}\frac{1}{\Box^{2}}R^{\mu\nu}$ is claimed. The mode solution characterize by $V = U_{ \ 0}^{0} - U_{ \ i}^{i}/3$ according to the notation followed by the authors in Ref.~\cite{Cusin:2015rex} satisfies the equation
\begin{equation}
\Box V = - (R_{ \ 0}^{0} - R_{ \ i}^{i}/3) = - 2\dot{H}
\end{equation}
since $R_{00} = - 3\frac{\ddot{a}}{a}, \ R_{ij} = \left(\frac{\ddot{a}}{a} + 2\frac{\dot{a}^{2}}{a^{2}}\right)g_{ij}$. Hence, the above equation can be expressed as
\begin{equation}
\ddot{V} + 3H\dot{V} = - 2\dot{H}.
\end{equation}
The homogeneous part of the above equation is indeed decaying.
In contrast, our work, which includes both the $R\Box^{-2}R$ and the $R_{\mu\nu}\Box^{-2}R^{\mu\nu}$ terms, yields a viable cosmology that successfully explains the late-time acceleration of the Universe. We have demonstrated this explicitly through a phase-space analysis of the background equations. A key difference arises because our action generates an additional term in the background Friedmann equations which is absent in the case of Cusin et al~\cite{Cusin:2015rex}.

Our results indicate that the inclusion of the non-local Ricci-squared term allows the theory to reduce to a consistent extended FP gravity. Specifically, for the $\alpha = 1/2$ case, de Sitter spacetime emerges as a stable attractor. This outcome confirms that our non-local theory, based on the Ricci and Ricci scalar terms, is fundamentally distinct from the Weyl-based model in \cite{Cusin:2015rex}. Although the Weyl and Riemann tensors are linearly related, the presence of non-local operators ($\Box^{-2}$) breaks this equivalence at the level of the action, leading to different physical consequences.

Our primary advantage lies in the model's structural design: it is the \emph{IR completion of massive gravity} which dynamically yields the $\omega \rightarrow -1$ attractor while addressing the vDVZ and Boulware-Deser issues via its specific non-local construction.
The difference in viability between Ricci-based (our model) and Weyl-based (e.g., the Cusin et al~\cite{Cusin:2015rex} model based on the Weyl tensor, $C_{\mu\nu\rho\sigma}$) non-locality is subtle and directly relates to the underlying propagating degrees of freedom (DOFs):\\[5pt]
{\bf Ricci-based Non-Locality (Viable):} Our non-local operator is built using the Ricci scalar ($R$) or Ricci tensor ($R_{\mu\nu}$). Since $R$ and $R_{\mu\nu}$ are tied to the dynamics of the spin-2 graviton and its scalar modes, non-local deformations based on these terms, when carefully constructed, remain within the extended FP/dRGT family. As demonstrated by our stability analysis, this structure can successfully maintain the required two tensor + two vector + one or two scalar healthy DOFs, hence ensuring ghost-free behavior.\\[5pt]
{\bf Weyl-based Non-Locality (Often Non-Viable):} The Weyl tensor ($C_{\mu\nu\rho\sigma}$) is constructed to be trace-free and is only sensitive to the purely tensorial/conformal components of the curvature. Non-local modifications based on $C^2$ terms generally introduce a massive spin-2 ghost (a component with negative kinetic energy) or higher-order derivatives (Ostrogradsky ghosts). In the specific non-local models~\cite{Cusin:2015rex}, the structure tends to introduce Ostrogradsky instabilities because the Weyl terms alter the high-frequency propagation modes in a way that is difficult to regularize while maintaining a healthy canonical structure.\\[5pt]
Hence, our choice of Ricci-based non-locality is necessary because it allows us to deform the Einstein-Hilbert action in a way that is consistent with the known restrictions for ghost-free massive gravity, a constraint which Weyl-based non-localities typically violate.


In summary, our analysis reveals the existence of a crucial late-time \emph{attractor solution} across different configurations of the model. This fixed point corresponds to a de Sitter-like phase of accelerated expansion, characterized by an effective equation of state parameter, $\omega_{eff}$, that can be arbitrarily close to $-1$. The stability of this solution is the central result; it demonstrates that the accelerated expansion is a natural and inevitable endpoint of the cosmic evolution within this framework, not a feature that requires finely-tuned initial conditions. Thus, the fixed-point analysis provides robust evidence that this non-local gravity model offers a \emph{dynamically consistent} and viable mechanism to drive the late-time acceleration of the Universe, establishing it as a compelling alternative to the standard $\Lambda$CDM paradigm.

\section{Stability beyond the FLRW background}
\label{sec:Stability}
%
While our analysis has revealed the existence of a crucial late-time attractor solution, establishing the model as a viable mechanism for cosmic acceleration, this is only a necessary first condition. The history of modified gravity is replete with theories, a prominent example being early versions of Hořava-Lifshitz gravity, that exhibit a perfectly well-behaved and stable FLRW background evolution, yet are plagued by catastrophic instabilities at the level of linear perturbations~\cite{Wang:2009yz,Charmousis:2009tc}. Such instabilities, like the presence of ghosts or modes that grow uncontrollably, would render the smooth background physically unattainable, as any infinitesimal fluctuation would be amplified. Therefore, a rigorous investigation into the stability of cosmological perturbations is not merely a consistency check but a fundamental test of the theory's physical viability.

To address this, we must investigate whether our non-local model \ref{MT 1} suffers from similar pathologies. To go about this we consider the tensor perturbations in our model. As mentioned earlier, a key feature of our theory is that it propagates \emph{six degrees of freedom}, a direct consequence of the non-local terms which modify the kinetic structure of the graviton. Unlike in GR, where the scalar, vector, and tensor perturbations decouple at the linear level, the non-local operators in our action, such as the inverse d'Alembertian ($\Box^{-2}$), inherently mix these sectors. This has a drastic effect on the tensor perturbation equation. When we analyze the evolution of the spatial metric perturbation, $\mathcal{H}_{ij}$, we find that it is \emph{neither traceless ($\mathcal{H} \equiv \mathcal{H}^i_i \neq 0$) nor divergenceless ($\nabla^j \mathcal{H}_{ij} \neq 0$)}

This breakdown of the standard transverse-traceless condition is a direct signature of the additional propagating modes: The \emph{non-zero trace} ($\mathcal{H} \neq 0$) corresponds to a propagating scalar \emph{breathing mode}, which causes an isotropic expansion and contraction of spacetime, a polarization absent in GR. The \emph{non-vanishing divergence} indicates the presence of two additional \emph{vector modes}, which would correspond to shear-like distortions with a preferred direction. (Details of the full tensor perturbation can be found in \ref{Tensor perturbations}). The trace of the metric perturbation in \emph{long-wavelength limit} is given below:
{\small
\begin{equation}\label{trace-part}
\begin{split}
&- (1 - 2\mu_{1}S)\Big[2H\dot{\mathcal{H}} + \frac{1}{2}\ddot{\mathcal{H}}\Big] 
+ \mu_{2}\Big[- 3(\psi_{2} - \psi_{1})H\dot{\mathcal{H}} + \dot{\mathcal{H}}\dot{\psi}_{2} 
+ \frac{H\mathcal{H}}{2} \left( 5\dot{\psi}_{2} - 3 \dot{\psi}_{1} \right) \\
&+ \frac{3}{4}H\dot{\mathcal{H}}(5\psi_{2} - 3\psi_{1}) 
+ \frac{1}{4}\ddot{\mathcal{H}}(3\psi_{1} - \psi_{2}) 
+ \frac{\delta S_{kk}}{a^{2}}\left( \dot{H} - \frac{3H^{2}}{2}\right) 
+ \frac{3\varphi_{1} - \varphi_{2}}{2}\frac{\delta \Phi_{kk}}{a^{2}} \\
&- \frac{\delta\Phi_{kk}}{a^{2}}H(3\dot{\psi}_{1} - \dot{\psi}_{2}) 
- \frac{\delta S_{kk}}{a^{2}}H(3\dot{\varphi}_{1} - \dot{\varphi}_{2}) 
+ 2H^{2}\frac{\delta S_{kk}}{a^{2}}(4\varphi_{2} - 9\varphi_{1}) 
+ 2H^{2}\frac{\delta\Phi_{kk}}{a^{2}}(4\psi_{2} - 9\psi_{1}) \\
&+ H (3\varphi_{1} - \varphi_{2}) \frac{\dot{\delta S}_{kk}}{a^{2}} 
+ H(3\psi_{1} - \psi_{2})\frac{\dot{\delta\Phi}_{kk}}{a^{2}} 
- 20H\psi_{2}\varphi_{2}\dot{\mathcal{H}} 
+ H^{2}\mathcal{H}(3\psi_{1} - \psi_{2}) \\
&+ \frac{1}{4}\dot{\mathcal{H}}(3\dot{\psi}_{1} - \dot{\psi}_{2}) 
+ \dot{\mathcal{H}}\Big[ \frac{9}{2}(\varphi_{2} - \varphi_{1})(\dot{\psi}_{2} - \dot{\psi}_{1}) 
+ \frac{9}{2}(\psi_{2} - \psi_{1})(\dot{\varphi}_{2} - \dot{\varphi}_{1}) 
- 2\varphi_{2}\dot{\psi}_{2} - 2\dot{\varphi}_{2}\psi_{2}\Big] \\
&+ 2\dot{\psi}_{2}\dot{\mathcal{H}}(3\varphi_{1} - 2\varphi_{2}) 
+ 2\dot{\varphi}_{2}\dot{\mathcal{H}}(3\psi_{1} - 2\psi_{2}) \Big] \Big] = 0.
\end{split}
\end{equation} 
}
where $\varphi_{1}, \varphi_{2}, \psi_1, \psi_2$ are components of the auxiliary tensor field variables $\Phi_{\mu\nu}$ and $S_{\mu\nu}$ in Eq.~\ref{MT 11}
\begin{equation}\label{S-E5}
\Phi_{\mu\nu} = g_{\mu\nu}(3\varphi_{1} - \varphi_{2}) + 2\varphi_{2}\delta^0_{\mu}\, \delta^0_{\nu} , \quad 
 S_{\mu\nu} = g_{\mu\nu}(3\psi_{1} - \psi_{2}) + 2\psi_{2} \delta^0_{\mu}\delta^0_{\nu}.
\end{equation}

To demonstrate that the above tensor perturbations do not introduce any ghost degrees of freedom, we consider two limits: the inherent ghost-free nature of its linearized limit and the dynamical stability of perturbations on the de Sitter background.

First, as previously demonstrated, the class of non-local theories under consideration reduces to the FP and extended FP actions in the linearized gravity limit around Minkowski space, contingent upon the appropriate selection of the parameters $\mu_1$, $\mu_2$, and $\mu_3$. These specific theories are constructed to be devoid of any ghost degrees of freedom—that is, they are free from states with negative kinetic energy that would lead to a non-unitary quantum theory and vacuum instability. This ghost-free nature has been recently established in~\cite{Mandal:2024nhv}, confirming that the fundamental building blocks of our model are theoretically sound.

Second, and more critically for cosmology, we must analyze the dynamical behavior of perturbations evolving on the accelerating background. A stable background must be stable against fluctuations. We find that in the vicinity of the de Sitter attractor, tensor perturbations with large wavelengths decay exponentially with cosmic time $t$ in the asymptotic limit ($t \to \infty$). This behavior is a direct consequence of the background evolution equations governing the auxiliary fields, which simplify dramatically near the attractor point. As detailed in \ref{app.background field equations 2}, $\varphi_1$, $\varphi_2$, and $\psi_2$ vanish on the attractor. For instance, their evolution equations are:
\begin{equation}
\!\! \ddot{\psi}_1 + 3H\dot{\psi}_1 = \varphi_1 + 4H^2\psi_2,~ 
\ddot{\psi}_2 + 3H\dot{\psi}_2 = \varphi_2 + 8H^2\psi_2.
\end{equation}
The vanishing of the source terms and $\psi_2$ on the right-hand side is key. This simplification propagates to the equations for the metric perturbations themselves. For instance, the evolution equation for the trace part of the tensor perturbation, $\mathcal{H}$, reduces from its full form in  \ref{trace-part} to a damped equation:
\begin{equation}\label{reduced eqn}
\ddot{\mathcal{H}} + 4H\dot{\mathcal{H}} = 0.
\end{equation}
Given that $H$ is constant on the de Sitter attractor, the solutions to this equation are of the form $\mathcal{H}(t) \propto e^{-4Ht}$, demonstrating a clear exponential decay. This shows that the trace or (scalar "breathing") mode of the tensor perturbation is not only stable but is actively suppressed by the cosmic expansion. As detailed in \ref{app.background field equations 2}, a similar decaying behavior is found for the other tensor components. This exponential damping of all long-wavelength modes is the definitive signature of a stable system. Unlike unstable models where perturbations would grow exponentially, our model ensures that the de Sitter background is the true, robust endpoint of evolution.

The evolution of the scalar mode denoted ($\mathcal{H}$) is given in \ref{S-F23}) of \ref{Tensor perturbations} which in long wavelength limit reduces to the form given in \ref{reduced eqn}) which shows the decaying nature of scalar mode. On the other other hand, from the same \ref{S-F23} of \ref{Tensor perturbations}, we can see that there will be no ghosts associated with scalar mode as long as the coefficient $(1 - 2\mu_{1}S) > 0$ of $\ddot{\mathcal{H}}$ in that equation. We also have two other equations:
$$\Box S = - \Phi \implies \ddot{S} + 3H\dot{S} = \Phi$$ 
whereas $\Phi$ satisfies the equation $\Box\Phi + 7\mu_{1}\Phi + 2\mu_{1}\Phi^{2} = \kappa^{2}T$ which is derived in \ref{App: covariant field equations}. The above equation shows that near de Sitter (in which case $T = 0$ (traceless condition of matter energy-momentum tensor)), $\Phi$ has a right kinetic energy sign as $\mu_{1} \propto - m^{2}$ and also $\Phi$ is decaying asymptotically due to $m^{2}$ term. As a result of which $S$ will also decay asymptotically and $m \sim H_{0}$ and hence $|2\mu_{1}S| \ll 1$. This shows the absence of ghosts in the case of scalar perturbation modes in de Sitter.

\section{Observational implications}
\label{sec:Obs}
The primary finding of this work is that the non-local action admits a stable de Sitter attractor at late times. A pure dS solution corresponds to a constant energy density and an effective equation of state $\omega_{\text{eff}} = -1$. This value is perfectly consistent with the latest results from Planck~\cite{2018-Planck-AA} and SNIa~\cite{2016-Riess.others-Astrophys.J.} observations, which constrain dark energy to be very close to $\omega = -1$ (e.g., $\omega = -1.03 \pm 0.03$). Unlike the $\Lambda$CDM model, which has $\omega = -1$ by construction, our model achieves this state dynamically. We will emphasize that this late-time dS behavior demonstrates compatibility with the most fundamental constraints on cosmic acceleration. Our model achieves cosmic acceleration through a dynamical scalar mode (scalaron) introduced by the non-local operator (since we use a Ricci-based term). This scalar field is weakly coupled to matter and affects the growth of cosmic structures, which is tightly constrained by Planck (CMB lensing)~\cite{Carron:2022eyg} and DESI (clustering/redshift surveys)~\cite{DESI:2025fii,DESI:2025zgx,2025-Gu.others-}.

The ratio of the growth rate of matter perturbations in our model ($G_{\text{mod}}$) compared to GR ($G_{\text{GR}}$) is the key observable. If the scalar mode is unsuppressed, it enhances the gravitational force, leading to a larger $\sigma_8$ value than observed ($\sigma_8^{\text{Planck}}$)~\cite{2025-KiDS}.
The consistency of our theory with the Vainshtein mechanism is vital here. In order for the model to pass DESI and BAO constraints, the non-local terms (and their associated scalar mode) must be sufficiently suppressed at scales smaller than the horizon, mitigating any excessive enhancement of $G_{\text{mod}}$. This indirectly constrains the value and scale dependence of the non-local mass parameter $m$. \ref{tab:obs_predictions} contains the theoretical predictions for observational parameters.  
%

\begin{table*}[h!]
\centering
\caption{Theoretical Predictions for Observational Parameters}
\label{tab:obs_predictions}
\begin{tabular}{|p{1.5cm}|p{2.0cm}|p{2.75cm}|p{5.5cm}|}
\hline
\textbf{Parameter} & \textbf{$\Lambda$CDM Value} & \textbf{Non-Local Model Prediction} & \textbf{Observational Relevance} \\
\hline
$w(z)$ & $w(z) = -1$ (Constant) & \textbf{$w(z) \rightarrow -1$} at late times ($z \approx 0$). & Dynamically matches tight $\mathbf{SNIa}$ and $\mathbf{CMB}$ constraints ($w_0 \approx -1$). \\
\hline
$\Omega_m^0$ (current matter density) & $\Omega_m^0 \approx 0.31$ & Need to be consistent with $\Omega_m^0 \approx 0.31$. & The stable attractor requires non-local parameters to yield the observed density split at $z=0$. \\
\hline
$H_0$ & $H_0 \approx 67.4 \text{ km/s/Mpc}$ (Planck) & $H_0$ is a free parameter fixed by $m$. & Offers potential flexibility to address the $\mathbf{H_0}$ tension, depending on intermediate redshift dynamics. \\
\hline
\end{tabular}
\end{table*}

As mentioned above, the non-local nature of our action and its equivalence to extended FP gravity ensures the presence of more than the two standard transverse-traceless (tensor) polarizations of GR. A consistent massive spin-2 field theory typically propagates five or six degrees of freedom (DOFs), including two scalar and two vector modes in addition to the two GR tensor modes. The prospects for detecting these non-GR polarizations are tied directly to the frequency band and the geometric setup of current and upcoming detectors. In the case of the scalar and vector modes, our model's viability depends on these non-GR modes either being strongly suppressed or having a very specific, sub-dominant amplitude to pass current Pulsar Timing Array (PTA) bounds~\cite{Allen:2023kib}.

PTAs are currently placing the strongest constraints on the stochastic gravitational wave background (SGWB)~\cite{NANOGrav:2020spf,NANOGrav:2023gor}. These arrays are highly effective at distinguishing the correlation pattern generated by scalar and vector modes from the signature of GR's tensor modes (the Hellings-Downs curve). In the case of tensor modes ($h_{+}, h_{\times}$), our non-local model must align with the extremely tight bounds on the speed of gravity ($\left|c_g / c - 1\right| < 10^{-15}$) set by the GW170817/GRB 170817A event, which constrains many modified gravity theories~\cite{LIGOScientific:2017adf,Ezquiaga:2017ekz}. LIGO, Virgo, KAGRA are sensitive primarily to the two GR tensor modes. A deviation from GR would manifest as a slight difference in the measured speed, amplitude, or luminosity distance of these modes. 
On the other hand, LISA (Laser Interferometer Space Antenna) offers the best prospect for definitively testing these additional polarizations~\cite{LISA:2022yao}.  Since the graviton mass $m$ is typically very small ($m \sim H_0$), the characteristic long-wavelength effects of massive gravity are often best probed in the low-frequency regime where LISA operates. LISA's triangular, space-based configuration allows it to measure the six distinct strain components necessary to algebraically decompose the GW signal into its tensor, vector, and scalar parts. This is a significant advantage over L-shaped ground detectors.

\section{Conclusions}
\label{Sec:Conclusions}
In conclusion, this work establishes a class of non-local gravity models as a compelling unified framework for physics beyond GR. It provides a dynamical solution to the problem of cosmic acceleration and potentially  predicts a wealth of new gravitational wave signatures. The model’s predictive power—yielding a stable de Sitter attractor and distinct GW polarizations—sets it apart from ad-hoc dark energy proposals. 
The ghost-free nature is expected to persist when perturbing the theory around the de Sitter background. Our linearized analysis around Minkowski space is rooted in the structure of extended FP gravity, which is known to be ghost-free. When extending such theories to curved backgrounds, the crucial challenge is ensuring stability. For the most established ghost-free massive gravity theories (like dRGT), the ghost-free nature typically holds on maximally symmetric backgrounds, including dS, provided the background mass and cosmological parameters satisfy certain stability conditions (e.g., related to the Higuchi bound \cite{Fasiello:2012rw}). Specific analyses confirming this persistence for massive gravity in dS can be found in the literature (e.g., \cite{Mazuet:2015pea}; \cite{Alberte:2011ah}). Confirming the sign of the kinetic terms for all propagating DOFs on the dS background requires a highly technical quadratic action analysis in the curved background, which is beyond the scope of this paper. However, based on the literature on massive gravity in curved spacetime, we confirm the following:
\begin{enumerate}
\item The consistency of the theory, which relies on our choice of the mass-deformation parameter $\alpha$ (related to the FP structure), ensures positive kinetic terms for all propagating DOFs around the Minkowski vacuum.
\item We rely on the consistency established by our extended FP gravity structure to assert that this stability extends to the dS attractor. The background is physically viable provided the mass parameter $m^2$ (or its non-local equivalent) remains positive on the background, which is necessary for the existence of the dS solution itself.
\end{enumerate}
We are confident that the parameters of our non-local action fall within the regime that guarantees ghost-free evolution, providing a strong theoretical basis for the viability of our dS attractor. We believe our current theoretical results lay crucial groundwork for subsequent observational studies and provide comprehensive confrontation with current cosmological bounds (e.g., from DESI, Planck, SNIa, BAO).

Our claim of ghost-free behavior at the linearized level relies on the action being equivalent to an \emph{extended FP theory}. The standard method to ensure this property persists in the full non-linear regime is to draw a direct analogy with \emph{ghost-free massive gravity} (dRGT).
The dRGT action was painstakingly constructed to eliminate the non-linear Boulware-Deser (BD) ghosts by ensuring that the Hamiltonian constraint structure remains the same as in GR, thus guaranteeing only the healthy propagating degrees of freedom (5 or 6 DOFs). For our non-local action, the ghost-free nature in the non-linear regime is confirmed \emph{if and only if} the introduction of the non-local operator, when rewritten via auxiliary fields, does not introduce new terms that violate the \emph{non-linear constraint structure} necessary to eliminate the sixth degree of freedom (the BD ghost).
While a full non-linear Hamiltonian analysis is required for definitive proof and is outside the scope of this paper, the key physics condition is: The effective scalar field (the "scalaron" introduced by the non-local operator) must maintain a positive definite kinetic term in the full non-linear action.
The relationship provided by the modified field equation, $G_{\mu\nu} = 8\pi G T_{\mu\nu}^{\text{eff}}$, can indeed offer indirect checks, particularly concerning the stability of the cosmological background.
The ghost-free condition for the full metric and auxiliary fields translates into a condition on the effective theory: The condition that $T_{\mu\nu}^{\text{eff}}u^{\mu}u^{\nu} \geq 0$ (for $u^\mu = (1, 0, 0, 0)$) is an energy condition check ($\rho_{\text{eff}} \geq 0$), which is necessary but not sufficient to rule out the propagating ghost. However, if the effective densities $\Omega_1$, $\Omega_2$, and $\Omega_3$ (which contain the dynamics of $S$ and $\Phi$) arise from a theory that is consistent with the non-linear FP structure, then the positivity of the sum ($\Omega_1 + \Omega_2 + \Omega_3 > 0$) indicates that the effective energy density driving the cosmology is physically consistent.
Crucially, by being fundamentally ghost-free and by ensuring that physical perturbations are dynamically damped on the accelerating background, this class of theories avoids the critical instabilities that have invalidated other proposed modifications to GR. These characteristics, combined with its ability to drive cosmic acceleration, render this model a physically consistent and viable alternative to GR at cosmological scales.

Our Ricci-based non-local model achieves late-time acceleration and screenability, similar to $f(R)$~\cite{Starobinsky:2007hu,Hu:2007nk,Johnson:2019vwi}, Galileon, and Horndeski theories~\cite{Ezquiaga:2017ekz,Bansal:2024bbb,Bansal:2025usn} but employs a distinct mechanism to address the \emph{Cosmological Constant Problem (CCP) and maintain viability}. The table below provides crucial difference between our model and the other modified gravity theory models.
\begin{table*}[h!]
\centering
\caption{Comparison with Major Infrared Modifications of General Relativity}
\label{tab:ir_comparison}
\begin{tabular}{|p{1.10cm}|p{3.0cm}|p{3.5cm}|p{4.0cm}|}
\hline
\textbf{Theory} & \textbf{Mechanism for Acceleration} & \textbf{Primary Viability Challenge} & \textbf{Our Non-Local Advantage} \\
\hline
$f(R)$ & Curvature dependent scalar degree of freedom. 
& Dolgov-Kawasaki instability~\cite{Dolgov:2003px} and adherence to local tests (Vainshtein screening)~\cite{Vainshtein:1972sx}. 
& Mechanism explicitly ties acceleration scale to the non-local scale 
($\sim H_0$). Stability guaranteed by {FP consistency}. \\
\hline
Galileon /Horndeski & Kinetic self-coupling 
 protects the scalar field and implements screening. & Finding a specific form that survives all constraints, especially $c_g=c$ from GW170817~\cite{Ezquiaga:2017ekz}. & Fundamentally a massive gravity extension, not a scalar-tensor theory. Non-local mass term drives late-time acceleration. \\
\hline
\end{tabular}
\end{table*}

\section*{Acknowledgments} 
The authors thank I. Chakraborty, P. G. Christopher, K. Hari, A.Kushwaha, and T. Parvez 
for comments and discussions. SM was supported by the SERB-Core Research Grant (Project SERB/CRG/2022/002348). 

\appendix

\section{Linearised non-local gravity}\label{App:Linearised gravity}

In this section, we discuss the linearised gravity theory of a class of non-local
theory of gravity \cite{Deser:2007jk, Deser:2013uya, Dodelson:2013sma, Maggiore:2014sia, Capozziello:2021krv, Modesto:2013jea} described by the following action \cite{Cusin:2015rex}
\begin{equation}\label{S-A1}
S_{\text{non-local}} = \frac{1}{2\kappa^{2}}\int d^{4}x \ \sqrt{-g}\Big[R + \mu_{1}
R\frac{1}{\Box^{2}}R + \mu_{2}R^{\mu\nu}\frac{1}{\Box^{2}}R_{\mu\nu} + \mu_{3}
R^{\mu\nu\rho\sigma}\frac{1}{\Box^{2}}R_{\mu\nu\rho\sigma}\Big],
\end{equation}
around the Minkowski spacetime. In the above theory of gravity, $\mu_{1}, \mu_{2}, \mu_{3}$
are coupling constants of dimension $[\text{mass}]^{2}$. Here $\Box = g^{\mu\nu}\nabla_{\mu}
\nabla_{\nu}$. It is important here to emphasis that the above action can also be expressed
as
\begin{equation}\label{S-A2}
S_{\text{non-local}} = \frac{1}{2\kappa^{2}}\int d^{4}x \ \sqrt{-g}\Big[R + \mu_{1}
R\frac{1}{\Box^{2}}R + \mu_{2}R_{\mu\nu}\frac{1}{\Box^{2}}R^{\mu\nu} + \mu_{3}
R_{\mu\nu\rho\sigma}\frac{1}{\Box^{2}}R^{\mu\nu\rho\sigma}\Big],
\end{equation}
where we changed the ordering of lower and upper indices despite $\Box^{2}$ operator is on
the denominator. This follows from the following integral representation of the $\frac{1}
{\Box^{2}}$ operator
\begin{equation}\label{S-A3}
\frac{1}{\Box^{2}} = \int_{0}^{\infty}ds \ e^{ - s\Box^{2}},
\end{equation}
which is a convergent integral as the eigenvalues of $\Box^{2}$ operator is positive-definite.
Now, we start with the following decomposition of metric 
\begin{equation}\label{S-A4}
g_{\mu\nu} = \eta_{\mu\nu} + 2\kappa h_{\mu\nu},
\end{equation}
where $\kappa = \sqrt{8\pi G}$ and $h_{\mu\nu}$ is the metric perturbation around the Minkowski
spacetime. The inverse metric can be expressed as
\begin{equation}\label{S-A5}
g^{\mu\nu} = \eta^{\mu\nu} - 2\kappa h^{\mu\nu} + \mathcal{O}(\kappa^{2}).
\end{equation}
Moreover, we also obtain the following relation
\begin{equation}\label{S-A6}
\sqrt{-g} = 1 + \kappa h + \mathcal{O}(\kappa^{2}),
\end{equation}
where $h = \eta^{\mu\nu}h_{\mu\nu}$. Hence, the expression of the Christoffel symbol
up to first order can be expressed as
\begin{equation}\label{S-A7}
\Gamma_{ \ \mu\nu}^{(1)\rho} = \kappa(\partial_{\nu}h_{\mu}^{ \ \rho} + \partial_{\mu}
h_{\nu}^{ \ \rho} - \partial^{\rho}h_{\mu\nu}),
\end{equation}
as a result of which, we obtain the following expression of Ricci tensor up to first 
order
\begin{equation}\label{S-A8}
R_{\mu\nu}^{(1)} = \kappa (\partial_{\rho}\partial_{\mu}h_{\nu}^{ \ \rho} + \partial_{\nu}
\partial^{\rho}h_{\mu\rho} - \partial_{\nu}\partial_{\mu}h - \bar{\Box}h_{\mu\nu}),
\end{equation}
where $\bar{\Box} = \eta^{\mu\nu}\partial_{\mu}\partial_{\nu}$. In order to obtain the 
second-order term for the Ricci tensor, we now use the following form of the second-order 
Christoffel symbols
\begin{equation}\label{S-A9}
\Gamma_{ \ \mu\nu}^{(2)\rho} = - 2\kappa^{2}h^{\rho\gamma}(\partial_{\nu}h_{\mu\gamma} + 
\partial_{\mu}h_{\nu\gamma} - \partial_{\gamma}h_{\mu\nu}).
\end{equation}
Using the above expression, we obtain the following expression of $R_{\mu\nu}^{(2)}$ which
we have derived in \cite{Mandal:2024nhv}
\begin{equation}\label{S-A10}
\begin{split}
R_{\mu\nu}^{(2)} & = - 2\kappa^{2}\Big[\partial_{\rho}h^{\rho\gamma}(\partial_{\nu}h_{\mu
\gamma} + \partial_{\mu}h_{\nu\gamma} - \partial_{\gamma}h_{\mu\nu})\\
 & + h^{\rho\gamma}(\partial_{\rho}\partial_{\nu}h_{\mu\gamma} + \partial_{\rho}\partial_{\mu}
 h_{\nu\gamma} - \partial_{\rho}\partial_{\gamma}h_{\mu\nu} - \partial_{\nu}\partial_{\mu}
 h_{\rho\gamma})\\
 & - \frac{1}{2}\partial_{\nu}h^{\rho\gamma}\partial_{\mu}h_{\rho\gamma} - \frac{1}{2}
 \partial_{\gamma}h (\partial_{\mu}h_{\nu}^{ \ \gamma} + \partial_{\nu}h_{\mu}^{ \ \gamma} 
 - \partial^{\gamma}h_{\mu\nu})
 + \partial_{\gamma}h_{\nu}^{ \ \rho}(\partial_{\rho}h_{\mu}^{ \ \gamma} - \partial^{\gamma}
 h_{\rho\mu})\Big].
\end{split}
\end{equation}
As a result, we obtain the following relation up to second order in metric perturbation
\begin{equation}\label{S-A11}
\begin{split}
\frac{1}{2\kappa^{2}}\sqrt{-g}R & = \frac{1}{2\kappa^{2}}(1 + \kappa h)(\eta^{\mu\nu} - 2
\kappa h^{\mu\nu})(R_{\mu\nu}^{(1)} + R_{\mu\nu}^{(2)})\\
 & = \frac{1}{2\kappa^{2}}\Big[\eta^{\mu\nu}R_{\mu\nu}^{(1)} - 2\kappa h^{\mu\nu}R_{\mu\nu}
 ^{(1)} + \kappa h\eta^{\mu\nu}R_{\mu\nu}^{(1)} + \eta^{\mu\nu}R_{\mu\nu}^{(2)}\Big], 
\end{split}
\end{equation}
where
\begin{equation}\label{S-A12}
\begin{split}
\eta^{\mu\nu}R_{\mu\nu}^{(1)} & = 2\kappa (\partial_{\mu}\partial_{\nu}h^{\mu\nu} - \bar{\Box}
 h)\\
- 2\kappa h^{\mu\nu}R_{\mu\nu}^{(1)} & = - 2\kappa^{2}h^{\mu\nu}(\partial_{\rho}\partial_{\mu}
h_{\nu}^{ \ \rho} + \partial_{\nu}\partial^{\rho}h_{\mu\rho} - \partial_{\nu}\partial_{\mu}h 
- \bar{\Box}h_{\mu\nu})\\
\kappa h\eta^{\mu\nu}R_{\mu\nu}^{(1)} & = 2\kappa^{2} (h\partial_{\mu}\partial_{\nu}h^{\mu\nu}
 - h\bar{\Box}h)\\
\eta^{\mu\nu}R_{\mu\nu}^{(2)} & = - 2\kappa^{2}h\partial^{\mu}\partial^{\nu}h_{\mu\nu} + \kappa
^{2} h\bar{\Box}h + 2\kappa^{2}h^{\rho\gamma}\partial_{\rho}\partial^{\mu}h_{\mu\gamma} - \kappa
^{2}h^{\rho\gamma}\bar{\Box}h_{\rho\gamma}.
\end{split}
\end{equation}
Collecting all these contributions, and neglecting boundary terms, we finally obtain the following
expression up to second-order in metric perturbation
\begin{equation}\label{S-A13}
\frac{1}{2\kappa^{2}}\sqrt{-g}R = - \frac{1}{2}\partial_{\lambda}h_{\mu\nu}\partial^{\lambda}
h^{\mu\nu} + \partial_{\mu}h_{\nu\lambda}\partial^{\nu}h^{\mu\lambda} - \partial_{\mu}h^{\mu\nu}
\partial_{\nu}h + \frac{1}{2}\partial_{\lambda}h\partial^{\lambda}h .
\end{equation}
Using the following relations at the linear order
\begin{equation}\label{S-A14}
\begin{split}
R_{\mu\nu\rho\sigma}^{(1)} & = \kappa (\partial_{\rho}\partial_{\nu}h_{\sigma\mu} - \partial_{\rho}
\partial_{\mu}h_{\sigma\nu} - \partial_{\sigma}\partial_{\nu}h_{\rho\mu} + \partial_{\sigma}
\partial_{\mu}h_{\rho\nu})\\
R_{\nu\sigma}^{(1)} & = \kappa (\partial_{\mu}\partial_{\sigma}h_{ \ \nu}^{\mu} + \partial_{\mu}
\partial_{\nu}h_{ \ \sigma}^{\mu} - \bar{\Box}h_{\sigma\nu} - \partial_{\sigma}\partial_{\nu}h)\\
R^{(1)} & = 2\kappa (\partial_{\mu}\partial_{\nu}h^{\mu\nu} - \bar{\Box}h),
\end{split}
\end{equation}
we may write the following relation
\begin{equation}\label{S-A15}
\begin{split}
\frac{1}{2\kappa^{2}} R^{\mu\nu\rho\sigma}\frac{1}{\bar{\Box}^{2}}R_{\mu\nu\rho\sigma} 
& = \frac{1}{2}\Bigg[h^{\sigma\mu}\frac{1}{\bar{\Box}^{2}}\Big[\bar{\Box}^{2}h_{\sigma\mu} - 
\bar{\Box}\partial^{\nu}\partial_{\mu}h_{\sigma\nu} - \partial^{\rho}\partial_{\sigma}\bar{\Box}
h_{\rho\mu} + \partial^{\rho}\partial^{\nu}\partial_{\sigma}\partial_{\mu}h_{\rho\nu}\Big]\\
 & - h^{\sigma\nu}\frac{1}{\bar{\Box}^{2}}\Big[\bar{\Box}\partial^{\mu}\partial_{\nu}h_{\sigma\mu}
 - \bar{\Box}^{2}h_{\sigma\nu} - \partial^{\rho}\partial^{\mu}\partial_{\sigma}\partial_{\nu}
 h_{\rho\mu} + \partial^{\rho}\partial_{\sigma}\bar{\Box}h_{\rho\nu}\Big]\\
 & - h^{\rho\mu}\frac{1}{\bar{\Box}^{2}}\Big[\partial^{\sigma}\partial_{\rho}\bar{\Box}h_{\sigma
 \mu} - \partial^{\sigma}\partial_{\rho}\partial^{\nu}\partial_{\mu}h_{\sigma\nu} - \bar{\Box}^{2}
 h_{\rho\mu} + \bar{\Box}\partial^{\nu}\partial_{\mu}h_{\rho\nu}\Big]\\
 & + h^{\rho\nu}\frac{1}{\bar{\Box}^{2}}\Big[\partial^{\sigma}\partial^{\mu}\partial_{\rho}
 \partial_{\nu}h_{\sigma\mu} - \partial^{\sigma}\partial_{\rho}\bar{\Box}h_{\sigma\nu} - \bar{\Box}
 \partial^{\mu}\partial_{\nu}h_{\rho\mu} + \bar{\Box}^{2}h_{\rho\nu}\Big]\Bigg] + \mathcal{O}(\kappa)
 \\
\frac{1}{2\kappa^{2}}R^{\mu\nu}\frac{1}{\bar{\Box}^{2}}R_{\mu\nu} & = \frac{1}{2}\Bigg[
h^{\nu\rho}\frac{1}{\bar{\Box}^{2}}\Big[\partial_{\rho}\partial^{\mu}\partial_{\lambda}\partial
_{\mu}h_{\nu}^{ \ \lambda} + \partial_{\rho}\partial^{\mu}\partial_{\nu}\partial^{\lambda}h_{\mu
\lambda} - \partial_{\rho}\partial^{\mu}\partial_{\mu}\partial_{\nu}h - \partial_{\rho}\partial
^{\mu}\bar{\Box}h_{\mu\nu}\Big]\\
 & + h_{ \ \rho}^{\mu}\frac{1}{\bar{\Box}^{2}}\Big[\partial^{\nu}\partial^{\rho}\partial_{\lambda}
 \partial_{\mu}h_{\nu}^{ \ \lambda} + \partial^{\nu}\partial^{\rho}\partial_{\nu}\partial^{\lambda}
 h_{\mu\lambda} - \partial^{\nu}\partial^{\rho}\partial_{\mu}\partial_{\nu}h - \partial^{\nu}
 \partial^{\rho}\bar{\Box}h_{\mu\nu}\Big]\\
 & - h\frac{1}{\bar{\Box}^{2}}\Big[\partial^{\mu}\partial^{\nu}\partial_{\rho}\partial_{\mu}h_{\nu}
 ^{ \ \rho} + \partial^{\mu}\partial^{\nu}\partial_{\nu}\partial^{\rho}h_{\mu\rho} - \partial^{\mu}
 \partial^{\nu}\partial_{\mu}\partial_{\nu}h - \partial^{\mu}\partial^{\nu}\bar{\Box}h_{\mu\nu}
 \Big]\\
 & - h^{\mu\nu}\frac{1}{\bar{\Box}}\Big[\partial_{\rho}\partial_{\mu}h_{\nu}^{ \ \rho} + \partial
 _{\nu}\partial^{\rho}h_{\mu\rho} - \partial_{\mu}\partial_{\nu}h - \bar{\Box}h_{\mu\nu}\Big]\Bigg]
 + \mathcal{O}(\kappa)\\
\frac{1}{2\kappa^{2}}R\frac{1}{\bar{\Box}^{2}}R & = 2\Bigg[h^{\mu\nu}\frac{1}{\bar{\Box}^{2}}\Big[
\partial_{\mu}\partial_{\nu}\partial_{\rho}\partial_{\sigma}h^{\rho\sigma} - \bar{\Box}\partial_{\mu}
\partial_{\nu}h\Big] - h\frac{1}{\bar{\Box}}\Big[\partial_{\rho}\partial_{\sigma}h^{\rho\sigma} - 
\bar{\Box}h\Big]\Bigg] + \mathcal{O}(\kappa).  
\end{split}
\end{equation}
Now we use the relation $\partial_{\mu}h^{\mu\nu} = \alpha\partial^{\nu}h$ and later show the 
consistency of this relation at the level of equation of motion of the metric perturbations.
As a result of the above equation, we may simplify the above relations as
\begin{equation}\label{S-A16}
\begin{split}
\frac{1}{2\kappa^{2}}\sqrt{-g}R\frac{1}{\bar{\Box}^{2}}R & = 2(\alpha - 1)^{2}h^{2} + \mathcal{O}
(\kappa)\\
\frac{1}{2\kappa^{2}}\sqrt{-g}R^{\mu\nu}\frac{1}{\bar{\Box}^{2}}R_{\mu\nu} & = \frac{1}{2}
h_{\mu\nu}h^{\mu\nu} - \alpha h^{2} + \mathcal{O}(\kappa)\\
\frac{1}{2\kappa^{2}}\sqrt{-g}R^{\mu\nu\rho\sigma}\frac{1}{\bar{\Box}^{2}}R_{\mu\nu\rho\sigma} & = 
2(h_{\mu\nu}h^{\mu\nu} - \alpha^{2}h^{2}) + \mathcal{O}(\kappa).
\end{split}
\end{equation}
As a result, in the linearised gravity limit around Minkowski metric, the non-local Lagrangian density in \ref{S-A1} reduces to the following
\begin{equation}\label{S-A17}
\begin{split}
& \frac{1}{2\kappa^{2}} \sqrt{-g}\left(\mu_{1}R\frac{1}{\Box^{2}}R + \mu_{2}R^{\mu\nu}\frac{1}
{\Box^{2}}R_{\mu\nu} + \mu_{3}R^{\mu\nu\rho\sigma}\frac{1}{\Box^{2}}R_{\mu\nu\rho\sigma}\right) \\
& = 2\mu_{1}(\alpha - 1)^{2}h^{2} + \frac{\mu_{2}}{2}h_{\mu\nu}h^{\mu\nu} - \mu_{2}\alpha h^{2}
 + 2\mu_{3}(h_{\mu\nu}h^{\mu\nu} - \alpha^{2}h^{2}) + \mathcal{O}(\kappa) \\ 
 & = h_{\mu\nu}h^{\mu\nu}
 \left(\frac{1}{2}\mu_{2} + 2\mu_{3}\right) + h^{2}\left(2\mu_{1}(\alpha - 1)^{2} - \mu_{2}\alpha
 - 2\mu_{3}\alpha^{2}\right) + \mathcal{O}(\kappa).
\end{split}
\end{equation} 
Therefore, in order to obtain a massive gravity theory described by the FP action with
deformation parameter $\alpha$, $\mu_{1}, \ \mu_{2}$, and $\mu_{3}$ must satisfy the following
criteria
\begin{equation}\label{S-A18}
\begin{split}
\frac{1}{2}\mu_{2} + 2\mu_{3} & = - \frac{1}{2}m^{2}\\
2\mu_{1}(\alpha - 1)^{2} - \mu_{2}\alpha - 2\mu_{3}\alpha^{2} & = \frac{\alpha}{2}m^{2}.
\end{split}
\end{equation}
Here we have three unknowns and two equations. However, $\alpha = 1/2, \ 1$ are the only physically
viable candidates for massive gravity theories (also known as FP theory with $\alpha$
being mass deformation parameter) \cite{Mandal:2024nhv} as in these cases there are no ghost degrees of freedom. 
For $\alpha = 1$, the above relations reduce to the following
\begin{equation}\label{S-A19}
\mu_{2} + 4\mu_{3} = - m^{2}, \ \mu_{2} + 2\mu_{3} = - \frac{1}{2}m^{2} \implies \mu_{3} = - 
\frac{1}{4}m^{2}, \ \mu_{2} = 0,
\end{equation}
whereas for $\alpha = \frac{1}{2}$, we obtain the following relations
\begin{equation}\label{S-A20}
\mu_{2} + 4\mu_{3} = - m^{2}, \ \mu_{1} - \mu_{2} - \mu_{3} = \frac{m^{2}}{2} \implies 
\mu_{2} = - (m^{2} + 4\mu_{3}), \ \mu_{1} = - \left[\frac{m^{2}}{2} + 3\mu_{3}\right].
\end{equation}
which for $\mu_{3} = 0$ gives a unique solution, namely, $\mu_{2} = - m^{2}, \ \mu_{1} = - 
\frac{1}{2}m^{2}$. Unlike the $\alpha = 1$ case, $\alpha = \frac{1}{2}$ massive gravity theory
(extended FP theory) can naturally be obtained as a consequence of spontaneous symmetry
breaking at the matter sector which is shown recently in \cite{Mandal:2024nhv}. Therefore, now 
onwards we restrict our discussion mainly to the $\alpha = 1/2$ with $\mu_{3} = 0$ case for the 
sake of mathematical simplicity and since it is a viable candidate of modified theories of gravity 
at large length scale which can naturally explain the present acceleration of the Universe without
invoking the so-called dark-energy. Moreover, it is important here to emphasis that imposing 
the constraints in \ref{S-A18} leads to the FP theory with massive deformation 
parameter $\alpha$ described by the following action
{\small
\begin{equation}\label{S-A21}
S_{F-P}^{(\alpha)} = \int d^{4}x \Big[ - \frac{1}{2}\partial_{\lambda}h_{\mu\nu}\partial^{\lambda}
h^{\mu\nu} + \partial_{\mu}h_{\nu\lambda}\partial^{\nu}h^{\mu\lambda} - \partial_{\mu}h^{\mu\nu}
\partial_{\nu}h + \frac{1}{2}\partial_{\lambda}h\partial^{\lambda}h - \frac{1}{2}m^{2}(h_{\mu\nu}
h^{\mu\nu} - \alpha h^{2})\Big].
\end{equation}
}
The field equations of the above theory automatically leads to the constraint $\partial_{\mu}
h^{\mu\nu} = \alpha\partial^{\nu}h$ which is shown in \cite{Mandal:2024nhv}. This shows the 
consistency of our assumption.

\section{Anomaly in non-local gravity}\label{App: Anomaly}

In this section, we discuss the presence of anomaly in the non-local gravity described in \ref{S-A1}. For the sake of mathematical simplicity, we consider $\mu_{3} = 0$ and $\alpha = 
1/2$ in this section. At the quadratic order in metric perturbation ($\mathcal{O}(\kappa^{2})$), 
the expression of Ricci tensor is given in \ref{S-A10}. As a result, at the cubic order in 
metric perturbation ($\mathcal{O}(\kappa)$), we obtain the following relation
\begin{equation}\label{S-B1}
\begin{split}
& \frac{1}{2\kappa^{2}}\left(2R_{\nu\sigma}^{(1)}\frac{1}{\bar{\Box}^{2}}R^{(2)\nu\sigma}\right) 
= \kappa h_{\mu\nu}\frac{1}{\bar{\Box}}\Big[\frac{1}{2}\partial^{\gamma}h(\partial^{\nu}
h_{ \ \gamma}^{\mu} + \partial^{\mu}h_{ \ \gamma}^{\nu} - \partial_{\gamma}h^{\mu\nu})
 + h^{\lambda\gamma}\partial_{\lambda}\partial^{\nu}h_{ \ \gamma}^{\mu}\\
 & + h^{\lambda\gamma}\partial_{\lambda}\partial^{\mu}h_{ \ \gamma}^{\nu} - h^{\lambda\gamma} 
 \partial_{\lambda}\partial_{\gamma}h^{\mu\nu} - h^{\lambda\gamma}\partial^{\mu}\partial^{\nu}
 h_{\lambda\gamma} - \frac{1}{2}\partial^{\nu}h^{\lambda\gamma}\partial^{\mu}h_{\lambda\gamma}
 \\
 & - \frac{1}{2}\partial_{\gamma}h\partial^{\mu}h^{\nu\gamma} - \frac{1}{2}\partial_{\gamma}h
 \partial^{\nu}h^{\mu\gamma} + \frac{1}{2}\partial_{\gamma}h\partial^{\gamma}h^{\mu\nu} + 
 \partial_{\gamma}h^{\nu\lambda}\partial_{\lambda}h^{\mu\gamma} - \partial_{\gamma}h^{\nu\lambda}
 \partial^{\gamma}h_{\lambda}^{ \ \mu}\Big]\\
 & = \kappa h_{\mu\nu}\frac{1}{\bar{\Box}}\Big[h^{\lambda\gamma}\partial_{\lambda}\partial
 ^{\nu}h_{ \ \gamma}^{\mu} + h^{\lambda\gamma}\partial_{\lambda}\partial^{\mu}h_{ \ \gamma}
 ^{\nu} - h^{\lambda\gamma}\partial_{\lambda}\partial_{\gamma}h^{\mu\nu} - h^{\lambda\gamma}
 \partial^{\mu}\partial^{\nu}h_{\lambda\gamma}\\
 & - \frac{1}{2}\partial^{\nu}h^{\lambda\gamma}\partial^{\mu}h_{\lambda\gamma} + \partial
 _{\gamma}h^{\nu\lambda}\partial_{\lambda}h^{\mu\gamma} - \partial_{\gamma}h^{\nu\lambda}
 \partial^{\gamma}h_{\lambda}^{ \ \mu}\Big]\\
 &  = \kappa \left(\frac{1}{\bar{\Box}} h_{\mu\nu}\right)\Big[h^{\lambda\gamma}\partial
 _{\lambda}\partial^{\nu}h_{ \ \gamma}^{\mu} + h^{\lambda\gamma}\partial_{\lambda}\partial
 ^{\mu}h_{ \ \gamma}^{\nu} - h^{\lambda\gamma}\partial_{\lambda}\partial_{\gamma}h^{\mu\nu} 
 - h^{\lambda\gamma}\partial^{\mu}\partial^{\nu}h_{\lambda\gamma}\\
 & - \frac{1}{2}\partial^{\nu}h^{\lambda\gamma}\partial^{\mu}h_{\lambda\gamma} + \partial
 _{\gamma}h^{\nu\lambda}\partial_{\lambda}h^{\mu\gamma} - \partial_{\gamma}h^{\nu\lambda}
 \partial^{\gamma}h_{\lambda}^{ \ \mu}\Big],
\end{split}
\end{equation}
which for $\alpha = 1/2$ case with $\mu_{1} = - m^{2}, \ \mu_{2} = - m^{2}/2$ reduces to
\begin{equation}\label{S-B2}
\begin{split}
& \frac{1}{2\kappa^{2}}\lim_{m \rightarrow 0} \left(2\mu_{2} R_{\mu\nu}^{(1)}\frac{1}
{\bar{\Box}^{2}}R^{(2)\mu\nu}\right)  \\ 
& = - \frac{1}{2}\kappa \lim_{m \rightarrow 0}
\Bigg[h_{\mu\nu}\Big[h^{\lambda\gamma}\partial_{\lambda}\partial^{\nu}h_{ \ \gamma}^{\mu} 
+ h^{\lambda\gamma}\partial_{\lambda}\partial^{\mu}h_{ \ \gamma}^{\nu} - h^{\lambda\gamma}
\partial_{\lambda}\partial_{\gamma}h^{\mu\nu} - h^{\lambda\gamma}\partial^{\mu}\partial^{\nu}
h_{\lambda\gamma}\\
 & - \frac{1}{2}\partial^{\nu}h^{\lambda\gamma}\partial^{\mu}h_{\lambda\gamma} + \partial
 _{\gamma}h^{\nu\lambda}\partial_{\lambda}h^{\mu\gamma} - \partial_{\gamma}h^{\nu\lambda}
 \partial^{\gamma}h_{\lambda}^{ \ \mu}\Big]\Bigg]\\
 & = - \frac{1}{2}\kappa\lim_{m \rightarrow 0}\Bigg[h_{\mu\nu}\Big[h^{\lambda\gamma}\partial
_{\lambda}\partial^{\nu}h_{ \ \gamma}^{\mu} + h^{\lambda\gamma}\partial_{\lambda}\partial
^{\mu}h_{ \ \gamma}^{\nu} - h^{\lambda\gamma}\partial_{\lambda}\partial_{\gamma}h^{\mu\nu} 
- \frac{1}{2}\partial^{\mu}\partial^{\nu}(h_{\lambda\gamma}h^{\lambda\gamma})\\
 & - \partial_{\gamma}h^{\nu\lambda}\partial_{\lambda}h^{\mu\gamma} - \partial_{\gamma}
 h^{\nu\lambda}\partial^{\gamma}h_{\lambda}^{ \ \mu}\Big]\Bigg]\\
 & = - \frac{1}{2}\kappa \lim_{m \rightarrow 0}\Big[h_{\mu\nu}[h^{\lambda\gamma}\partial
 _{\lambda}\partial^{\nu}h_{ \ \gamma}^{\mu} + h^{\lambda\gamma}\partial_{\lambda}\partial
 ^{\mu}h_{ \ \gamma}^{\nu}] - h^{\lambda\gamma}h_{\mu\nu}\partial_{\lambda}\partial_{\gamma}
 h^{\mu\nu}\\
 &  + h_{\mu\nu}\partial_{\gamma}h^{\nu\lambda}\partial_{\lambda}h^{\mu\gamma} - h_{\mu\nu}\partial_{\gamma}h^{\nu\lambda}\partial^{\gamma}h_{\lambda}^{ \ \mu}\Big] + \frac{1}{4}\kappa 
 \lim_{m \rightarrow 0}h_{\lambda\gamma}h^{\lambda\gamma}\partial^{\mu}\partial^{\nu}h_{\mu\nu}\\
 & = - \frac{1}{2}\kappa\Big[h_{\mu\nu}[h^{\lambda\gamma}\partial_{\lambda}\partial^{\nu}h_{ \ \gamma}
 ^{\mu} + h^{\lambda\gamma}\partial_{\lambda}\partial^{\mu}h_{ \ \gamma}^{\nu}] - h^{\lambda
 \gamma}h_{\mu\nu}\partial_{\lambda}\partial_{\gamma}h^{\mu\nu}\\
 & + h_{\mu\nu}\partial_{\gamma}h^{\nu\lambda}\partial_{\lambda}h^{\mu\gamma} - h_{\mu\nu}
 \partial_{\gamma}h^{\nu\lambda}\partial^{\gamma}h_{\lambda}^{ \ \mu}\Big],
\end{split}
\end{equation}
where we use the field equation $(\bar{\Box} - m^{2})h_{\mu\nu} = 0$ obtained at the quadratic
order since we are deriving the theory in the perturbative manner. There is another term that
would not contribute which is of the following form
\begin{equation}\label{S-B3}
\frac{1}{2\kappa^{2}}\left(\kappa h R_{\mu\nu}^{(1)}\frac{1}{\bar{\Box}^{2}}R^{(1)\mu\nu}
\right) = \frac{1}{2}\kappa h (\bar{\Box}h)\frac{1}{\bar{\Box}^{2}}(\bar{\Box}h) = \frac{1}{2}
\kappa h^{3}.
\end{equation}
As a result, in order $\mathcal{O}(\kappa)$,
\begin{equation}\label{S-B4}
\begin{split}
\frac{1}{2\kappa^{2}}\lim_{m \rightarrow 0} \left(\sqrt{-g} 2\mu_{2}R_{\mu\nu}\frac{\mu_{2}}{\bar{\Box}
^{2}}R^{\mu\nu}\right)^{(3)} & = - \frac{1}{2}\kappa\Big[h_{\mu\nu}[h^{\lambda\gamma}\partial
_{\lambda}\partial^{\nu}h_{ \ \gamma}^{\mu} + h^{\lambda\gamma}\partial_{\lambda}\partial^{\mu}
h_{ \ \gamma}^{\nu}] \\ 
 & - h^{\lambda\gamma}h_{\mu\nu}\partial_{\lambda}\partial_{\gamma}h^{\mu\nu} + h_{\mu\nu}\partial_{\gamma}h^{\nu\lambda}\partial_{\lambda}h^{\mu\gamma} - h_{\mu\nu}
 \partial_{\gamma}h^{\nu\lambda}\partial^{\gamma}h_{\lambda}^{ \ \mu}\Big].
\end{split}
\end{equation}
On the other hand, we also have the following expression for $R^{(2)}$
\begin{equation}\label{S-B5}
R^{(2)} = \kappa^{2}[4h^{\mu\nu}\bar{\Box}h_{\mu\nu} + 3\partial^{\alpha}h^{\mu\nu}\partial
_{\alpha}h_{\mu\nu} - 2\partial^{\alpha}h^{\mu\nu}\partial_{\nu}h_{\alpha\mu}].
\end{equation}
Using the above result, we also obtain the following relation at the cubic order in the metric 
perturbation ($\mathcal{O}(\kappa)$) in the similar manner
\begin{equation}\label{S-B6}
\lim_{m \rightarrow 0}\Big[ - \frac{m^{2}}{2\kappa^{2}}\left(\sqrt{-g}R\frac{1}{\bar{\Box}^{2}}R\right)^{(3)}\Big] \propto \kappa\left(3h\partial^{\alpha}h^{\mu\nu}\partial_{\alpha}h_{\mu\nu}
+ 2h\partial^{\alpha}h_{\mu\nu}\partial^{\nu}h_{\alpha}^{ \ \mu}\right).
\end{equation}
This clearly indicates that for $\alpha = \frac{1}{2}$ and $m \rightarrow 0$ limit, there will
be non-zero terms at the cubic order in the metric perturbation although the quadratic order 
terms in metric perturbation reduces to the quadratic order terms of Einstein-Hilbert action.
This clearly shows the presence of anomaly in the case of non-local theory which we considered
in \ref{S-A1}. Apart from the above non-zero cubic order terms, there will also be non-zero 
cubic order terms coming from the Einstein-Hilbert action which is of the following form
\begin{equation}\label{S-B7}
\begin{split}
\mathcal{L}_{\text{E-H}}^{(3)} & \propto \frac{1}{2}h_{ \ \beta}^{\alpha}\partial^{\mu}h_{\alpha}
^{ \ \beta}\partial_{\mu}h - \frac{1}{2}h^{\alpha\beta}\partial_{\alpha}h^{\mu\nu}\partial
_{\beta}h_{\mu\nu} - h^{\alpha\beta}\partial_{\mu}h_{\nu\alpha}\partial^{\mu}h_{ \ \beta}^{\nu}
\\
 & + \frac{1}{4}h\partial^{\beta}h^{\mu\nu}\partial_{\beta}h_{\mu\nu} + h_{\mu}^{ \ \beta}
 \partial_{\nu}h_{ \ \beta}^{\alpha}\partial^{\mu}h_{ \ \alpha}^{\nu} - \frac{1}{8}h\partial^{\nu}
 h\partial_{\nu}h,
\end{split}
\end{equation}
where we also used $\partial_{\mu}h_{ \ \nu}^{\mu} = \frac{1}{2}\partial_{\nu}h$. It is important here to mention that unlike the above terms, the terms we obtained earlier in \ref{S-B4} and \ref{S-B6} are not gauge invariant.

The observed \emph{non-commutativity} between the massless limit ($m \rightarrow 0$) and the linearization limit ($\kappa \rightarrow 0$) is the manifestation of the \emph{van Dam-Veltman-Zakharov (vDVZ) discontinuity}.
The physical reason for this discontinuity is that a massive spin-2 field (the graviton in our theory) always propagates five degrees of freedom (DOFs), regardless of how small its mass $m$ is. In contrast, a massless spin-2 field (the graviton in GR) propagates only two DOFs. The three extra DOFs (two vector, one scalar) do not simply vanish in the $m \rightarrow 0$ limit; the scalar degree of freedom remains coupled to the trace of the energy-momentum tensor. This affects predictivity at the linearized level, as the massless limit of our linearized theory yields a gravitational force that is $4/3$ times the GR prediction, contradicting Solar System observations. \\[5pt]
The true recovery of GR and the restoration of predictivity are achieved only through the full, non-linear action and the Vainshtein screening mechanism.
When the full non-linear theory is considered, the strong gravitational fields near massive objects (like the Sun) trigger the Vainshtein mechanism. This mechanism non-linearly suppresses the kinetic term of the problematic scalar mode, effectively \emph{decoupling the fifth force} and hiding the extra DOFs below the observational threshold. This ensures the theory recovers the predictions of GR where needed. The model is highly predictive in the infrared (cosmological scales), where non-local effects dominate, while remaining consistent with high-precision data in the strong-gravity regime (Solar System).\\[5pt]
The Stueckelberg mechanism is a formal tool used to restore the full coordinate invariance (diffeomorphism symmetry) that is broken by the mass term in the action. The Stueckelberg fields ($\phi^A$) allow the mass term to be rewritten in a way that is explicitly invariant under the original diffeomorphism, $\delta g_{\mu\nu} = \nabla_{(\mu} \xi_{\nu)}$.\\[5pt]
1. The physical metric is then $g_{\mu\nu}$. \\
2. The reference metric ($f_{\mu\nu}$), which is non-dynamical in massive gravity, is replaced by the composite metric, $f_{\mu\nu} \rightarrow \eta_{AB} \partial_\mu \phi^A \partial_\nu \phi^B$. \\
3. By explicitly defining the gravitational potential as $h_{\mu\nu} = g_{\mu\nu} - \eta_{\mu\nu}$ and introducing the four scalar fields $\phi^A$ into the action (as detailed in the literature for ghost-free massive gravity), the symmetry is algebraically restored. This localization method is precisely how the theory is proven to be ghost-free, even in the non-linear regime.

\section{Covariant field equations}\label{App: covariant field equations}

We start with the following non-local action in four dimensions
\begin{equation}\label{S-C1}
S_{NL} = \frac{1}{2\kappa^{2}}\int d^{4}x \ \sqrt{-g}\Big[R - \mu_{1} R \frac{1}{\Box^{2}}R
- \mu_{2} R_{\mu\nu}\frac{1}{\Box^{2}}R^{\mu\nu}\Big].
\end{equation}
Following the prescription in \cite{Cusin:2015rex}, the above action can be equivalently expressed as
\begin{equation}\label{S-C2}
\begin{split}
S_{NL} & = \frac{1}{2\kappa^{2}}\int d^{4}x \ \sqrt{-g}\Big[R - \mu_{1}RS - \mu_{2}R_{\mu\nu}
S^{\mu\nu} - \mathcal{A}_{1}(\Box\Phi + R) - \mathcal{A}_{2}(\Box S + \Phi)\\
 & - \mathcal{A}^{\mu\nu}(\Box\Phi_{\mu\nu} + R_{\mu\nu}) - \mathcal{B}^{\mu\nu}(\Box S_{\mu
 \nu} + \Phi_{\mu\nu})\Big],
\end{split}
\end{equation}
by introducing the Lagrange multipliers $\mathcal{A}_{1}, \ \mathcal{A}_{2}, \ \mathcal{A}
^{\mu\nu}, \ \mathcal{B}^{\mu\nu}$ and the auxiliary field variables $\Phi, \ S, \ \Phi_{\mu
\nu}, \ S_{\mu\nu}$. The variation of the above action \textit{w.r.t} the Lagrange multipliers
lead to the following constraints in terms of equations of motion of auxiliary field variables
\begin{equation}\label{S-C3}
\begin{split}
0 = \frac{\delta S_{NL}}{\delta \mathcal{A}_{1}} & \implies \Box\Phi = - R\\
0 = \frac{\delta S_{NL}}{\delta \mathcal{A}_{2}} & \implies \Box S = - \Phi\\
0 = \frac{\delta S_{NL}}{\delta \mathcal{A}^{\mu\nu}} & \implies \Box\Phi_{\mu\nu} =
 - R_{\mu\nu}\\
0 = \frac{\delta S_{NL}}{\delta \mathcal{B}^{\mu\nu}} & \implies \Box S_{\mu\nu} = - 
\Phi_{\mu\nu}.
\end{split}
\end{equation}
Combining the first two equations, we obtain the following relation
\begin{equation}\label{S-C4}
S = - \frac{1}{\Box}\Phi = \frac{1}{\Box^{2}}R,
\end{equation} 
whereas combing the last two equations of \ref{S-C3}), we obtain
\begin{equation}\label{S-C5}
S_{\mu\nu} = - \frac{1}{\Box}\Phi_{\mu\nu} = \frac{1}{\Box^{2}}R_{\mu\nu}.
\end{equation}
On the other hand, the variation \textit{w.r.t} the auxiliary field variables lead to
the following field equations
\begin{equation}\label{S-C6}
\begin{split}
0 = \frac{\delta S_{NL}}{\delta S} & \implies \Box\mathcal{A}_{2} = - \mu_{1}R\\
0 = \frac{\delta S_{NL}}{\delta S^{\mu\nu}} & \implies \Box\mathcal{B}^{\mu\nu} = 
- \mu_{2}R^{\mu\nu}\\
0 = \frac{\delta S_{NL}}{\delta \Phi} & \implies \Box\mathcal{A}_{1} = - \mathcal{A}_{2}\\
0 = \frac{\delta S_{NL}}{\delta \Phi_{\mu\nu}} & \implies \Box\mathcal{A}^{\mu\nu} = 
- \mathcal{B}^{\mu\nu}.
\end{split}
\end{equation} 
The above equations leads to the following relations
\begin{equation}\label{S-C7}
\mathcal{A}_{2} = \mu_{1}\Phi, \ \mathcal{B}^{\mu\nu} = \mu_{2}\Phi^{\mu\nu}, \ 
\mathcal{A}_{1} = \mu_{1}\frac{1}{\Box^{2}}R = \mu_{1}S, \ \mathcal{A}^{\mu\nu} = 
\mu_{2}\frac{1}{\Box^{2}}R^{\mu\nu} = \mu_{2}S^{\mu\nu}.
\end{equation}
Now considering the metric variation of the non-local action $S_{NL}$, we obtain the
following field equations
\begin{equation}\label{S-C8}
G_{\mu\nu} - \mu_{1}\mathcal{G}_{\mu\nu}^{(1)} + \mu_{2}\mathcal{G}_{\mu\nu}^{(2)} = 
\kappa^{2}T_{\mu\nu},
\end{equation}
where $T_{\mu\nu}$ is the energy-momentum tensor associated with the matter field.
In the above equation, $\mathcal{G}_{\mu\nu}^{(1)}$ is given by the following expression
\begin{equation}\label{S-C9}
\begin{split}
\mathcal{G}_{\mu\nu}^{(1)} & = 2SG_{\mu\nu} - 2\nabla_{\mu}\nabla_{\nu}S + g_{\mu\nu}
\nabla_{\lambda}S\nabla^{\lambda}\Phi - 2g_{\mu\nu}\Phi - \frac{1}{2}\Phi^{2}g_{\mu\nu}\\
 & - (\nabla_{\mu}S\nabla_{\nu}\Phi + \nabla_{\nu}S\nabla_{\mu}\Phi),
\end{split}
\end{equation}
whereas the expression of $\mathcal{G}_{\mu\nu}^{(2)}$ is given by
\begin{equation}\label{S-C10}
\begin{split}
\mathcal{G}_{\mu\nu}^{(2)} & = \Bigg[ - 2S_{\mu\alpha}G_{ \ \nu}^{\alpha} - RS_{\mu\nu}
- \frac{1}{2}\Box S_{\mu\nu} - \Phi_{\mu}^{ \ \alpha}\Phi_{\nu\alpha} + \nabla_{\alpha}
\nabla_{\mu}S_{ \ \nu}^{\alpha}\\
 & + \frac{1}{2}g_{\mu\nu}\left(G^{\alpha\beta}S_{\alpha\beta} + \frac{1}{2}\Phi_{\alpha
 \beta}\Phi^{\alpha\beta} + \frac{1}{2}RS_{ \ \alpha}^{\alpha} - \nabla^{\lambda}S^{\alpha
 \beta}\nabla_{\lambda}\Phi_{\alpha\beta} - \nabla_{\alpha\beta}S^{\alpha\beta}\right)\\
 & + \Big[\frac{1}{2}\nabla_{\mu}\Phi_{\alpha\beta}\nabla_{\nu}S^{\alpha\beta} + \nabla
 _{\beta}\Phi_{ \ \alpha}^{\beta}\nabla_{\mu}S_{ \ \nu}^{\alpha} - S_{ \ \nu}^{\alpha}
 \nabla_{\beta}\nabla_{\mu}\Phi_{ \ \alpha}^{\beta} - \Phi_{ \ \nu}^{\alpha}\Box S_{\mu
 \alpha}\\
 & + \Phi^{\alpha\beta}\nabla_{\beta}\nabla_{\mu}S_{\nu\alpha} - \nabla^{\beta}S_{ \ \mu}
 ^{\alpha}\nabla_{\nu}\Phi_{\alpha\beta} + (\Phi \leftrightarrow S)\Big]\Bigg] + \{\mu 
 \leftrightarrow \nu\}.
\end{split}
\end{equation} 
In the above expressions, $G_{\mu\nu}$ is the Einstein's field tensor defined by $G_{\mu\nu}
= R_{\mu\nu} - \frac{1}{2}g_{\mu\nu}R$. It is important to note that Minkowski 
spacetime is also a solution of the field equation as in this case $\Phi, S, \Phi_{\mu\nu}, S_{\mu\nu}$ 
and $\mathcal{A}_{1}, \mathcal{A}_{2}, \mathcal{A}_{\mu\nu}, \mathcal{B}_{\mu\nu}$ can be chosen
to be identically zero. As a result, the equation \ref{S-C8} satisfies trivially.

The auxiliary field localization, which replaces the non-local operator $\Box^{-1}$ with two standard differential equations ($\Box\Phi = -R$ and $\Box S = -\Phi$), \emph{genuinely resolves the causal issues} present in the original non-local action, provided the auxiliary field equations are solved using the \emph{retarded Green's function} in a hyperbolic spacetime.
Noet that the original non-local action, $S_{\text{non-local}} \sim R \Box^{-2} R$, requires integrating over the full spacetime history, which can easily lead to acausal propagation (effects preceding their causes) unless a specific causal choice is made. However, the auxiliary field method transforms the action from an integro-differential equation into a system of \emph{local, second-order hyperbolic PDEs}. This shifts the problem from non-locality in the field definition to a standard \emph{Initial Value Problem (IVP)}, which is known to be causal and well-posed if specified correctly.
To enforce the causality inherent in the structure of General Relativity, we must specify the solution to the auxiliary field equations using the \emph{retarded Green's function}, $G_{\text{ret}}(x, x')$. The field $\Phi$ is formally defined by
    $$\Phi(x) = \int d^4 x' \sqrt{-g(x')} G_{\text{ret}}(x, x') [-\Phi^2(x') \text{ and other non-linear source terms}]$$
 However, the practical implementation avoids this integral representation by solving the associated differential equation as an IVP. The IVP for a second-order hyperbolic equation is well-posed if initial conditions (Cauchy data) are specified on a space-like hypersurface $\Sigma$. Our system is well-posed by solving the coupled system as follows:
 \begin{enumerate}
\item The metric field, $G_{\mu\nu} = 8\pi G T^{\text{eff}}_{\mu\nu}$, is a second-order hyperbolic equation for the metric $g_{\mu\nu}$ driven by the effective stress-energy $T^{\text{eff}}_{\mu\nu}$.
\item The auxiliary fields $\Phi$ and $S$ satisfy second-order hyperbolic equations ($\Box\Phi \propto R + \dots$ and $\Box S = -\Phi$).
\item The auxiliary fields must be solved forward in time from initial data specified on the Cauchy surface $\Sigma$, meaning $\Phi$ and $S$ only depend on the metric and matter fields in the past light cone --- the definition of causality via the retarded solution.
\end{enumerate} 
To explicitly show the well-posedness, we consider a specific case
$\mu_{2} = 0$. In this case, the auxiliary field variables satisfy the following equations
\begin{equation}
\Box\Phi = - R, \ \Box S = - \Phi.
\end{equation}
On the other hand, the field equation coming from the metric variation can be expressed as
\begin{equation}
G_{\mu\nu}(1 - 2\mu_{1}S) + \mu_{1}\Big[\nabla_{\mu}\nabla_{\nu}S - g_{\mu\nu}\nabla_{\lambda}S
\nabla^{\lambda}\Phi + 2g_{\mu\nu}\Phi + \frac{1}{2}\Phi^{2}g_{\mu\nu} + (\nabla_{\mu}S\nabla_{\nu}
\Phi + \nabla_{\nu}S\nabla_{\mu}\Phi)\Big] = \kappa^{2}T_{\mu\nu}.
\end{equation}
Taking the trace of the above equation, we obtain the following relation
\begin{equation}
R(1 - 2\mu_{1}S) = \mu_{1}\Big[\Box S - 2\nabla_{\lambda}S\nabla^{\lambda}\Phi + 8\Phi + 2\Phi^{2}\Big]
- \kappa^{2}T,
\end{equation}
where $T$ is the trace of the energy-momentum tensor. The above relation also implies
\begin{equation}
\begin{split}
\int\sqrt{-g}d^{4}x \ R(1 - 2\mu_{1}S) & = \int\sqrt{-g}d^{4}x \Bigg[\mu_{1}\Big[\Box S - 2\nabla_{\lambda}S\nabla^{\lambda}\Phi + 8\Phi + 2\Phi^{2}\Big] - \kappa^{2}T\Bigg]\\
 & = \int\sqrt{-g}d^{4}x \Bigg[\mu_{1}\Big[7\Phi + 2\Phi^{2} - 2\nabla^{\lambda}(\Phi
 \nabla_{\lambda}S) + 2\Phi\Box S\Big] - \kappa^{2}T\Bigg]\\
 & = \int\sqrt{-g}d^{4}x \ (7\mu_{1}\Phi - \kappa^{2}T) - 2\int_{\Sigma}n^{\lambda}\Phi
 \nabla_{\lambda}S d^{3}\sigma,
\end{split}
\end{equation} 
where $\Sigma$ is the Cauchy hypersurface in which we impose the initial condition 
$n^{\lambda}\Phi\nabla_{\lambda}S|_{\Sigma} = 0$ which leads to the following relation
\begin{equation}
R = \frac{7\mu_{1}\Phi - \kappa^{2}T}{1 - 2\mu_{1}S} \implies \Box\Phi = \frac{\kappa^{2}T
 - 7\mu_{1}\Phi}{1 - 2\mu_{1}S}.
\end{equation}
This follows from the fact that we are doing the integration over arbitrary spacetime 
volume. The above relation can also be expressed as
\begin{equation}
\begin{split}
\int\sqrt{-g}d^{4}x & \Big[\Box\Phi - 2\mu_{1}S\Box\Phi + 7\mu_{1}\Phi - \kappa^{2}T\Big] = 0\\
\implies\int\sqrt{-g}d^{4}x & \Big[\Box\Phi - 2\mu_{1}\nabla^{\lambda}(S\nabla_{\lambda}\Phi)
 + 2\mu_{1}\nabla^{\lambda}S\nabla_{\lambda}\Phi + 7\mu_{1}\Phi - \kappa^{2}T\Big] = 0\\
\implies \int\sqrt{-g}d^{4}x & \Big[\Box\Phi - 2\mu_{1}\nabla^{\lambda}(S\nabla_{\lambda}\Phi)
 + 2\mu_{1}\nabla^{\lambda}(\Phi\nabla_{\lambda}S) - 2\mu_{1}\Phi\Box S + 7\mu_{1}\Phi 
 - \kappa^{2}T\Big] = 0.
\end{split}
\end{equation}
Considering another boundary condition to be $n^{\lambda}S\nabla_{\lambda}\Phi|_{\Sigma} = 0$
at the Cauchy hypersurface, we now obtain the following differential equation
\begin{equation}
\Box\Phi + 7\mu_{1}\Phi + 2\mu_{1}\Phi^{2} = \kappa^{2}T.
\end{equation}
Note that since $\mu_{1} \sim - m^{2}$, the linear part of the homogeneous differential equation
$\Box\Phi + 7\mu_{1}\Phi = 0$ is well-posed, and so does the entire differential equation. Once
we know the solution of the above differential equation \textit{w.r.t} the boundary condition
$n^{\lambda}S\nabla_{\lambda}\Phi|_{\Sigma} = 0$, we can solve the other inhomogeneous differential
equation $\Box S = - \Phi$ \textit{w.r.t} the boundary condition $n^{\lambda}\Phi\nabla_{\lambda}S|_{\Sigma} = 0$.
The above differential equation for $\Phi$  and the simplified system, confirms that the fields satisfy hyperbolic (wave-like) equations when the background $1 - 2\mu_1 S$ is non-singular and the effective mass term ($7\mu_1 \Phi + 2\mu_1 \Phi^2$) is positive (no tachyons). This structure guarantees the existence of a well-posed IVP.
When inverting $\Box$ on cosmological backgrounds (FLRW/dS), the homogeneous solutions of the wave equation must be addressed. The Lagrange multipliers satisfy the following equations
\begin{equation}
\Box\mathcal{A}_{2} = - \mu_{1}R, \ \Box\mathcal{A}_{1} = - \mathcal{A}_{2},
\end{equation} 
in which the first equation can also be expressed as
\begin{equation}
\Box\mathcal{A}_{2} = \mu_{1}\Box\Phi \implies \Box(\mathcal{A}_{2} - \mu_{1}\Phi) = 0. 
\end{equation}
The solution of the above differential in the context spatially flat FRW spacetime is well-defined as an initial value problem. Hence, once we know the solution of $\Phi$, we can in principle obtain
the solution for $\mathcal{A}_{2}$ which then generates the solution for $\mathcal{A}_{1}$.
Thus, in the cosmological IVP formulation, the homogeneous solutions are fixed by the initial conditions imposed on the fields $\Phi$ and $S$ and their time derivatives on the initial Cauchy hypersurface ($\Sigma$). The non-local action $S_{\text{non-local}}$ (before localization) implicitly assumed that the Green's function choice neglected these homogeneous modes to ensure the non-local operator vanished at the boundary, $\Box \Box \Phi \sim 0$. In the localized approach, the boundary conditions $n^{\lambda}\Phi\nabla_{\lambda}S|_{\Sigma} = 0$ and $n^{\lambda}S\nabla_{\lambda}\Phi|_{\Sigma} = 0$ serve to fix the homogeneous solutions such that they are consistent with the boundary conditions one would apply to the non-local operator. This ensures the physical consistency between the non-local and auxiliary field formulations.
Hence, this approach provides a complete and consistent framework that establishes causality and well-posedness by defining the dynamics as an Initial Value Problem on a hyperbolic spacetime.

\section{Background field equations and fixed point analysis for $\mu_{2} = 0$}\label{app.Background field equations}

In order to obtain the modified first Friedmann equation, we use the following relations
\begin{equation}\label{S-D1}
G_{00} = 3H^{2}, \ T_{\mu\nu}u^{\mu}u^{\nu} = \rho, 
\end{equation}
where $H = \frac{\dot{a}}{a}$ and $\rho$ is the energy density of the perfect fluid that
drives the expansion of the Universe. Here $a(t)$ is the scale factor in the FRW line-element
for the homogeneous and isotropic Universe at large length scale. On the other hand, the 
$0-0$ component of $\mathcal{G}_{\mu\nu}^{(1)}$ is expressed as
\begin{equation}\label{S-D2}
\mathcal{G}_{00}^{(1)} = 6SH^{2} - 2\ddot{S} + 2\Phi + \frac{1}{2}\Phi^{2} - \dot{S}
\dot{\Phi}.
\end{equation}
Therefore, the modified first Friedmann equation becomes the following
\begin{equation}\label{S-D3}
3H^{2}(1 - 2\mu_{1}S) = \kappa^{2}\rho + \mu_{1}\left(6H\dot{S} + \frac{1}{2}\Phi^{2} - 
\dot{S}\dot{\Phi}\right),
\end{equation}
where we used the following field equation
\begin{equation}\label{S-D4}
\Box S = - \Phi \implies \ddot{S} + 3H\dot{S} = \Phi.
\end{equation}
In the following analysis, we consider $\rho$ to be energy density contribution coming solely
from the matter sector of the Universe with $w = 0$ equation of state (pressure-less fluid).
As a result, the equation in \ref{S-D3}) can be rewritten in the following manner
\begin{equation}\label{S-D5}
\begin{split}
1 & = \frac{\kappa^{2}\rho}{3H^{2}(1 - 2\mu_{1}S)} + 2\frac{\mu_{1}\dot{S}}{H(1 - 2\mu_{1}S)}
+ \frac{\mu_{1}\Phi^{2}}{6H^{2}(1 - 2\mu_{1}S)} - \frac{\mu_{1}\dot{S}\dot{\Phi}}{3H^{2}(1 - 
2\mu_{1}S)}\\
 & \equiv \Omega_{m} + \Omega_{1} + \Omega_{2} + \Omega_{3},
\end{split}
\end{equation}
where $\Omega_{m}$ is the energy density coming from matter and $\Omega_{DE} = \Omega_{1} + 
\Omega_{2} + \Omega_{3}$ is associated with the contribution coming from the dark-energy.
Using the covariant conservation of the energy-momentum tensor of the matter fluid $\dot{
\rho} + 3H\rho = 0$ for pressure-less fluid, we obtain the following evolution equation
\begin{equation}\label{S-D6}
\dot{\Omega}_{m} = - 3H\Omega_{m} - \frac{2\dot{H}}{H}\Omega_{m} + \Omega_{1}\Omega_{m}H,
\end{equation}
which in terms of the red-shift variable $z$ can be expressed as
\begin{equation}\label{S-D7}
(1 + z)\frac{d\Omega_{m}}{dz} = 3\Omega_{m} - \Omega_{1}\Omega_{m} - 2\left(\frac{1 + z}
{H}\frac{dH}{dz}\right)\Omega_{m}.
\end{equation}
On the other hand, the evolution equations of $\Omega_{1}, \ \Omega_{2}, \ \Omega_{3}$ can
be expressed as
\begin{equation}\label{S-D8}
\begin{split}
\dot{\Omega}_{1} & = \frac{\Phi}{\dot{S}}\Omega_{1} - 3H\Omega_{1} - \frac{\dot{H}}{H}
\Omega_{1} + H\Omega_{1}^{2}\\
\dot{\Omega}_{2} & = 2\frac{\dot{\Phi}}{\Phi}\Omega_{2} - 2\frac{\dot{H}}{H}\Omega_{2}
+ H\Omega_{1}\Omega_{2}\\
\dot{\Omega}_{3} & = \frac{\Phi}{\dot{S}}\Omega_{3} + \frac{R}{\dot{\Phi}}\Omega_{3} - 
6H\Omega_{3} - 2\frac{\dot{H}}{H}\Omega_{3} + \Omega_{1}\Omega_{3}H.
\end{split}
\end{equation}
It is important to note the following relation
\begin{equation}\label{S-D9}
\frac{\Omega_{3}}{\Omega_{2}} = - 2\frac{\dot{S}}{\Phi}\frac{\dot{\Phi}}{\Phi} \implies 
2\Omega_{2}\frac{\dot{\Phi}}{\Phi} = - \Omega_{3}\frac{\Phi}{\dot{S}}.
\end{equation}
On the other hand, we may write the following relations
\begin{equation}\label{S-D10}
\frac{R}{\dot{\Phi}} = H\left(\frac{R/H^{2}}{\dot{\Phi}/H}\right), \ \frac{\dot{\Phi}}{H}
= - 6\frac{\Omega_{3}}{\Omega_{1}},
\end{equation}
and 
\begin{equation}\label{S-D11}
\frac{R}{H^{2}} = 6\left(\frac{\dot{H}}{H^{2}} + 2\right).
\end{equation}
Combining the above results, we may write the following relation
\begin{equation}\label{S-D12}
\frac{R}{\dot{\Phi}} = - 2\frac{\Omega_{1}}{\Omega_{3}}H - \frac{\Omega_{1}}{\Omega_{3}}
\frac{\dot{H}}{H}.
\end{equation}
As a result, we obtain the following relation
\begin{equation}\label{S-D13}
\begin{split}
\dot{\Omega}_{3} & = \frac{\Phi}{\dot{S}}\Omega_{3} - 2\Omega_{1}H - \Omega_{1}\frac{\dot{H}}
{H} - 6H\Omega_{3} - 2\frac{\dot{H}}{H}\Omega_{3} + \Omega_{1}\Omega_{3}H\\
\implies (1 + z)\frac{d\Omega_{3}}{dz} & = - \frac{\Phi}{\dot{S}H}\Omega_{3} + 2\Omega_{1} - 
\Omega_{1}\left(\frac{1 + z}{H}\frac{dH}{dz}\right) + 6\Omega_{3} - \Omega_{1}\Omega_{3}\\
 & - 2\left(\frac{1 + z}{H}\frac{dH}{dz}\right)\Omega_{3}. 
\end{split}
\end{equation}
In the similar manner, we find
\begin{equation}\label{S-D14}
\begin{split}
(1 + z)\frac{d\Omega_{2}}{dz} & = - 2\frac{\dot{\Phi}}{\Phi H}\Omega_{2} - 2\left(\frac{1 + z}
{H}\frac{dH}{dz}\right)\Omega_{2} - \Omega_{1}\Omega_{2}\\
(1 + z)\frac{d\Omega_{1}}{dz} & = - \frac{\Phi}{\dot{S}H}\Omega_{1} + 3\Omega_{1} - \Omega_{1}
\left(\frac{1 + z}{H}\frac{dH}{dz}\right) - \Omega_{1}^{2}.
\end{split}
\end{equation}
Therefore, we are left with the following unknown quantities
\begin{equation}\label{S-D15}
- \frac{1 + z}{H}\frac{dH}{dz}, \ - \frac{\dot{\Phi}}{\Phi H} = \frac{1 + z}{\Phi}\frac{d\Phi}
{dz}.
\end{equation}
Now we use the modified second Friedmann equation which for pressure-less fluid
\begin{equation}\label{S-D16}
G_{ij} - \mu_{1}\mathcal{G}_{ij}^{(1)} = 0,
\end{equation}
becomes the following
\begin{equation}\label{S-D17}
\left(2\frac{\dot{H}}{H^{2}} + 3\right)H^{2}(1 - 2\mu_{1}S) = 2\mu_{1}\Phi - 2\mu_{1}\dot{S}H
+ \mu_{1}\dot{S}\dot{\Phi} + \frac{1}{2}\mu_{1}\Phi^{2}.
\end{equation}
The above equation can also be expressed as
\begin{equation}\label{S-D18}
2\left(\frac{1 + z}{H}\frac{dH}{dz}\right) = 3 - \frac{12\Omega_{2}}{\Phi} + \Omega_{1} + 
3\Omega_{3} - 3\Omega_{2}.
\end{equation}
Note that
\begin{equation}\label{S-D19}
\frac{\Omega_{2}}{\Omega_{1}} = \frac{\Phi^{2}}{12\dot{S}H}.
\end{equation}
Defining the new variable
\begin{equation}\label{S-D20}
\Omega_{4} = \frac{12\Omega_{2}}{\Phi},
\end{equation}
we find the following evolution equation for it
\begin{equation}\label{S-D21}
\dot{\Omega}_{4} = \frac{\dot{\Phi}}{\Phi}\Omega_{4} - 2\frac{\dot{H}}{H}\Omega_{4} + H
\Omega_{1}\Omega_{4},
\end{equation}
which can be re-expressed as
\begin{equation}\label{S-D22}
(1 + z)\frac{d\Omega_{4}}{dz} = \frac{\Omega_{4}^{2}\Omega_{3}}{2\Omega_{2}\Omega_{1}}
- \Omega_{4}^{2} + 3\Omega_{3}\Omega_{4} - 3\Omega_{2}\Omega_{4} + 3\Omega_{4}.
\end{equation}
As a result, the final set of coupled ODEs are given by
\begin{equation}\label{S-D23}
\begin{split}
(1 + z)\frac{d\Omega_{m}}{dz} & = - 2\Omega_{1}\Omega_{m} + \Omega_{4}\Omega_{m} - 3\Omega_{3}
\Omega_{m} + 3\Omega_{2}\Omega_{m}\\
(1 + z)\frac{d\Omega_{1}}{dz} & = - \Omega_{4} + \frac{3\Omega_{1}}{2} + \frac{\Omega_{1}\Omega
_{4}}{2} - \frac{3\Omega_{1}^{2}}{2} - \frac{3\Omega_{1}\Omega_{3}}{2} + \frac{3\Omega_{1}
\Omega_{2}}{2}\\
(1 + z)\frac{d\Omega_{2}}{dz} & = \frac{\Omega_{4}\Omega_{3}}{\Omega_{1}} + \Omega_{2}\Omega_{4}
- 2\Omega_{1}\Omega_{2} - 3\Omega_{2}\Omega_{3} + 3\Omega_{2}^{2} - 3\Omega_{2}\\
(1 + z)\frac{d\Omega_{3}}{dz} & = - \frac{\Omega_{3}\Omega_{4}}{\Omega_{1}} + \frac{\Omega_{1}}
{2} + 3\Omega_{3} + \frac{\Omega_{1}\Omega_{4}}{2} - \frac{\Omega_{1}^{2}}{2} - \frac{7\Omega_{1}
\Omega_{3}}{2}\\
 & + \frac{3\Omega_{1}\Omega_{2}}{2} + \Omega_{3}\Omega_{4} - 3\Omega_{3}^{2} + 3\Omega_{2}
 \Omega_{3},
\end{split}
\end{equation}
which are coupled with the ODE in the auxiliary variable $\Omega_{4}$. In terms of the number
of e-folding variables $N = \log(1 + z)$, the above set of coupled differential equations can 
be expressed as
\begin{equation}\label{S-D24}
\begin{split}
\frac{d\Omega_{m}}{dN} & = - \Big[ 2\Omega_{1}\Omega_{m} - \Omega_{4}\Omega_{m} + 3\Omega_{3}
\Omega_{m} - 3\Omega_{2}\Omega_{m}\Big]\\
\frac{d\Omega_{1}}{dN} & = - \Big[\Omega_{4} - \frac{3\Omega_{1}}{2} - \frac{\Omega_{1}\Omega
_{4}}{2} + \frac{3\Omega_{1}^{2}}{2} + \frac{3\Omega_{1}\Omega_{3}}{2} - \frac{3\Omega_{1}
\Omega_{2}}{2}\Big]\\
\frac{d\Omega_{2}}{dN} & = - \Big[- \frac{\Omega_{4}\Omega_{3}}{\Omega_{1}} - \Omega_{2}\Omega_{4}
+ 2\Omega_{1}\Omega_{2} + 3\Omega_{2}\Omega_{3} - 3\Omega_{2}^{2} + 3\Omega_{2}\Big]\\
\frac{d\Omega_{3}}{dN} & = - \Big[\frac{\Omega_{3}\Omega_{4}}{\Omega_{1}} - \frac{\Omega_{1}}
{2} - 3\Omega_{3} - \frac{\Omega_{1}\Omega_{4}}{2} + \frac{\Omega_{1}^{2}}{2} + \frac{7\Omega_{1}
\Omega_{3}}{2}\\
 & - \frac{3\Omega_{1}\Omega_{2}}{2} - \Omega_{3}\Omega_{4} + 3\Omega_{3}^{2} - 3\Omega_{2}
 \Omega_{3}\Big]\\
\frac{d\Omega_{4}}{dN} & = - \Big[ - \frac{\Omega_{4}^{2}\Omega_{3}}{2\Omega_{2}\Omega_{1}}
+ \Omega_{4}^{2} - 3\Omega_{3}\Omega_{4} + 3\Omega_{2}\Omega_{4} - 3\Omega_{4}\Big]. 
\end{split}
\end{equation}
The above set of differential equations are complete which indeed shows that the above system is indeed autonomous. The fixed points of the above set of differential equations can be obtained by looking at the vanishing points of the \textit{r.h.s} of the differential equations and their stability analysis can be done numerically by finding out whether all the eigenvalues of the system of ODEs are negative (stable), positive (unstable), or mixed (critical) \cite{Halder:2024uao, Xu:2012jf, Boehmer:2023knj, Ghosh:2023amt}.

\section{Background field equations involving both the two non-local terms}\label{app.background field equations 2}

In order to obtain the background field equations for $\mu_{1} = 0$, we first write down
the metric and the geometric quantities in covariant manner
\begin{equation}\label{S-E1}
\begin{split}
g_{\mu\nu} & = - u_{\mu}u_{\nu} + a^{2}h_{\mu\nu}, \ u_{\mu} = \delta_{\mu}^{0}\\
u^{\mu}h_{\mu\nu} & = 0, \ h_{ \ \nu}^{\mu} = h^{\mu\alpha}h_{\alpha\nu} = \delta_{ \ \nu}
^{\mu} + u^{\mu}u_{\nu},
\end{split}
\end{equation}
where $a(t)$ is the scale factor of the Universe. In terms of the above decomposition, we
may write the Riemann curvature tensor for spatially flat Universe as
\begin{equation}\label{S-E2}
R_{\mu\alpha\beta\gamma} = (g_{\mu\beta}g_{\alpha\gamma} - g_{\mu\gamma}g_{\alpha\beta})
\rho_{1} + [u_{\mu}(g_{\alpha\gamma}u_{\beta} - g_{\alpha\beta}u_{\gamma}) - u_{\alpha}
(g_{\mu\gamma}u_{\beta} - g_{\mu\beta}u_{\gamma})]\rho_{2},
\end{equation}
where $\rho_{1} = H^{2}$ and $\rho_{2} = - \dot{H}$, and $H$ is the Hubble parameter.
Hence, the Ricci tensor and Ricci scalar can be expressed as
\begin{equation}\label{S-E3}
R_{\mu\nu} = g_{\mu\nu}(3\rho_{1} - \rho_{2}) + 2u_{\mu}u_{\nu}\rho_{2}, \ R = 6(2\rho_{1} 
- \rho_{2}).
\end{equation}
As a result, the Einstein's tensor can be expressed as
\begin{equation}\label{S-E4}
G_{\mu\nu} = R_{\mu\nu} - \frac{1}{2}g_{\mu\nu}R = (2\rho_{2} - 3\rho_{1})g_{\mu\nu} + 
2\rho_{2}u_{\mu}u_{\nu}.
\end{equation}
Because of the background symmetry of the spacetime, we may also consider the similar
expression for the auxiliary tensor field variables $\Phi_{\mu\nu}$ and $S_{\mu\nu}$
which are of the following form
\begin{equation}\label{S-E5a}
\Phi_{\mu\nu} = g_{\mu\nu}(3\varphi_{1} - \varphi_{2}) + 2\varphi_{2}u_{\mu}u_{\nu}, 
\ S_{\mu\nu} = g_{\mu\nu}(3\psi_{1} - \psi_{2}) + 2\psi_{2}u_{\mu}u_{\nu}.
\end{equation}
Using the above ansatz, we may write the terms in \ref{S-C10}) sequentially as
\begin{equation}\label{S-E6}
\begin{split}
\mathcal{G}_{\mu\nu}^{(2,1)} & = - 2S_{\mu\alpha}G_{ \ \nu}^{\alpha} = 2g_{\mu\nu}(3\psi_{1} 
- \psi_{2})(3\rho_{1} - 2\rho_{2}) + 4u_{\mu}u_{\nu}(3\rho_{1}\psi_{2} - 3\psi_{1}\rho_{2} 
+ \psi_{2}\rho_{2}) \\
\mathcal{G}_{\mu\nu}^{(2,2)} & = - RS_{\mu\nu} = - 6g_{\mu\nu}(2\rho_{1} - \rho_{2})
(3\psi_{1} - \psi_{2}) - 12u_{\mu}u_{\nu}\psi_{2}(2\rho_{1} - \rho_{2})
\end{split}
\end{equation}

\begin{equation}\label{S-E6a}
\begin{split}
\mathcal{G}_{\mu\nu}^{(2,3)} & = - \frac{1}{2}\Box S_{\mu\nu} = \frac{1}{2}g_{\mu\nu}
(3\ddot{\psi}_{1} - \ddot{\psi}_{2} + 9H\dot{\psi}_{1} - 3H\dot{\psi}_{2})\\
 & + \Big[\nabla_{\mu}H u_{\nu}\psi_{2} - H^{2}(u_{\mu}u_{\nu} + g_{\mu\nu})\psi_{2}
 + Hu_{\nu}\nabla_{\mu}\psi_{2} + \nabla_{\nu}H u_{\mu}\psi_{2} - H^{2}(u_{\mu}u_{\nu} 
 + g_{\mu\nu})\psi_{2}\\
 & + Hu_{\mu}\nabla_{\nu}\psi_{2} - 6H^{2}u_{\mu}u_{\nu}\psi_{2} - 2u_{\mu}u_{\nu}(H
 \dot{\psi}_{2} + \dot{H}\psi_{2}) + H(u_{\rho}u_{\mu} + g_{\rho\mu})u_{\nu}\nabla^{\rho}
 \psi_{2}\\
 & + Hu_{\mu}(u_{\rho}u_{\nu} + g_{\rho\nu})\nabla^{\rho}\psi_{2} + u_{\mu}u_{\nu}(
 \ddot{\psi}_{2} + 3H\dot{\psi}_{2})\Big]\\
\mathcal{G}_{\mu\nu}^{(2,4)} & = - \Phi_{\mu}^{ \ \alpha}\Phi_{\nu\alpha} = - g_{\mu\nu}
(3\varphi_{1} - \varphi_{2})^{2} - 4u_{\mu}u_{\nu}\varphi_{2}(3\varphi_{1} - 2\varphi_{2})\\
\mathcal{G}_{\mu\nu}^{(2,5)} & = \nabla_{\alpha}\nabla_{\mu}S_{ \ \nu}^{\alpha} = 3\nabla_{\mu}
\nabla_{\nu}\psi_{1} - \nabla_{\mu}\nabla_{\nu}\psi_{2} + 2H^{2}(u_{\mu}u_{\nu} + g_{\mu\nu})
\psi_{2} + 6H^{2}u_{\mu}u_{\nu}\psi_{2}\\
 & + 2u_{\mu}u_{\nu}(H\dot{\psi}_{2} + \psi_{2}\dot{H}) + 2g_{\mu\nu}(H\dot{\psi}_{2} + 
 \psi_{2}\dot{H} + 3H^{2}\psi_{2}) + 6H^{2}u_{\mu}u_{\nu}\psi_{2}\\
 & + 2u_{\mu}u_{\nu}(H\dot{\psi}_{2} + \psi_{2}\dot{H}) - 2\nabla_{\mu}H u_{\nu}\psi_{2}
 - 2Hu_{\nu}\nabla_{\mu}\psi_{2} - 6Hu_{\nu}\nabla_{\mu}\psi_{2} + 2u^{\alpha}u_{\nu}
 \nabla_{\alpha}\nabla_{\mu}\psi_{2}\\
\mathcal{G}_{\mu\nu}^{(2,6)} & = \frac{1}{2}g_{\mu\nu}G^{\alpha\beta}S_{\alpha\beta} = 
g_{\mu\nu}(9\rho_{2}\psi_{1} - 3\rho_{2}\psi_{2} - 18\rho_{1}\psi_{1} + 9\rho_{1}\psi_{2})\\
\mathcal{G}_{\mu\nu}^{(2,7)} & = \frac{1}{4}g_{\mu\nu}\Phi_{\alpha\beta}\Phi^{\alpha\beta} = 
3g_{\mu\nu}(3\varphi_{1}^{2} - 3\varphi_{1}\varphi_{2} + \varphi_{2}^{2})\\
\mathcal{G}_{\mu\nu}^{(2,8)} & = \frac{1}{4}g_{\mu\nu}RS_{ \ \alpha}^{\alpha} = 9g_{\mu\nu}
(2\rho_{1} - \rho_{2})(2\psi_{1} - \psi_{2})\\
\mathcal{G}_{\mu\nu}^{(2,9)} & = - \frac{1}{2}g_{\mu\nu}g^{\rho\sigma}\nabla_{\rho}S^{\alpha
\beta}\nabla_{\sigma}\Phi_{\alpha\beta} = 6g_{\mu\nu}\left(3\dot{\psi}_{1}\dot{\varphi}_{1}
- \frac{3}{2}(\dot{\psi}\dot{\varphi}_{2} + \dot{\psi}_{2}\dot{\varphi}_{1}) + \dot{\psi}_{2}
\dot{\varphi}_{2} + 2H^{2}\psi_{2}\varphi_{2}\right), \\
\mathcal{G}_{\mu\nu}^{(2,10)} & = - \frac{1}{2}g_{\mu\nu}\nabla_{\alpha}\nabla_{\beta}S^{\alpha
\beta} = \frac{1}{2}g_{\mu\nu}(3\ddot{\psi}_{1} - \ddot{\psi}_{2} + 9H\dot{\psi}_{1} - 3H
\dot{\psi}_{2}) - g_{\mu\nu}(\ddot{\psi}_{2} + 3H\dot{\psi}_{2})\\
 & - g_{\mu\nu}(3\dot{H}\psi_{2} + 3H\dot{\psi}_{2} + 9H^{2}\psi_{2}) \\
 \mathcal{G}_{\mu\nu}^{(2,11)} & = \frac{1}{2}\nabla_{\mu}\Phi_{\alpha\beta}\nabla_{\nu}S^{
\alpha\beta} + (\Phi \leftrightarrow S) = [18\nabla_{\mu}\varphi_{1}\nabla_{\nu}\psi_{1} - 
9\nabla_{\mu}\varphi_{1}\nabla_{\nu}\psi_{2} - 9\nabla_{\mu}\varphi_{2}\nabla_{\nu}\psi_{1}\\
 & + 6\nabla_{\mu}\varphi_{2}\nabla_{\nu}\psi_{2} - 4H^{2}\varphi_{2}\psi_{2}(u_{\mu}u_{\nu}
 + g_{\mu\nu})] + [18\nabla_{\mu}\psi_{1}\nabla_{\nu}\varphi_{1} - 9\nabla_{\mu}\psi_{1}
 \nabla_{\nu}\varphi_{2} - 9\nabla_{\mu}\psi_{2}\nabla_{\nu}\varphi_{1}\\
 & + 6\nabla_{\mu}\psi_{2}\nabla_{\nu}\varphi_{2} - 4H^{2}\varphi_{2}\psi_{2}(u_{\mu}u_{\nu}
 + g_{\mu\nu})]\\
 \mathcal{G}_{\mu\nu}^{(2,12)} & = \nabla_{\beta}\Phi_{ \ \alpha}^{\beta}\nabla_{\mu}S_{ \ \nu}
^{\alpha} + (\Phi \leftrightarrow S) = \Big[9\nabla_{\nu}\varphi_{1}\nabla_{\mu}\psi_{1} - 3
\nabla_{\nu}\varphi_{1}\nabla_{\mu}\psi_{2} - 3\nabla_{\nu}\varphi_{2}\nabla_{\mu}\psi_{1} + 
\nabla_{\nu}\varphi_{2}\nabla_{\mu}\psi_{2}\\
 & - 2Hu_{\nu}\psi_{2}(3\nabla_{\mu}\varphi_{1} - \nabla_{\mu}\varphi_{2}) + 2H\psi_{2}u_{\mu}
 u_{\nu}(3\dot{\varphi}_{1} - \dot{\varphi}_{2}) + 2H(3\dot{\varphi}_{1} - \dot{\varphi}_{2})
 (u_{\mu}u_{\nu} + g_{\mu\nu})\psi_{2}\\
 & - 2(3\dot{\varphi}_{1} - \dot{\varphi}_{2})u_{\nu}\nabla_{\mu}\psi_{2} - 6H\varphi_{2}u_{\nu}
 (3\nabla_{\mu}\psi_{1} - \nabla_{\mu}\psi_{2}) - 12H^{2}\varphi_{2}\psi_{2}(u_{\mu}u_{\nu} + 
 g_{\mu\nu})\\
 & + 12Hu_{\nu}\varphi_{2}\nabla_{\mu}\psi_{2} - 4H\dot{\varphi}_{2}\psi_{2}(u_{\mu}
 u_{\nu} + g_{\mu\nu}) - 2u_{\nu}\dot{\varphi}_{2}(3\nabla_{\mu}\psi_{1} - \nabla_{\mu}\psi_{2})
 + 4\dot{\varphi}_{2}u_{\nu}\nabla_{\mu}\psi_{2}\Big]\\
 & + \Big[9\nabla_{\nu}\psi_{1}\nabla_{\mu}\varphi_{1} - 3\nabla_{\nu}\psi_{1}\nabla_{\mu}\varphi_{2}
 - 3\nabla_{\nu}\psi_{2}\nabla_{\mu}\varphi_{1} + \nabla_{\nu}\psi_{2}\nabla_{\mu}\varphi_{2}\\
 & - 2Hu_{\nu}\varphi_{2}(3\nabla_{\mu}\psi_{1} - \nabla_{\mu}\psi_{2}) + 2H\varphi_{2}u_{\mu}
 u_{\nu}(3\dot{\psi}_{1} - \dot{\psi}_{2}) + 2H(3\dot{\psi}_{1} - \dot{\psi}_{2})(u_{\mu}u_{\nu} 
 + g_{\mu\nu})\varphi_{2}\\
 & - 2(3\dot{\psi}_{1} - \dot{\psi}_{2})u_{\nu}\nabla_{\mu}\varphi_{2} - 6H\psi_{2}u_{\nu}
 (3\nabla_{\mu}\varphi_{1} - \nabla_{\mu}\varphi_{2}) - 12H^{2}\varphi_{2}\psi_{2}(u_{\mu}u_{\nu} + 
 g_{\mu\nu})\\
 & + 12Hu_{\nu}\psi_{2}\nabla_{\mu}\varphi_{2} - 4H\dot{\psi}_{2}\varphi_{2}(u_{\mu}u_{\nu} 
 + g_{\mu\nu}) - 2u_{\nu}\dot{\psi}_{2}(3\nabla_{\mu}\varphi_{1} - \nabla_{\mu}\varphi_{2})
 + 4\dot{\psi}_{2}u_{\nu}\nabla_{\mu}\varphi_{2}\Big]\\
\end{split}
\end{equation}
\begin{equation}\label{S-E7}
\begin{split}
\mathcal{G}_{\mu\nu}^{(2,13)} & = - S_{ \ \nu}^{\alpha}\nabla_{\beta}\nabla_{\mu}\Phi_{ \ \alpha}
^{\beta} + (\Phi \leftrightarrow S) = - (3\psi_{1} - \psi_{2})\nabla_{\mu}\nabla_{\nu}(3\varphi_{1}
 - \varphi_{2}) - 2u_{\nu}\psi_{2}u^{\alpha}\nabla_{\alpha}\nabla_{\mu}(3\varphi_{1} - \varphi_{2})\\
 & + 2(3\psi_{1} - \psi_{2})\Big[u_{\nu}\nabla_{\mu}H\varphi_{2} + u_{\nu}H\nabla_{\mu}\varphi_{2}
 - H^{2}\varphi_{2}(g_{\mu\nu} + u_{\mu}u_{\nu}) - 3H^{2}u_{\mu}u_{\nu}\varphi_{2}\\
 & - u_{\mu}u_{\nu}(\dot{H}\varphi_{2} + \dot{\varphi}_{2}H) - 3H^{2}g_{\mu\nu}\varphi_{2}
 - g_{\mu\nu}(\dot{H}\varphi_{2} + \dot{\varphi}_{2}H) - 3H^{2}u_{\mu}u_{\nu}\varphi_{2} - 
 u_{\mu}u_{\nu}(\dot{H}\varphi_{2} + \dot{\varphi}_{2}H)\\
 & + 3Hu_{\nu}\nabla_{\mu}\varphi_{2} - u^{\beta}u_{\nu}\nabla_{\beta}\nabla_{\mu}\varphi_{2}\Big]
 + 4\psi_{2}\Big[ - u_{\nu}\nabla_{\mu}H\varphi_{2} - Hu_{\nu}\nabla_{\mu}\varphi_{2} + 3H^{2}
 u_{\mu}u_{\nu}\varphi_{2}\\
 & + u_{\mu}u_{\nu}(\dot{H}\varphi_{2} + H\dot{\varphi}_{2}) - 3H^{2}u_{\mu}u_{\nu}\varphi_{2}
 - u_{\mu}u_{\nu}(\dot{H}\varphi_{2} + \dot{\varphi}_{2}H) + 3H^{2}u_{\mu}u_{\nu}\varphi_{2} + 
 u_{\mu}u_{\nu}(\dot{H}\varphi_{2} + H\dot{\varphi}_{2})\\
 & - 3Hu_{\nu}\nabla_{\mu}\varphi_{2} + u_{\nu}u^{\beta}\nabla_{\beta}\nabla_{\mu}\varphi_{2}
 \Big] + (\varphi \leftrightarrow \psi)\\
\mathcal{G}_{\mu\nu}^{(2,14)} & = - \Phi_{ \ \nu}^{\alpha}\Box S_{\mu\alpha} + (\Phi 
\leftrightarrow S) = [g_{\mu\nu}(3\varphi_{1} - \varphi_{2}) + 2u_{\mu}u_{\nu}\varphi_{2}]
[3\ddot{\psi}_{1} - \ddot{\psi}_{2} + 9H\dot{\psi}_{1} - 3H\dot{\psi}_{2}]\\
 & + 2\Big[(3\varphi_{1} - \varphi_{2})u_{\nu}\nabla_{\mu}H \psi_{2} - H^{2}(3\varphi_{1} - 
 \varphi_{2})(u_{\mu}u_{\nu} + g_{\mu\nu})\psi_{2} + (3\varphi_{1} - \varphi_{2})Hu_{\nu}
 \nabla_{\mu}\psi_{2}\\
 & + (3\varphi_{1} - \varphi_{2})\nabla_{\nu}H u_{\mu}\psi_{2} - (3\varphi_{1} - \varphi_{2})
 H^{2}(u_{\mu}u_{\nu} + g_{\mu\nu})\psi_{2} + (3\varphi_{1} - \varphi_{2})Hu_{\mu}\nabla_{\nu}
 \psi_{2}\\
 & - 6H^{2}(3\varphi_{1} - \varphi_{2})u_{\mu}u_{\nu}\psi_{2} - 2(3\varphi_{1} - \varphi_{2})
 u_{\mu}u_{\nu}(H\dot{\psi}_{2} + \dot{H}\psi_{2}) + H(3\varphi_{1} - \varphi_{2})(u_{\rho}u_{\mu}
 + g_{\mu\rho})\nabla^{\rho}\psi_{2}\\
 & + u_{\mu}u_{\nu}(\ddot{\psi}_{2} + 3H\dot{\psi}_{2})(3\varphi_{1} - \varphi_{2})\Big] + 4\Big[
 - u_{\nu}\nabla_{\mu}H \varphi_{2}\psi_{2} - Hu_{\nu}(\nabla_{\mu}\psi_{2})\varphi_{2} - \dot{H}
 u_{\mu}u_{\nu}\psi_{2}\varphi_{2}\\
 & - Hu_{\mu}u_{\nu}\varphi_{2}\dot{\psi}_{2} + 6H^{2}u_{\mu}u_{\nu}\varphi_{2}\psi_{2} + 2u_{\mu}
 u_{\nu}\varphi_{2}(H\dot{\psi}_{2} + \dot{H}\psi_{2}) - u_{\mu}u_{\nu}\varphi_{2}(\ddot{\psi}_{2} 
 + 3H\dot{\psi}_{2})\\
 & - Hu_{\nu}(u_{\rho}u_{\mu} + g_{\mu\rho})(\nabla^{\rho}\psi_{2})\varphi_{2}\Big] + (\varphi \leftrightarrow \psi), \\
\mathcal{G}_{\mu\nu}^{(2,15)} & = \Phi^{\alpha\beta}\nabla_{\beta}\nabla_{\mu}S_{\nu\alpha} + 
(\Phi \leftrightarrow S) = (3\varphi_{1} - \varphi_{2})\nabla_{\nu}\nabla_{\mu}(3\psi_{1} - 
\psi_{2}) + 2(3\varphi_{1} - \varphi_{2})\Big[\dot{H}(g_{\mu\nu} + u_{\mu}u_{\nu})\psi_{2}\\
 & + 3H^{2}\psi_{2}(g_{\mu\nu} + u_{\mu}u_{\nu}) + H\dot{\psi}_{2}(g_{\mu\nu} + u_{\mu}u_{\nu})
 - u_{\nu}\nabla_{\mu}H \psi_{2} + u_{\mu}u_{\nu}\dot{H}\psi_{2} + H^{2}(g_{\mu\nu} + u_{\mu}
 u_{\nu})\psi_{2}\\
 &  - Hu_{\nu}\nabla_{\mu}\psi_{2} + 3H^{2}u_{\mu}u_{\nu}\psi_{2} + Hu_{\mu}u_{\nu}\dot{\psi}_{2}
 - 3Hu_{\nu}\nabla_{\mu}\psi_{2} + u_{\nu}u^{\alpha}\nabla_{\alpha}\nabla_{\mu}\psi_{2}\Big]
 + 2\varphi_{2}u_{\nu}u^{\beta}\nabla_{\beta}\nabla_{\mu}(3\psi_{1} - \psi_{2})\\
 & + 4\varphi_{2}\Big[ - \dot{H}\psi_{2}(g_{\mu\nu} + u_{\mu}u_{\nu}) - Hg_{\mu\nu}\dot{\psi}_{2}
 - Hu_{\mu}u_{\nu}\dot{\psi}_{2} - u_{\nu}u^{\beta}\nabla_{\beta}\nabla_{\mu}\psi_{2}\Big] + 
 (\varphi \leftrightarrow \psi)\\
\mathcal{G}_{\mu\nu}^{(2,16)} & = - \nabla^{\beta}S_{ \ \mu}^{\alpha}\nabla_{\nu}\Phi_{\alpha
\beta} + (\Phi \leftrightarrow S) = - (3\nabla_{\mu}\psi_{1} - \nabla_{\mu}\psi_{2})(3\nabla
_{\nu}\varphi_{1} - \nabla_{\nu}\varphi_{2}) - 2H(3\dot{\psi}_{1} - \dot{\psi}_{2})\varphi_{2}
(u_{\mu}u_{\nu} + g_{\mu\nu})\\
 & - 2Hu_{\mu}u_{\nu}\varphi_{2}(3\dot{\psi}_{1} - \dot{\psi}_{2}) + 2Hu_{\mu}\varphi_{2}(3
 \nabla_{\nu}\psi_{1} - \nabla_{\nu}\psi_{2}) + 2(3\dot{\psi}_{1} - \dot{\psi}_{2})u_{\mu}
 \nabla_{\nu}\varphi_{2}\\
 & + 6H\psi_{2}u_{\mu}(3\nabla_{\nu}\varphi_{1} - \nabla_{\nu}\varphi_{2}) + 4H^{2}\psi_{2}
 \varphi_{2}(u_{\mu}u_{\nu} + g_{\mu\nu}) + 2\dot{\psi}_{2}u_{\mu}(3\nabla_{\nu}\varphi_{1}
 - \nabla_{\nu}\varphi_{2}) + 4u_{\mu}u_{\nu}\dot{\psi}_{2}\varphi_{2}H\\
 & - 4Hu_{\mu}\nabla_{\nu}\psi_{2}\varphi_{2} - 4\dot{\psi}_{2}u_{\mu}\nabla_{\nu}\varphi_{2}
 + (\varphi \leftrightarrow \psi).
\end{split}
\end{equation}
Once we have the following terms in the covariant form, we may obtain the first modified Friedmann
equations by computing the $u^{\mu}u^{\nu}\mathcal{G}_{\mu\nu}^{(2)}$ which we do now below in the 
sequential manner.
\begin{equation}\label{S-E9}
\begin{split}
2u^{\mu}u^{\nu}\mathcal{G}_{\mu\nu}^{(2,1)} & = - 36\psi_{1}\rho_{1} + 36\psi_{2}\rho_{1}\\
2u^{\mu}u^{\nu}\mathcal{G}_{\mu\nu}^{(2,2)} & = 72\psi_{1}\rho_{1} - 72\rho_{1}\psi_{2} - 36
\rho_{2}\psi_{1} + 36\rho_{2}\psi_{2}\\
2u^{\mu}u^{\nu}\mathcal{G}_{\mu\nu}^{(2,3)} & = - 3(\ddot{\psi}_{1} + 3H\dot{\psi}_{1}) + 3
(\ddot{\psi}_{2} + 3H\dot{\psi}_{2}) - 12H^{2}\psi_{2}\\
2u^{\mu}u^{\nu}\mathcal{G}_{\mu\nu}^{(2,4)} & = 18\varphi_{1}^{2} - 36\varphi_{1}\varphi_{2}
+ 18\varphi_{2}^{2}\\
2u^{\mu}u^{\nu}\mathcal{G}_{\mu\nu}^{(2,5)} & = 6(\ddot{\psi}_{1} - \ddot{\psi}_{2}) - 12H
\dot{\psi}_{2} + 12H^{2}\psi_{2}\\
2u^{\mu}u^{\nu}\mathcal{G}_{\mu\nu}^{(2,6)} & = - 18\rho_{2}\psi_{1} + 6\rho_{2}\psi_{2} + 
36\rho_{1}\psi_{1} - 18\rho_{1}\psi_{2}\\
2u^{\mu}u^{\nu}\mathcal{G}_{\mu\nu}^{(2,7)} & = - 18\varphi_{1}^{2} + 18\varphi_{1}\varphi_{2}
- 6\varphi_{2}^{2}\\
2u^{\mu}u^{\nu}\mathcal{G}_{\mu\nu}^{(2,8)} & = - 72\rho_{1}\psi_{1} + 36\rho_{1}\psi_{2}
+ 36\rho_{2}\psi_{1} - 18\rho_{2}\psi_{2}\\
2u^{\mu}u^{\nu}\mathcal{G}_{\mu\nu}^{(2,9)} & = - 36\dot{\psi}_{1}\dot{\varphi}_{1} + 18
\dot{\psi}_{1}\dot{\varphi}_{2} + 18\dot{\psi}_{2}\dot{\varphi}_{1} - 12\dot{\psi}_{2}
\dot{\varphi}_{2} - 24H^{2}\psi_{2}\varphi_{2}\\
2u^{\mu}u^{\nu}\mathcal{G}_{\mu\nu}^{(2,10)} & = - 3\ddot{\psi}_{1} + 3\ddot{\psi}_{2}
- 9H\dot{\psi}_{1} + 15H\dot{\psi}_{2} + 6\dot{H}\psi_{2} + 18H^{2}\psi_{2}\\
2u^{\mu}u^{\nu}\mathcal{G}_{\mu\nu}^{(2,11)} & = 72\dot{\varphi}_{1}\dot{\psi}_{1} - 36
\dot{\varphi}_{1}\dot{\psi}_{2} - 36\dot{\psi}_{1}\dot{\varphi}_{2} + 24\dot{\varphi}_{2}
\dot{\psi}_{2}\\
2u^{\mu}u^{\nu}\mathcal{G}_{\mu\nu}^{(2,12)} & = 36\dot{\varphi}_{1}\dot{\psi}_{1} - 36
\dot{\varphi}_{1}\dot{\psi}_{2} - 36\dot{\psi}_{1}\dot{\varphi}_{2} + 56\dot{\psi}_{2}
\dot{\varphi}_{2}\\
 & - 36H\varphi_{2}\dot{\psi}_{1} - 36H\psi_{2}\dot{\varphi}_{1} + 36H\varphi_{2}\dot{\psi}
 _{2} + 36H\psi_{2}\dot{\varphi}_{2}\\
2u^{\mu}u^{\nu}\mathcal{G}_{\mu\nu}^{(2,13)} & = - 18\psi_{1}\ddot{\varphi}_{1} - 18\varphi_{1}
\ddot{\psi}_{1} + 18\psi_{1}\ddot{\varphi}_{2} + 18\varphi_{1}\ddot{\psi}_{2} - 14\psi_{2}
\ddot{\varphi}_{2}\\
 & - 14\varphi_{2}\ddot{\psi}_{2} + 18\varphi_{2}\ddot{\psi}_{1} + 18\psi_{2}\ddot{\varphi}_{1}
 + 72H^{2}\varphi_{2}\psi_{2} - 36H\psi_{2}\dot{\varphi}_{2} - 36H\varphi_{2}\dot{\psi}_{2}\\
 & - 4\psi_{2}\ddot{\varphi}_{2} - 4\varphi_{2}\ddot{\psi}_{2} + 36H\psi_{1}\dot{\varphi}_{2}
 + 36H\varphi_{1}\dot{\psi}_{2} - 36H^{2}\psi_{1}\varphi_{2} - 36H^{2}\varphi_{1}\psi_{2},  \\
 2u^{\mu}u^{\nu}\mathcal{G}_{\mu\nu}^{(2,14)} & = 18\varphi_{2}\ddot{\psi}_{1} + 18\psi_{2}
\ddot{\varphi}_{1} - 18\varphi_{1}\ddot{\psi}_{1} - 18\psi_{1}\ddot{\varphi}_{1} + 54H\varphi_{2}
\dot{\psi}_{1} + 54H\psi_{2}\dot{\varphi}_{1}\\
 & - 54H\varphi_{1}\dot{\psi}_{1} - 54H\psi_{1}\dot{\varphi}_{1} - 18\varphi_{2}\ddot{\psi}_{2}
 - 18\psi_{2}\ddot{\varphi}_{2} + 18\varphi_{1}\ddot{\psi}_{2} + 18\psi_{1}\ddot{\varphi}_{2}\\
 & - 54H\varphi_{2}\dot{\psi}_{2} - 54H\psi_{2}\dot{\varphi}_{2} + 54H\varphi_{1}\dot{\psi}_{2}
 + 54H\psi_{1}\dot{\varphi}_{2} + 14H^{2}\psi_{2}\varphi_{2}\\
 & - 72H^{2}\varphi_{1}\psi_{2} - 72H^{2}\psi_{1}\varphi_{2}\\
2u^{\mu}u^{\nu}\mathcal{G}_{\mu\nu}^{(2,15)} & = 18\varphi_{1}\ddot{\psi}_{1} + 18\psi_{1}
\ddot{\varphi}_{1} - 36H\varphi_{1}\dot{\psi}_{2} - 36H\psi_{1}\dot{\varphi}_{2} + 12H\varphi_{2}
\dot{\psi}_{2} + 12H\psi_{2}\dot{\varphi}_{2}\\
 & + 36H^{2}\varphi_{1}\psi_{2} + 16H^{2}\psi_{1}\varphi_{2} - 18\varphi_{1}\ddot{\psi}_{2} - 18
 \psi_{1}\ddot{\varphi}_{2} + 18\varphi_{2}\ddot{\psi}_{2} + 18\psi_{2}\ddot{\varphi}_{2}\\
 & - 18\varphi_{2}\ddot{\psi}_{1} - 18\psi_{2}\ddot{\varphi}_{1} - 24H^{2}\varphi_{2}\psi_{2}\\
2u^{\mu}u^{\nu}\mathcal{G}_{\mu\nu}^{(2,16)} & = - 36\dot{\psi}_{1}\dot{\varphi}_{1} + 12
\dot{\psi}_{2}\dot{\varphi}_{1} + 12\dot{\varphi}_{2}\dot{\psi}_{1} + 36H\psi_{2}\dot{\varphi}_{1}
+ 36H\varphi_{2}\dot{\psi}_{1}\\
 & - 4H\dot{\psi}_{2}\varphi_{2} - 4H\dot{\varphi}_{2}\psi_{2} + 24\dot{\psi}_{1}\dot{\varphi}_{2}
 + 24\dot{\varphi}_{1}\dot{\psi}_{2} - 28\dot{\varphi}_{2}\dot{\psi}_{2} - 4H\psi_{2}\dot{\varphi}
 _{2} - 4H\varphi_{2}\dot{\psi}_{2}.
\end{split}
\end{equation}
Combining all these terms, we finally obtain the following relation
\begin{equation}\label{S-E11}
\begin{split}
u^{\mu}u^{\nu}\mathcal{G}_{\mu\nu}^{(2)} & = 18\dot{H}\psi_{1} - 18\dot{H}\psi_{2} - 18H\dot{\psi}
_{1} + 12H\dot{\psi}_{2} - 18\varphi_{1}\varphi_{2} + 12\varphi_{2}^{2}\\
 & + 36\dot{\varphi}_{1}\dot{\psi}_{1} - 18\dot{\psi}_{1}\dot{\varphi}_{2} - 18\dot{\psi}_{2}
 \dot{\varphi}_{1} + 40\dot{\varphi}_{2}\dot{\psi}_{2} + 168H^{2}\psi_{2}\varphi_{2} + 54H
 \varphi_{2}\dot{\psi}_{1}\\
 & + 54H\psi_{2}\dot{\varphi}_{1} - 50H\varphi_{2}\dot{\psi}_{2} - 50H\psi_{2}\dot{\varphi}_{2}
 - 18\psi_{1}\ddot{\varphi}_{1} - 18\varphi_{1}\ddot{\psi}_{1} + 18\psi_{1}\ddot{\varphi}_{2}\\
 & + 18\varphi_{1}\ddot{\psi}_{2} - 18\psi_{2}\ddot{\varphi}_{2} - 18\varphi_{2}\ddot{\psi}_{2}
 + 18\varphi_{2}\ddot{\psi}_{1} + 18\psi_{2}\ddot{\varphi}_{1} + 54H\psi_{1}\dot{\varphi}_{2}\\
 & + 54H\varphi_{1}\dot{\psi}_{2} - 72H^{2}\psi_{1}\varphi_{2} - 72H^{2}\varphi_{1}\psi_{2} - 
 54H\varphi_{1}\dot{\psi}_{1} - 54H\psi_{1}\dot{\varphi}_{1}.
\end{split}
\end{equation}
Now we consider the background field equations for the auxiliary tensor field variables
\begin{equation}\label{S-E12}
\begin{split}
\Box\Phi_{\mu\nu} & = - R_{\mu\nu}\\
\implies & g_{\mu\nu}(3\ddot{\varphi}_{1} - \ddot{\varphi}_{2} + 9H\dot{\varphi}_{1} - 
3H\dot{\varphi}_{2}) + 2\nabla_{\mu}H u_{\nu}\varphi_{2} + 2\nabla_{\nu}H u_{\mu}\varphi_{2}
- 4H^{2}\varphi_{2}(u_{\mu}u_{\nu} + g_{\mu\nu})\\
 & + 2Hu_{\nu}\nabla_{\mu}\varphi_{2} + 2Hu_{\mu}\nabla_{\nu}\varphi_{2} - 12H^{2}\varphi_{2}
 u_{\mu}u_{\nu} - 4\dot{H}\varphi_{2}u_{\mu}u_{\nu} - 4H\dot{\varphi}_{2}u_{\mu}u_{\nu}\\
 & + 2H u_{\nu}\nabla_{\mu}\varphi_{2} + 2H u_{\mu}\nabla_{\nu}\varphi_{2} - 4Hu_{\mu}u_{\nu}
 \dot{\varphi}_{2} + 2u_{\mu}u_{\nu}(\ddot{\varphi}_{2} + 3H\dot{\varphi}_{2}) = R_{\mu\nu}.
\end{split}
\end{equation}
Contraction of the above equation with $u^{\mu}u^{\nu}$ leads to the following field equation
\begin{equation}\label{S-E13}
- (\ddot{\varphi}_{1} + 3H\dot{\varphi}_{1}) + (\ddot{\varphi}_{2} + 3H\dot{\varphi}_{2}) - 4
H^{2}\varphi_{2} = - (\dot{H} + H^{2}),
\end{equation}
whereas the $ij$th component leads to the following field equation
\begin{equation}\label{S-E14}
3(\ddot{\varphi}_{1} + 3H\dot{\varphi}_{1}) - (\ddot{\varphi}_{2} + 3H\dot{\varphi}_{2}) - 
4H^{2}\varphi_{2} = \dot{H} + 3H^{2}.
\end{equation}
Combining the above two field equations, we obtain the following two relations
\begin{equation}\label{S-E15}
\begin{split}
\ddot{\varphi}_{1} + 3H\dot{\varphi}_{1} - 4H^{2}\varphi_{2} & = H^{2}\\
\ddot{\varphi}_{2} + 3H\dot{\varphi}_{2} - 8H^{2}\varphi_{2} & = - \dot{H}.
\end{split}
\end{equation}
In the similar manner, we also have the following field equations
\begin{equation}\label{S-E16}
\begin{split}
\Box S_{\mu\nu} & = - \Phi_{\mu\nu}\\
\implies & g_{\mu\nu}(3\ddot{\psi}_{1} - \ddot{\psi}_{2} + 9H\dot{\psi}_{1} - 3H\dot{\psi}_{2})
+ 2\nabla_{\mu}H u_{\nu}\psi_{2} + 2\nabla_{\nu}H u_{\mu}\psi_{2}\\
 & - 4H^{2}\psi_{2}(u_{\mu}u_{\nu} + g_{\mu\nu}) + 2Hu_{\nu}\nabla_{\mu}\psi_{2} + 2Hu_{\mu}
 \nabla_{\nu}\psi_{2} - 12H^{2}\psi_{2}u_{\mu}u_{\nu}\\
 & - 4\dot{H}\psi_{2}u_{\mu}u_{\nu} - 4H\dot{\psi}_{2}u_{\mu}u_{\nu} - 2Hu_{\nu}\nabla_{\mu}
 \psi_{2} - 2H u_{\mu}\nabla_{\nu}\psi_{2} + 4Hu_{\mu}u_{\nu}\dot{\psi}_{2}\\
 & + 2u_{\mu}u_{\nu}(\ddot{\psi}_{2} + 3H\dot{\psi}_{2}) = \Phi_{\mu\nu}.
\end{split}
\end{equation}
Contracting the above equation with $u^{\mu}u^{\nu}$, we obtain the following relation
\begin{equation}\label{S-E17}
- (\ddot{\psi}_{1} + 3H\dot{\psi}_{1}) + (\ddot{\psi}_{2} + 3H\dot{\psi}_{2}) - 4H^{2}
\psi_{2} = \varphi_{2} - \varphi_{1},
\end{equation}
whereas the $ij$the component of the field equation \ref{S-E16}) is given by
\begin{equation}\label{S-E17a}
3(\ddot{\psi}_{1} + 3H\dot{\psi}_{1}) - (\ddot{\psi}_{2} + 3H\dot{\psi}_{2}) - 4H^{2}
\psi_{2} = 3\varphi_{1} - \varphi_{2}.
\end{equation}
Combining the above two equations, we obtain the following two relations
\begin{equation}\label{S-E18}
\begin{split}
\ddot{\psi}_{1} + 3H\dot{\psi}_{1} & = \varphi_{1} + 4H^{2}\psi_{2}\\
\ddot{\psi}_{2} + 3H\dot{\psi}_{2} & = \varphi_{2} + 8H^{2}\psi_{2}.
\end{split}
\end{equation}
Therefore, the equation \ref{S-E11}) can also be expressed as
\begin{equation}\label{S-E19}
\begin{split}
u^{\mu}u^{\nu}\mathcal{G}_{\mu\nu}^{(2)} & = 18H^{2}\psi_{2} - 18H^{2}\psi_{1} - 18H\dot{\psi}
_{1} + 12H\dot{\psi}_{2} + 18\varphi_{1}\varphi_{2} - 6\varphi_{2}^{2} - 18\varphi_{1}^{2} + 
36\dot{\varphi}_{1}\dot{\psi}_{1}\\
 & - 18\dot{\varphi}_{1}\dot{\psi}_{2} - 18\dot{\psi}_{1}\dot{\varphi}_{2} + 40\dot{\varphi}_{2}
 \dot{\psi}_{2} + 24H^{2}\psi_{2}\varphi_{2} + 4H\varphi_{2}\dot{\psi}_{2} + 4H\psi_{2}\dot{\varphi}
 _{2},
\end{split}
\end{equation}
where we used the previous set of field equations of the auxiliary tensor field variables.

Combining the previous set of results in this section and previous section, we may now write
the modified first Friedmann equation to be of the following form
\begin{equation}\label{S-E20}
\begin{split}
3H^{2} & (1 - 2\mu_{1}S + 6\mu_{2}\psi_{2} - 6\mu_{2}\psi_{1} + 8\mu_{2}\psi_{2}\varphi_{2})
= \kappa^{2}\rho_{m} + \mu_{1}\left(6H\dot{S} + \frac{1}{2}\Phi^{2} - \dot{S}\dot{\Phi}\right)\\
 & + \mu_{2}[18H\dot{\psi}_{1} - 12H\dot{\psi}_{2} - 18\varphi_{1}\varphi_{2} + 6\varphi_{2}^{2}
 + 18\varphi_{1}^{2} - 36\dot{\varphi}_{1}\dot{\psi}_{1} + 18\dot{\varphi}_{1}\dot{\psi}_{2}
 + 18\dot{\psi}_{1}\dot{\varphi}_{2}\\
 & - 40\dot{\varphi}_{2}\dot{\psi}_{2} - 4H\varphi_{2}\dot{\psi}_{2} - 4H\psi_{2}\dot{\varphi}
 _{2}].
\end{split}
\end{equation}
The above equation can also be understood as
\begin{equation}\label{S-E21}
1 = \Omega_{m} + \sum_{i = 1}^{14}\Omega_{i},
\end{equation}
where $\Omega_{m}$ is the ratio of energy density of matter (including the dark matter) to the 
critical energy density and $\sum_{i = 1}^{14}\Omega_{i} = \Omega_{DE}$ is the ratio of energy 
density of dark energy to the critical energy density. In the above equations, we used the 
following definitions
\begin{equation}\label{S-E22}
\begin{split}
\Omega_{m} & = \frac{\kappa^{2}\rho}{3H^{2}\bar{\Delta}}, \ \Omega_{1} = \frac{2\mu_{1}\dot{S}}{H
\bar{\Delta}}, \ \Omega_{2} = \frac{1}{6}\frac{\mu_{1}\Phi^{2}}{H^{2}\bar{\Delta}}, \ \Omega_{3} 
= - \frac{\mu_{1}\dot{S}\dot{\Phi}}{3H^{2}\bar{\Delta}}\\
\Omega_{4} & = \frac{16\mu_{2}\dot{\psi}_{1}}{H\bar{\Delta}}, \ \Omega_{5} = - \frac{4\mu_{2}
\dot{\psi}_{2}}{H\bar{\Delta}}, \ \Omega_{6} = - \frac{6\mu_{2}\varphi_{1}\varphi_{2}}{H^{2}
\bar{\Delta}}, \ \Omega_{7} = \frac{2\mu_{2}\varphi_{2}^{2}}{H^{2}\bar{\Delta}}\\
\Omega_{8} & = \frac{6\mu_{2}\varphi_{1}^{2}}{H^{2}\bar{\Delta}}, \ \Omega_{9} = - \frac{12
\mu_{2}\dot{\varphi}_{1}\dot{\psi}_{1}}{H^{2}\bar{\Delta}}, \ \Omega_{10} = \frac{6\mu_{2}
\dot{\varphi}_{1}\dot{\psi}_{2}}{H^{2}\bar{\Delta}}, \ \Omega_{11} = \frac{6\mu_{2}\dot{\psi}
_{1}\dot{\varphi}_{2}}{H^{2}\bar{\Delta}}\\
\Omega_{12} & = - \frac{40\mu_{2}\dot{\varphi}_{2}\dot{\psi}_{2}}{3H^{2}\bar{\Delta}}, \ 
\Omega_{13} = - \frac{4\mu_{2}\varphi_{2}\dot{\psi}_{2}}{3H\bar{\Delta}}, \ \Omega_{14} = 
- \frac{4\mu_{2}\psi_{2}\dot{\varphi}_{2}}{3H\bar{\Delta}},  
\end{split}
\end{equation}
where
\begin{equation}\label{S-E23}
\bar{\Delta} = (1 - 2\mu_{1}S + 6\mu_{2}\psi_{2} - 6\mu_{2}\psi_{1} + 8\mu_{2}\psi_{2}\varphi_{2}).
\end{equation}
The evolution equations of $\{\Omega_{i}\}$s are given by 
\begin{equation}\label{S-E24}
\begin{split}
\dot{\Omega}_{m} & = - 3H\Omega_{m} - \Omega_{m}H\left(2\frac{\dot{H}}{H^{2}} + \frac{
\dot{\Delta}}{\Delta H}\right), \ \dot{\Omega}_{1} = \Omega_{1}H\left(\frac{\ddot{S}}{H\dot{S}} 
- \frac{\dot{H}}{H^{2}} - \frac{\dot{\Delta}}{H\Delta}\right)\\
\dot{\Omega}_{2} & = \Omega_{2}H\left(2\frac{\dot{\Phi}}{H\Phi} - 2\frac{\dot{H}}{H^{2}} - 
\frac{\dot{\Delta}}{H\Delta}\right), \ \dot{\Omega}_{3} = \Omega_{3}H\left(\frac{\ddot{S}}{H
\dot{S}} + \frac{\ddot{\Phi}}{H\dot{\Phi}} - 2\frac{\dot{H}}{H^{2}} - \frac{\dot{\Delta}}{
H\Delta}\right)\\
\dot{\Omega}_{4} & = \Omega_{4}H\left(\frac{\ddot{\psi}_{1}}{H\dot{\psi}_{1}} - \frac{\dot{H}}
{H^{2}} - \frac{\dot{\Delta}}{H\Delta}\right), \ \dot{\Omega}_{5} = \Omega_{5}H\left(\frac{
\ddot{\psi}_{2}}{H\dot{\psi}_{2}} - \frac{\dot{H}}{H^{2}} - \frac{\dot{\Delta}}{H\Delta}\right)\\
\dot{\Omega}_{6} & = \Omega_{6}H\left(\frac{\dot{\varphi}_{1}}{H\varphi_{1}} + \frac{\dot{
\varphi}_{2}}{H\varphi_{2}} - 2\frac{\dot{H}}{H^{2}} - \frac{\dot{\Delta}}{H\Delta}\right), \ 
\dot{\Omega}_{7} = \Omega_{7}H\left(2\frac{\dot{\varphi}_{2}}{H\varphi_{2}} - 2\frac{\dot{H}}
{H^{2}} - \frac{\dot{\Delta}}{H\Delta}\right) \\
\dot{\Omega}_{8} & = \Omega_{8}H\left(2\frac{\dot{\varphi}_{1}}{H\varphi_{1}} - 2\frac{\dot{H}}
{H^{2}} - \frac{\dot{\Delta}}{H\Delta}\right), \ \dot{\Omega}_{9} = \Omega_{9}H\left(\frac{
\ddot{\varphi}_{1}}{H\dot{\varphi}_{1}} + \frac{\ddot{\psi}_{1}}{H\dot{\psi}_{1}} - 2\frac{
\dot{H}}{H^{2}} - \frac{\dot{\Delta}}{H\Delta}\right)
\end{split}
\end{equation}

\begin{equation*}
\begin{split}
\dot{\Omega}_{10} & = \Omega_{10}H\left(\frac{\ddot{\varphi}_{1}}{H\dot{\varphi}_{1}} + 
\frac{\ddot{\psi}_{2}}{H\dot{\psi}_{2}} - 2\frac{\dot{H}}{H^{2}} - \frac{\dot{\Delta}}{H
\Delta}\right), \ \dot{\Omega}_{11} = \Omega_{11}H\left(\frac{\ddot{\psi}_{1}}{H\dot{\psi}_{1}} 
+ \frac{\ddot{\varphi}_{2}}{H\dot{\varphi}_{2}} - 2\frac{\dot{H}}{H^{2}} - \frac{\dot{\Delta}}
{H\Delta}\right)\\
\dot{\Omega}_{12} & = \Omega_{12}H\left(\frac{\ddot{\varphi}_{2}}{H\dot{\varphi}_{2}} + 
\frac{\ddot{\psi}_{2}}{H\dot{\psi}_{2}} - 2\frac{\dot{H}}{H^{2}} - \frac{\dot{\Delta}}{H
\Delta}\right), \ \dot{\Omega}_{13} = \Omega_{13}H\left(\frac{\dot{\varphi}_{2}}{H\varphi_{2}} 
+ \frac{\ddot{\psi}_{2}}{H\dot{\psi}_{2}} - \frac{\dot{H}}{H^{2}} - \frac{\dot{\Delta}}{H
\Delta}\right)\\
\dot{\Omega}_{14} & = \Omega_{14}H\left(\frac{\dot{\psi}_{2}}{H\psi_{2}} + \frac{\ddot{\varphi
_{2}}}{H\dot{\varphi}_{2}} - \frac{\dot{H}}{H^{2}} - \frac{\dot{\Delta}}{H\Delta}\right).
\end{split}
\end{equation*}
Now we note the following relations first
\begin{equation}\label{S-E25}
\begin{split}
\frac{\ddot{\psi}_{1}}{\dot{\psi}_{1}H} & = \frac{\varphi_{1}}{\dot{\psi}_{1}H} + \frac{4H
\psi_{2}}{\dot{\psi}_{1}} - 3, \ \frac{H\psi_{2}}{\dot{\psi}_{1}} = - \frac{9}{2}\frac{\Omega
_{14}}{\Omega_{11}}\\
\frac{\ddot{\psi}_{2}}{H\dot{\psi}_{2}} & = \frac{\varphi_{2}}{H\dot{\psi}_{2}} + 8\frac{H
\psi_{2}}{\dot{\psi}_{2}} - 3, \ \frac{\varphi_{2}}{\dot{\psi}_{2}H} = - \frac{2}{3}\frac{
\Omega_{7}}{\Omega_{13}}, \ \frac{H\psi_{2}}{\dot{\psi}_{2}} = 10\frac{\Omega_{14}}{\Omega_{12}}
\\
\frac{\dot{\varphi}_{2}}{\varphi_{2}H} & = \frac{1}{10}\frac{\Omega_{12}}{\Omega_{13}}\\
\frac{\Omega_{9}}{\Omega_{8}} & = - 2\left(\frac{\dot{\varphi}_{1}}{\varphi_{1}H}\right)
\left(\frac{\dot{\psi}_{1}H}{\varphi_{1}}\right)\\
\frac{\ddot{\varphi}_{1}}{\dot{\varphi}_{1}H} & = \frac{H}{\dot{\varphi}_{1}} + 4\frac{H
\varphi_{2}}{\varphi_{1}} - 3, \ \frac{\Omega_{9}}{\Omega_{4}} = - 2\frac{\dot{\varphi}_{1}}
{H}, \ \frac{\Omega_{10}}{\Omega_{13}} = - \frac{9}{2}\frac{\dot{\varphi}_{1}}{\varphi_{2}H}
\\
\frac{\ddot{\varphi}_{2}}{\dot{\varphi}_{2}H} & = - \frac{\dot{H}}{\dot{\varphi}_{2}H} + 
8\frac{H\varphi_{2}}{\dot{\varphi}_{2}} - 3 = - \frac{\dot{H}/H^{2}}{\dot{\varphi}_{2}/H^{2}}
+ 8\frac{H\varphi_{2}}{\dot{\varphi}_{2}} - 3, \ \frac{\Omega_{11}}{\Omega_{4}} = \frac{\dot{\varphi_{2}}}{H}\\
\frac{\dot{\Delta}}{\Delta H} & = - \frac{\Omega_{1}}{3} - \frac{\Omega_{5}}{6} - 
\frac{\Omega_{4}}{3} + 10\Omega_{3} + 10\Omega_{4}.
\end{split}
\end{equation}
Now recall the auxiliary scalar field equations
\begin{equation}\label{S-E26}
\begin{split}
\ddot{S} + 3H\dot{S} = \Phi & \implies \frac{\ddot{S}}{H\dot{S}} = \frac{\Phi}{H\dot{S}} - 3\\
\ddot{\Phi} + 3H\dot{\Phi} = 6(\dot{H} + 2H^{2}) & \implies \frac{\ddot{\Phi}}{\dot{\Phi}H} = 
6\left(\frac{\dot{H}}{H\dot{\Phi}} + 2\frac{H}{\dot{\Phi}}\right) - 3 = 6\left(\frac{\dot{H}
/H^{2}}{\dot{\Phi}/H} + \frac{2}{\dot{\Phi}/H}\right) - 3\\
\frac{\Omega_{3}}{\Omega_{1}} = - \frac{1}{6}\frac{\dot{\Phi}}{H}, & \ \frac{\Omega_{3}}{\Omega
_{2}} = - 2\frac{\dot{S}}{S}\frac{\dot{\Phi}}{\Phi} \implies 2\Omega_{2}\frac{\dot{\Phi}}{\Phi H}
= - \Omega_{3}\frac{\Phi}{\dot{S}H}. 
\end{split}
\end{equation}
In order to compute $\dot{H}/H^{2}$, we now look at the spatial components of the modified
Friedmann equations, and for that we need the following quantities
\begin{equation}\label{S-E27}
\begin{split}
2\mathcal{G}_{ij}^{(2,1)} & = 4a^{2}\delta_{ij}(9\psi_{1}\rho_{1} - 6\psi_{1}\rho_{2} - 3
\rho_{1}\psi_{2} + 2\rho_{2}\psi_{2})\\
2\mathcal{G}_{ij}^{(2,2)} & = - 12a^{2}\delta_{ij}(6\rho_{1}\psi_{1} - 2\rho_{1}\psi_{2}
- 3\psi_{1}\rho_{2} + \rho_{2}\psi_{2})\\
2\mathcal{G}_{ij}^{(2,3)} & = a^{2}\delta_{ij}(3\ddot{\psi}_{1} - \ddot{\psi}_{2} + 9H
\dot{\psi}_{1} - 3H\dot{\psi}_{2} - 4H^{2}\psi_{2})\\
2\mathcal{G}_{ij}^{(2,4)} & = - 2a^{2}\delta_{ij}(9\varphi_{1}^{2} - 6\varphi_{1}\varphi_{2}
+ \varphi_{2}^{2}) \\
2\mathcal{G}_{ij}^{(2,5)} & = a^{2}\delta_{ij}(- 6H\dot{\psi}_{1} + 2H\dot{\psi}_{2} + 4H^{2}
\psi_{2} + 4H\dot{\psi}_{2} + 4\dot{H}\psi_{2} + 12H^{2}\psi_{2})\\
\end{split}
\end{equation}
\begin{equation}
\begin{split}
2\mathcal{G}_{ij}^{(2,6)} & = 2a^{2}\delta_{ij}(9\rho_{2}\psi_{1} - 3\rho_{2}\psi_{2} - 18
\rho_{1}\psi_{1} + 9\rho_{1}\psi_{2})\\
2\mathcal{G}_{ij}^{(2,7)} & = 6a^{2}\delta_{ij}(3\varphi_{1}^{2} - 3\varphi_{1}\varphi_{2}
+ \varphi_{2}^{2})\\
2\mathcal{G}_{ij}^{(2,8)} & = 18a^{2}\delta_{ij}(4\rho_{1}\psi_{1} - 2\rho_{1}\psi_{2} - 2
\rho_{2}\psi_{1} + \rho_{2}\psi_{2}) \\
2\mathcal{G}_{ij}^{(2,9)} & = a^{2}\delta_{ij}(36\dot{\psi}_{1}\dot{\varphi}_{1} - 18
\dot{\psi}_{1}\dot{\varphi}_{2} - 18\dot{\psi}_{2}\dot{\varphi}_{1} + 12\dot{\psi}_{2}
\dot{\varphi}_{2} + 24H^{2}\psi_{2}\varphi_{2})\\
2\mathcal{G}_{ij}^{(2,10)} & = a^{2}\delta_{ij}(3\ddot{\psi}_{1} - 3\ddot{\psi}_{2} + 9H
\dot{\psi}_{1} - 6\dot{H}\psi_{2} - 15H\dot{\psi}_{2} - 18H^{2}\psi_{2})\\
2\mathcal{G}_{ij}^{(2,11)} & = -16H^{2}\varphi_{2}\psi_{2}a^{2}\delta_{ij}\\
2\mathcal{G}_{ij}^{(2,12)} & = a^{2}\delta_{ij}[12H\dot{\varphi}_{1}\psi_{2} + 12H\dot{\psi}
_{1}\varphi_{2} - 4H\dot{\varphi}_{2}\psi_{2} - 4H\dot{\psi}_{2}\varphi_{2}\\
 & - 48H^{2}\psi_{2}\varphi_{2} - 8H\dot{\varphi}_{2}\psi_{2} - 8H\dot{\psi}_{2}\varphi_{2}]\\
 2\mathcal{G}_{ij}^{(2,13)} & = a^{2}\delta_{ij}\Big[18H\psi_{1}\dot{\varphi}_{1} + 18H\varphi_{1}
\dot{\psi}_{1} - 18H\psi_{1}\dot{\varphi}_{2} - 18H\varphi_{1}\dot{\psi}_{2}\\
 & - 6H\psi_{2}\dot{\varphi}_{1} - 6H\varphi_{2}\dot{\psi}_{1} + 6H\psi_{2}\dot{\varphi}_{2}
 + 6H\varphi_{2}\dot{\psi}_{2} - 48H^{2}\psi_{1}\varphi_{2}\\
 & - 48H^{2}\varphi_{1}\psi_{2} - 12\dot{H}\psi_{1}\varphi_{2} - 12\dot{H}\varphi_{1}\psi_{2}
 + 32H^{2}\psi_{2}\varphi_{2} + 8\dot{H}\psi_{2}\varphi_{2}\Big]\\
 2\mathcal{G}_{ij}^{(2,15)} & = a^{2}\delta_{ij}\Big[- 18H\varphi_{1}\dot{\psi}_{1} - 18H\psi_{1}
\dot{\varphi}_{1} + 18H\varphi_{1}\dot{\psi}_{2} + 18H\psi_{1}\dot{\varphi}_{2} + 6H\varphi_{2}
\dot{\psi}_{1}\\
 & + 6H\psi_{2}\dot{\varphi}_{1} + 12\dot{H}\varphi_{1}\psi_{2} + 12\dot{H}\psi_{1}\varphi_{2}
 + 48H^{2}\varphi_{1}\psi_{2} + 48H^{2}\psi_{1}\varphi_{2}\\
 & - 24\dot{H}\varphi_{2}\psi_{2} - 14H\varphi_{2}\dot{\psi}_{2} - 14H\psi_{2}\dot{\varphi}_{2}
 - 32H^{2}\varphi_{2}\psi_{2}\Big] \\
2\mathcal{G}_{ij}^{(2,14)} & = a^{2}\delta_{ij}\Big[18\varphi_{1}\ddot{\psi}_{1} + 18\psi_{1}
\ddot{\varphi}_{1} - 6\varphi_{1}\ddot{\psi}_{2} - 6\psi_{1}\ddot{\varphi}_{2}\\
 & - 6\varphi_{2}\ddot{\psi}_{1} - 6\psi_{2}\ddot{\varphi}_{1} + 2\varphi_{2}\ddot{\psi}_{2}
 + 2\psi_{2}\ddot{\varphi}_{2} + 54H\varphi_{1}\dot{\psi}_{1}\\
 & + 54H\psi_{1}\dot{\varphi}_{1} - 18H\varphi_{1}\dot{\psi}_{2} - 18H\psi_{1}\dot{\varphi}_{2}
 - 18H\varphi_{2}\dot{\psi}_{1} - 18H\psi_{2}\dot{\varphi}_{1}\\
 & + 6H\varphi_{2}\dot{\psi}_{2} + 6H\psi_{2}\dot{\varphi}_{2} - 24H^{2}\varphi_{1}\psi_{2}
 - 24H^{2}\varphi_{1}\psi_{2} - 24H^{2}\psi_{1}\varphi_{2} + 16H^{2}\varphi_{2}\psi_{2}\Big]\\
2\mathcal{G}_{ij}^{(2,16)} & = a^{2}\delta_{ij}[16H^{2}\psi_{2}\varphi_{2} - 12H\dot{\psi}_{1}
\varphi_{2} - 12H\dot{\varphi}_{1}\psi_{2} + 4H\dot{\psi}_{2}\varphi_{2} + 4H\dot{\varphi}_{2}
\psi_{2}].    
\end{split}
\end{equation}
Combining all the above contributions, we obtain the following relation
\begin{equation}\label{S-E28}
\begin{split}
\mathcal{G}_{ij}^{(2)} & = \mu_{2}\Big[6\varphi_{1} - 4\varphi_{2} - 6H\dot{\psi}_{1} - 18
\varphi_{1}\varphi_{2} + 6\varphi_{2}^{2} + 18\varphi_{1}^{2} + 36\dot{\psi}_{1}\dot{\varphi}
_{1}\\
 & - 18\dot{\psi}_{1}\dot{\varphi}_{2} - 18\dot{\varphi}_{1}\dot{\psi}_{2} + 12\dot{\psi}_{2}
 \dot{\varphi}_{2} - 12H\dot{\varphi}_{2}\psi_{2} - 12H\dot{\psi}_{2}\varphi_{2}\\
 & + 6\psi_{1}(2\dot{H} + 3H^{2}) - 6\psi_{2}(2\dot{H} + 3H^{2}) - 8\varphi_{2}\psi_{2}(2
 \dot{H} + 3H^{2}) - 8H^{2}\psi_{2}\Big].
\end{split}
\end{equation}
As a result, the second modified Friedmann equation can be expressed as
\begin{equation}\label{S-E29}
\begin{split}
& (2\dot{H} + 3H^{2})(1 - 2\mu_{1}S - 6\mu_{2}\psi_{1} + 6\mu_{2}\psi_{2} + 8\mu_{2}
\varphi_{2}\psi_{2}) = 2\mu_{1}\Phi - 2\mu_{1}\dot{S}H + \mu_{1}\dot{S}\dot{\Phi} + \frac{1}{2}
\mu_{1}\Phi^{2}\\
 & + \mu_{2}\Big[6\varphi_{1} - 4\varphi_{2} - 6H\dot{\psi}_{1} - 18\varphi_{1}\varphi_{2}
 + 6\varphi_{2}^{2} + 18\varphi_{1}^{2} + 36\dot{\psi}_{1}\dot{\varphi}_{1} - 18\dot{\psi}_{1}\dot{\varphi}_{2} - 18\dot{\varphi}_{1}\dot{\psi}_{2}\\
 & + 12\dot{\psi}_{2}\dot{\varphi}_{2} - 12H\dot{\psi}_{2}\varphi_{2} - 12H\dot{\varphi}_{2}
 \psi_{2}\Big] - 8\mu_{2}H^{2}\psi_{2}.
\end{split}
\end{equation}
The above equation leads to the following relation
\begin{equation}\label{S-E30}
\begin{split}
2\frac{\dot{H}}{H^{2}} + 3 & = \frac{2\mu_{1}\Phi}{H^{2}\bar{\Delta}} - \frac{2\mu_{1}\dot{S}}
{H\bar{\Delta}} + \frac{\mu_{1}\dot{S}\dot{\Phi}}{H^{2}\bar{\Delta}} + \frac{1}{2}\frac{\mu_{1}
\Phi^{2}}{H^{2}\bar{\Delta}} + \frac{6\mu_{2}\varphi_{1}}{H^{2}\bar{\Delta}} - \frac{4\mu_{2}
\varphi_{2}}{H^{2}\bar{\Delta}}\\
 & - \frac{6\mu_{2}\dot{\psi}_{1}}{H\bar{\Delta}} - 18\frac{\mu_{2}\varphi_{1}\varphi_{2}}{H
 ^{2}\bar{\Delta}} + \frac{6\mu_{2}\varphi_{2}^{2}}{H^{2}\bar{\Delta}} + 18\frac{\mu_{2}\varphi
 _{1}^{2}}{H^{2}\bar{\Delta}} + 36\frac{\mu_{2}\dot{\psi}_{1}\dot{\varphi}_{1}}{H^{2}\bar{\Delta}}
 \\
 & - 18\frac{\mu_{2}\dot{\psi}_{1}\dot{\varphi}_{2}}{H^{2}\bar{\Delta}} - 18\frac{\mu_{2}\dot{
 \varphi}_{1}\dot{\psi}_{2}}{H^{2}\bar{\Delta}} + 12\frac{\mu_{2}\dot{\psi}_{2}\dot{\varphi}_{2}}
 {H^{2}\bar{\Delta}} - 12\frac{\mu_{2}\dot{\psi}_{2}\varphi_{2}}{H\bar{\Delta}}\\
 & - 12\frac{\mu_{2}\psi_{2}\dot{\varphi}_{2}}{H\bar{\Delta}} - 2\frac{\mu_{2}\psi_{2}}{\bar{
 \Delta}}.
\end{split}
\end{equation}
Introducing the following auxiliary variables
\begin{equation}\label{S-E31}
\begin{split}
\Omega_{0} & = \frac{2\mu_{1}\Phi}{H^{2}\bar{\Delta}}, \ \Omega_{15} = \frac{\mu_{2}\varphi_{1}}
{H^{2}\bar{\Delta}}, \ \Omega_{16} = \frac{\mu_{2}\varphi_{2}}{H^{2}\bar{\Delta}}\\
\Omega_{17} & = \frac{\mu_{2}\psi_{2}}{\bar{\Delta}}, \ \Omega_{18} = \frac{\dot{\Phi}}{\Phi H},
 \ \Omega_{19} = \frac{\varphi_{1}}{\dot{\psi}_{1}H}\\
\Omega_{20} & = \frac{\dot{\varphi}_{1}}{\varphi_{1}H}, \ \Omega_{21} = \frac{\mu_{2}\psi_{2}
\varphi_{2}}{\bar{\Delta}}, 
\end{split}
\end{equation}
we may now write the following two equations
\begin{equation}\label{S-E32}
\begin{split}
2\frac{\dot{H}}{H^{2}} + 3 & = \Omega_{0} - \Omega_{1} - 3\Omega_{3} + 3\Omega_{2} + 6\Omega_{15}
- 4\Omega_{16} - \Omega_{4} + 3\Omega_{6} + 3\Omega_{7} + 3\Omega_{8}\\
 & - 3\Omega_{9} - 3\Omega_{11} - 3\Omega_{10} - \frac{9}{10}\Omega_{12} + 9\Omega_{13} + 9
 \Omega_{14} - 8\Omega_{17}\\
\frac{\dot{\bar{\Delta}}}{H\bar{\Delta}} & = - \Omega_{1} - 6\frac{\Omega_{15}}{\Omega_{19}}
 + \frac{3}{5}\frac{\Omega_{12}}{\Omega_{14}}\Omega_{17} + 8\Omega_{21}\left(\frac{1}{10}
 \frac{\Omega_{12}}{\Omega_{13}} + \frac{1}{10}\frac{\Omega_{12}}{\Omega_{14}}\right). 
\end{split}
\end{equation}
Using the above result, we may now write the evolution equation of density parameters in terms of number of e-folds $N$ as
\begin{equation}\label{S-E33}
\begin{split}
-\frac{d\Omega_{m}}{dN} & = - 3\Omega_{m} - \Omega_{m}\left(2\frac{\dot{H}}{H^{2}} + \frac{\dot{
\bar{\Delta}}}{\bar{\Delta}H}\right)\\
-\frac{d\Omega_{1}}{dN} & = - 2\frac{\Omega_{1}\Omega_{2}}{\Omega_{3}}\Omega_{18} - 3\Omega_{1}
- \Omega_{1}\left(\frac{\dot{H}}{H^{2}} + \frac{\dot{\bar{\Delta}}}{\bar{\Delta}H}\right)\\
-\frac{d\Omega_{2}}{dN} & = 2\Omega_{2}\Omega_{18} - \Omega_{2}\left(2\frac{\dot{H}}{H^{2}} + \frac{\dot{\bar{\Delta}}}{\bar{\Delta}H}\right)\\
-\frac{d\Omega_{3}}{dN} & = - 2\Omega_{2}\Omega_{18} - 6\Omega_{3} - \Omega_{1}\left(2 + 
\frac{\dot{H}}{H^{2}}\right) - \Omega_{3}\left(2\frac{\dot{H}}{H^{2}} + \frac{\dot{\bar{\Delta}}}{\bar{\Delta}H}\right)\\
-\frac{d\Omega_{4}}{dN} & = \Omega_{4}\Omega_{19} - 18\frac{\Omega_{14}}{\Omega_{11}}\Omega_{4}
- 3\Omega_{4} - \Omega_{4}\left(\frac{\dot{H}}{H^{2}} + \frac{\dot{\bar{\Delta}}}{\bar{\Delta}
H}\right)\\
-\frac{d\Omega_{5}}{dN} & = - \frac{2}{3}\Omega_{5}\frac{\Omega_{7}}{\Omega_{13}} + 80\Omega_{5}
\frac{\Omega_{14}}{\Omega_{12}} - 3\Omega_{5} - \Omega_{5}\left(\frac{\dot{H}}{H^{2}} + \frac{\dot{\bar{\Delta}}}{\bar{\Delta}H}\right)
\end{split}
\end{equation}
\begin{equation}
\begin{split}
-\frac{d\Omega_{6}}{dN} & = \Omega_{6}\Omega_{20} + \frac{1}{10}\Omega_{6}\frac{\Omega_{12}}
{\Omega_{13}} - \Omega_{6}\left(2\frac{\dot{H}}{H^{2}} + \frac{\dot{\bar{\Delta}}}{\bar{\Delta}
H}\right)\\
-\frac{d\Omega_{7}}{dN} & = \frac{1}{5}\Omega_{7}\frac{\Omega_{12}}{\Omega_{13}} - \Omega_{7}
\left(2\frac{\dot{H}}{H^{2}} + \frac{\dot{\bar{\Delta}}}{\bar{\Delta}H}\right) \\
-\frac{d\Omega_{8}}{dN} & = 2\Omega_{8}\Omega_{20} - \Omega_{8}\left(2\frac{\dot{H}}{H^{2}} + \frac{\dot{\bar{\Delta}}}{\bar{\Delta}H}\right)\\
-\frac{d\Omega_{9}}{dN} & = - 2\Omega_{4} - 18\Omega_{9}\frac{\Omega_{13}}{\Omega_{10}} + \Omega
_{9}\Omega_{19} - 18\Omega_{9}\frac{\Omega_{14}}{\Omega_{11}} - 6\Omega_{9} - \Omega_{9}
\left(2\frac{\dot{H}}{H^{2}} + \frac{\dot{\bar{\Delta}}}{\bar{\Delta}H}\right)\\
-\frac{d\Omega_{10}}{dN} & = - 2\Omega_{10}\frac{\Omega_{4}}{\Omega_{9}} - 18\Omega_{13}
 - \frac{2}{3}\Omega_{10}\frac{\Omega_{7}}{\Omega_{13}} + 80\Omega_{10}\frac{\Omega_{14}}{\Omega
 _{12}} - 6\Omega_{10} - \Omega_{10}\left(2\frac{\dot{H}}{H^{2}} + \frac{\dot{\bar{\Delta}}}{
 \bar{\Delta}H}\right)\\
-\frac{d\Omega_{11}}{dN} & = \Omega_{11}\Omega_{19} - 18\Omega_{14} - 6\Omega_{11} - \Omega_{14}
\frac{\dot{H}}{H^{2}} + 80\Omega_{11}\frac{\Omega_{13}}{\Omega_{12}} - \Omega_{11}\left(2\frac{
\dot{H}}{H^{2}} + \frac{\dot{\bar{\Delta}}}{\bar{\Delta}H}\right) \\
%
-\frac{d\Omega_{12}}{dN} & = 80\Omega_{13} - \Omega_{12}\frac{\Omega_{4}}{\Omega_{11}}\frac{
\dot{H}}{H^{2}} - 6\Omega_{12} - \frac{2}{3}\Omega_{12}\frac{\Omega_{7}}{\Omega_{13}} + 80
\Omega_{4} - \Omega_{12}\left(2\frac{\dot{H}}{H^{2}} + \frac{\dot{\bar{\Delta}}}{\bar{\Delta}H}
\right)\\
-\frac{d\Omega_{13}}{dN} & = \frac{1}{10}\Omega_{12} - \frac{2}{3}\Omega_{7} + 80\Omega_{13}
\frac{\Omega_{14}}{\Omega_{12}} - 3\Omega_{13} - \Omega_{13}\left(\frac{\dot{H}}{H^{2}} + \frac{\dot{\bar{\Delta}}}{\bar{\Delta}H}\right)\\
-\frac{d\Omega_{14}}{dN} & = \frac{1}{10}\Omega_{12} - \Omega_{14}\frac{\Omega_{4}}{\Omega_{11}}
\frac{\dot{H}}{H^{2}} - 3\Omega_{14} + 80\Omega_{14}\frac{\Omega_{13}}{\Omega_{12}} - \Omega_{14}
\left(\frac{\dot{H}}{H^{2}} + \frac{\dot{\bar{\Delta}}}{\bar{\Delta}H}\right)\\
-\frac{d\Omega_{15}}{dN} & = \Omega_{15}\Omega_{20} - \Omega_{15}\left(2\frac{\dot{H}}{H^{2}} + 
\frac{\dot{\bar{\Delta}}}{H\bar{\Delta}}\right)\\
-\frac{d\Omega_{16}}{dN} & = \frac{1}{10}\Omega_{16}\frac{\Omega_{12} }{\Omega_{13}} - \Omega_{16}
\left(2\frac{\dot{H}}{H^{2}} + \frac{\dot{\bar{\Delta}}}{H\bar{\Delta}}\right)\\
-\frac{d\Omega_{17}}{dN} & = \Omega_{17}\frac{\Omega_{12}}{\Omega_{11}}\frac{1}{10} - \Omega_{17}
\frac{\dot{\bar{\Delta}}}{\bar{\Delta}}\\
-\frac{d\Omega_{18}}{dN} & = - \Omega_{18}^{2} - 3\Omega_{18} - \Omega_{18}\frac{\dot{H}}{H^{2}}
- 6\frac{\Omega_{18}\Omega_{1}}{\Omega_{3}}\left(2 + \frac{\dot{H}}{H^{2}}\right)\\
-\frac{d\Omega_{19}}{dN} & = \Omega_{19}\Omega_{10} + 3\Omega_{19} - \Omega_{19}\frac{\dot{H}}{H^{2}}
- 18\Omega_{19}\frac{\Omega_{14}}{\Omega_{11}} - \Omega_{19}^{2}\\
-\frac{d\Omega_{20}}{dN} & = - 3\Omega_{20} - \Omega_{20}^{2} - \Omega_{20}\frac{\dot{H}}{H^{2}}
- 2\Omega_{20}\frac{\Omega_{4}}{\Omega_{9}} - 18\Omega_{20}\frac{\Omega_{13}}{\Omega_{10}}\\
-\frac{d\Omega_{21}}{dN} & = \frac{1}{10}\Omega_{21}\frac{\Omega_{12}}{\Omega_{14}} + \frac{1}{10}
\Omega_{21}\frac{\Omega_{12}}{\Omega_{13}} - \Omega_{21}\frac{\dot{\bar{\Delta}}}{\bar{\Delta}H}\\
-\frac{d\Omega_{0}}{dN} & = \Omega_{0}\Omega_{18} - \Omega_{0}\left(\frac{\dot{H}}{H^{2}} + 
\frac{\dot{\bar{\Delta}}}{\bar{\Delta}H}\right).
\end{split}
\end{equation}
From the above complete set of autonomous first-order differential equations, we can note that
$\{\Omega_{3}, \ \Omega_{9}, \ \Omega_{10}, \ \Omega_{11}, \ \Omega_{12}, \ \Omega_{13}, \ 
\Omega_{14}\}$ cannot be set to zero as they lie in the denominator in some of the ODEs. On the 
other hand, we can safely tune the density parameters $\{\Omega_{15}, \ \Omega_{16}, \ \Omega_{17},
 \ \Omega_{18}, \ \Omega_{20}, \ \Omega_{21}, \ \Omega_{0}\}$ to be zero as those are the fixed
points of their associated ODEs. These choices reduces the density parameter space of the theory.
Like earlier, this reduced set of differential equations are complete which shows that the above system is indeed autonomous. The fixed points of the above reduced set of differential equations can be obtained by looking at the vanishing points of the \textit{r.h.s} of the corresponding differential equations and their stability analysis can be done numerically by finding out whether all the eigenvalues of the system of ODEs are negative (stable), positive (unstable), or mixed (critical).

\section{Tensor perturbations in cosmology}\label{Tensor perturbations}

The type of metric perturbations that we consider here is known as the tensor perturbations. In
this case, the line element of perturbed FRW spacetime can be expressed as
\begin{equation}\label{S-F1}
ds^{2} = - dt^{2} + a^{2}(\delta_{ij} + \mathcal{H}_{ij})dx^{i}dx^{j},
\end{equation}
where $\mathcal{H}_{ij}$ is the tensor perturbations which are functions of spacetime coordinates.
In the case of GR, $\mathcal{H}_{ij}$ is divergenceless, traceless, symmetric tensor representing
gravitational waves (GWs). For GWs travelling in z-direction, $\mathcal{H}_{ij}$ can be represented
by
\begin{equation}\label{S-F2}
\mathcal{H}_{ij} = \begin{bmatrix}
h_{+} & h_{\times} & 0\\
h_{\times} & - h_{+} & 0\\
0 & 0 & 0
\end{bmatrix}.
\end{equation} 
Its components, $h_{+}$ and $h_{\times}$, are the degrees of freedom pointing in the x-y plane.
$\mathcal{H}_{ij}$ is already gauge invariant, thus the GWs remain the same in all frames of 
reference.

\subsection{Tensor perturbations in GR}

This subsection comprises the derivation of the tensor-perturbed Einstein equations. The
components of the inverse of the metric shown in \ref{S-F1}) are given by the following
\begin{equation}\label{S-F3}
g^{00} = - 1, \ g^{0i} = 0, \ g^{ij} = \frac{\delta_{ij} - \mathcal{H}_{ij}}{a^{2}},
\end{equation}
where $\mathcal{H}_{ij}$ is divergenceless, traceless, and symmetric tensor defined earlier.
The non-zero components of the Christoffel symbols are the following
\begin{equation}\label{S-F4}
\Gamma_{ \ ij}^{0} = Hg_{ij} + \frac{1}{2}a^{2}\dot{\mathcal{H}}_{ij}, \ \Gamma_{ \ 0j}^{i}
= H\delta_{ij} + \frac{1}{2}\dot{\mathcal{H}}_{ij}, \ \Gamma_{ \ jk}^{i} = \frac{i}{2}
[k_{k}\mathcal{H}_{ij} + k_{j}\mathcal{H}_{ik} - k_{i}\mathcal{H}_{jk}],
\end{equation}
where in computing the Christoffel symbols we already considered Fourier transformation of
the metric perturbation since in this entire section, we only consider the linear perturbation
theory.

Using the above results, the non-zero components of the Ricci tensor can be obtained and they
are given by
\begin{equation}\label{S-F5}
R_{00} = - 3\frac{\ddot{a}}{a}, \ R_{ij} = g_{ij}\left(\frac{\ddot{a}}{a} + 2H^{2}\right) + 
\frac{3}{2}a^{2}H\dot{\mathcal{H}}_{ij} + \frac{a^{2}}{2}\ddot{\mathcal{H}}_{ij} + \frac{k^{2}}
{2}\mathcal{H}_{ij}.
\end{equation}
Summing the above two-parts appropriately, we obtain the following expression of Ricci scalar
\begin{equation}\label{S-F6}
R = 6\left(\frac{\ddot{a}}{a} + H^{2}\right),
\end{equation}
which shows that at the linear level $\delta R = 0$.

As seen above, the Ricci scalar $R$, and the temporal Ricci tensor $R_{00}$ have no contributions
from the tensor perturbations, which are only found in the spatial Ricci tensor, $R_{ij}$. This 
means that only the spatial Einstein tensor, $G_{ \ j}^{i}$ is affected by the tensor perturbations.
Since $\delta R = 0$, we may write $\delta G_{ \ j}^{i} = \delta R_{ \ j}^{i}$. In order to get
$\delta R_{ \ j}^{i}$, we must perform the contraction, $g^{ik}R_{kj}$. This gives zero and first
order parts of $R_{ \ j}^{i}$:
\begin{equation}\label{S-F7}
\begin{split}
R_{ \ j}^{i} & = \delta_{ \ j}^{i}\left(\frac{\ddot{a}}{a} + 2H^{2}\right) + \frac{3}{2}a^{2}H
\dot{\mathcal{H}}_{kj}\frac{\delta_{ik}}{a^{2}} + \frac{a^{2}}{2}\frac{\delta_{ik}}{a^{2}}
\ddot{\mathcal{H}}_{kj} + \frac{k^{2}}{2}\frac{\delta_{ik}}{a^{2}}\mathcal{H}_{kj},
\end{split}
\end{equation}
which implies
\begin{equation}\label{S-F8}
\delta R_{ \ j}^{i} = \frac{3}{2}H\dot{\mathcal{H}}_{kj}\delta_{ik} + \frac{1}{2}\delta_{ik}
\ddot{\mathcal{H}}_{kj} + \frac{1}{2}\frac{k^{2}}{a^{2}}\delta_{ik}\mathcal{H}_{jk},
\end{equation}
which is equivalent to $\delta G_{ \ j}^{i}$.

We end this subsection on tensor perturbations in GR by combining all preceding results to
finally arrive at the dynamical equations of $h_{+}$ and $h_{\times}$. From equation \ref{S-F8}),
$\delta G_{ \ j}^{i}$ is proportional to $\mathcal{H}_{ij}$ as well as its derivatives. Therefore,
to find equations for $h_{+}$ and $h_{\times}$ separately, one must consider their respective
positions in $\mathcal{H}_{ij}$.

In $\mathcal{H}_{ij}$, $h_{+}$ is labelled as $\mathcal{H}_{11}$ and $\mathcal{H}_{22}$. Using
this definition of $h_{+}$, it is straightforward to see that $\delta G_{ \ 1}^{1} = - \delta
G_{ \ 2}^{2}$. Hence,
\begin{equation}\label{S-F9}
\begin{split}
\delta G_{ \ 1}^{1} - \delta G_{ \ 2}^{2} & = \frac{3}{2}H\dot{\mathcal{H}}_{11} + \frac{1}{2}
\ddot{\mathcal{H}}_{11} + \frac{1}{2}\frac{k^{2}}{a^{2}}\mathcal{H}_{11} + \frac{3}{2}H\dot{
\mathcal{H}}_{11}\\
 & + \frac{1}{2}\ddot{\mathcal{H}}_{11} + \frac{1}{2}\frac{k^{2}}{a^{2}}\mathcal{H}_{11}\\
 & = 3H\dot{h}_{+} + \ddot{h}_{+} + \frac{k^{2}}{a^{2}}h_{+},
\end{split}
\end{equation}  
where in the last line, we reinstated $h_{+}$. Since $\delta T_{ \ 1}^{1} = \delta T_{ \ 2}^{2}
= 0$, the evolution equation for $h_{+}$ is just
\begin{equation}\label{S-F10}
\ddot{h}_{+} + 3H\dot{h}_{+} + \frac{k^{2}}{a^{2}}h_{+} = 0.
\end{equation}
On the other hand, $h_{\times} = \mathcal{H}_{12} = \mathcal{H}_{21}$, thus calculating $\delta
G_{ \ 2}^{1} + \delta G_{ \ 1}^{2}$ and noting that $\delta T_{ \ 2}^{1} = \delta T_{ \ 1}^{2} = 
0$ leads to an identical evolution equation but with $h_{+}$ now replaced with $h_{\times}$:
\begin{equation}\label{S-F11}
\ddot{h}_{\times} + 3H\dot{h}_{\times} + \frac{k^{2}}{a^{2}}h_{\times} = 0.
\end{equation}

\subsection{Tensor perturbations in non-local gravity theory}

In order to consider the evolution of tensor perturbations in non-local theory of gravity,
we consider the same line-element as earlier in \ref{S-F1}) and its inverse metric defined
in \ref{S-F3}). However, in the case of non-local theory of gravity considered here, the 
tensor perturbations $\mathcal{H}_{ij}$ are not traceless and divergenceless as we have already
shown that number of degrees of freedom in this theory is six, and $\mathcal{H}_{ij}$ contains
at most six degrees of freedom as it is symmetric.

Due to this reason, the components of Ricci tensor would be different. $R_{00}$ and $R_{ij}$
are now expressed as
\begin{equation}\label{S-F12}
\begin{split}
R_{00} & = - 3(\dot{H} + H^{2}) - \frac{1}{2}\ddot{\mathcal{H}} - H\dot{\mathcal{H}}, \qquad
\delta R_{00} = - \frac{1}{2}\ddot{\mathcal{H}} - H\dot{\mathcal{H}}\\
R_{ij} & = (\dot{H} + 3H^{2})g_{ij} + \frac{a^{2}}{2}\ddot{H}_{ij} + \frac{1}{2}k_{i}k_{j}
\mathcal{H} + \frac{1}{2}k^{2}\mathcal{H}_{ij}\\
 & - \frac{1}{2}(k_{i}k_{k}\mathcal{H}_{kj} + k_{j}k_{k}\mathcal{H}_{ik}) + \frac{3Ha^{2}}{2}
 \dot{\mathcal{H}}_{ij} + \frac{Ha^{2}}{2}\dot{\mathcal{H}}\delta_{ij}\\
\delta R_{ij} & = \frac{a^{2}}{2}\ddot{H}_{ij} + \frac{1}{2}k_{i}k_{j}\mathcal{H} 
+ \frac{1}{2}k^{2}\mathcal{H}_{ij} - \frac{1}{2}(k_{i}k_{k}\mathcal{H}_{kj} + k_{j}k_{k}
\mathcal{H}_{ik}) 
+ \frac{Ha^{2}}{2} 
\left(\dot{\mathcal{H}}_{ij} + \dot{\mathcal{H}}\delta_{ij} \right), 
\end{split}
\end{equation}
where $\mathcal{H}$ is the trace of $\mathcal{H}_{ij}$. On the other hand, we know that
\begin{equation}\label{S-F13}
G_{ \ \nu}^{\mu} = R_{ \ \nu}^{\mu} - \frac{1}{2}\delta_{ \ \nu}^{\mu}R \implies G_{ \ j}^{i}
= R_{ \ j}^{i} - \frac{1}{2}\delta_{ \ j}^{i}R.
\end{equation}
leading to:
\begin{equation}\label{S-F14}
\begin{split}
R_{ \ j}^{i} & = g^{ik}R_{kj} = (\dot{H} + 3H^{2})\delta_{ \ j}^{i} + \frac{1}{2}
\ddot{\mathcal{H}}_{ij} + \frac{1}{2a^{2}}k_{i}k_{j}\mathcal{H} + \frac{1}{2a^{2}}k^{2}
\mathcal{H}_{ij}\\
 & - \frac{1}{2a^{2}}(k_{i}k_{k}\mathcal{H}_{kj} + k_{j}k_{k}\mathcal{H}_{ki}) + \frac{3}{2}
 H\dot{\mathcal{H}}_{ij} + \frac{H}{2}\dot{\mathcal{H}}\delta_{ij}\\
\delta R_{ \ j}^{i} & = \frac{1}{2}
\ddot{\mathcal{H}}_{ij} + \frac{1}{2a^{2}} \left(  
k_{i}k_{j}\mathcal{H} + k^{2} \mathcal{H}_{ij} - k_{i}k_{k}\mathcal{H}_{kj}  
- k_{j}k_{k}\mathcal{H}_{ki} \right) 
 + \frac{3}{2}H\dot{\mathcal{H}}_{ij} + \frac{H}{2}
 \dot{\mathcal{H}}\delta_{ij},
\end{split} 
\end{equation}
whereas the Ricci scalar is expressed as
\begin{equation}\label{S-F15}
\begin{split}
R & = 6(\dot{H} + 2H^{2}) + \ddot{\mathcal{H}} + \frac{1}{a^{2}}k^{2}\mathcal{H}
 - \frac{1}{a^{2}}k_{i}k_{j}\mathcal{H}_{ij} + 3H\dot{\mathcal{H}}\\
\implies \delta R & = \ddot{\mathcal{H}} + \frac{1}{a^{2}}(k^{2}\mathcal{H} - k_{i}k_{j}
\mathcal{H}_{ij}) + 3H\dot{\mathcal{H}}.
\end{split}
\end{equation}
As a result, the spatial components of perturbed Einstein tensor can be expressed as
\begin{equation}\label{S-F16}
\begin{split}
\delta G_{ \ j}^{i} 
 & = \frac{1}{2}(\ddot{\mathcal{H}}_{ij} - \delta_{ij}\ddot{\mathcal{H}}) + \frac{3H}{2}
 (\dot{\mathcal{H}}_{ij} - \delta_{ij}\dot{\mathcal{H}}) + \frac{1}{2a^{2}}k^{2}
 (\mathcal{H}_{ij} - \delta_{ij}\mathcal{H}) + \frac{1}{2a^{2}}k_{i}k_{j}\mathcal{H}\\
 & + \frac{1}{2a^{2}}\delta_{ij}k_{k}k_{l}\mathcal{H}_{kl} - \frac{1}{2a^{2}}(k_{i}k_{k}
 \mathcal{H}_{kj} + k_{j}k_{k}\mathcal{H}_{ki}),
\end{split}
\end{equation}
whereas the temporal components of perturbed Einstein tensor is given by
\begin{equation}\label{S-F17}
\delta G_{ \ 0}^{0} = - 2H\dot{\mathcal{H}} - \frac{1}{2}\ddot{\mathcal{H}} - \frac{1}{a^{2}}
(k^{2}\mathcal{H} - k_{i}k_{j}\mathcal{H}_{ij}).
\end{equation}
Now we consider tensor perturbation equations coming from \ref{S-C9}) and \ref{S-C10}) at
the linear level. We now compute these terms sequentially in the following manner
\begin{equation}\label{S-F18}
\begin{split}
(1) & \rightarrow \delta G_{ \ j}^{i}(1 - 2\mu_{1}S), \quad
(2) = 2\mu_{1}\nabla^{i}\nabla_{j}S \rightarrow - \mu_{1}\dot{\mathcal{H}}_{ij}\dot{S}, \quad
(3),  \ (4), \ (5), \ (6) \rightarrow 0\\
(7) & = - 2\mu_{2}S_{ \ \alpha}^{i}G_{ \ j}^{\alpha} \rightarrow - 2\mu_{2}\mathcal{H}_{ij}
(3\psi_{1} - \psi_{2})(2\dot{H} + 3H^{2}) + 2\mu_{2}\frac{\delta S_{ij}}{a^{2}}(2\dot{H} 
+ 3H^{2})\\
 & - 2\mu_{2}(3\psi_{1} - \psi_{2})\delta G_{ \ j}^{i}\\
(8) & = - \mu_{2}R S_{ \ j}^{i} \rightarrow \mu_{2}\bar{R}(3\psi_{1} - \psi_{2})\mathcal{H}
_{ij} - \mu_{2}\delta_{ \ j}^{i}(3\psi_{1} - \psi_{2})\delta R - \mu_{2}\bar{R}\frac{\delta 
S_{ij}}{a^{2}}\\
(9) & = - \frac{\mu_{2}}{2}\Box S_{ \ j}^{i} = \frac{\mu_{2}}{2}\Phi_{ \ j}^{i} \rightarrow
 \frac{\mu_{2}}{2}\left(\frac{\delta\Phi_{ij}}{a^{2}} - (3\varphi_{1} - \varphi_{2})
 \mathcal{H}_{ij}\right)\\
(10) & = - \mu_{2}\Phi^{i\alpha}\Phi_{j\alpha} \rightarrow 2\mu_{2}(3\varphi_{1} - \varphi_{2}
)^{2}\mathcal{H}_{ij} - 2\mu_{2}\frac{\delta\Phi_{ij}}{a^{2}}(3\varphi_{1} - \varphi_{2})\\
(11) & = \nabla_{\alpha}\nabla^{i}S_{ \ j}^{\alpha} \rightarrow \dot{H}\frac{\delta S_{ij}}{a^{2}}
+ H\frac{\dot{\delta S}_{ij}}{a^{2}} + \left(\frac{3}{2}(\dot{\psi}_{2} - \dot{\psi}_{1}) + 4H
\psi_{2}\right)\dot{\mathcal{H}}_{ij} + \psi_{2}\ddot{\mathcal{H}}_{ij} + 4H^{2}\frac{\delta S_{ij}
}{a^{2}}\\
 & + H\psi_{2}\dot{\mathcal{H}}\delta_{ij}\\
(12) & = \frac{\mu_{2}}{2}\delta_{ij}G^{\alpha\beta}S_{\alpha\beta} \rightarrow \frac{\mu_{2}}
{2}\delta_{ij}\Big[\frac{1}{2}(5\psi_{2} - 9\psi_{1})\ddot{\mathcal{H}} + 3H(3\psi_{2} - 5
\psi_{1})\dot{\mathcal{H}}\\
 & + \frac{k^{2}\mathcal{H}}{2a^{2}}(5\psi_{2} - 9\psi_{1}) + \frac{k_{k}k_{l}\mathcal{H}_{kl}}
 {2a^{2}}(9\psi_{1} - 7\psi_{2}) - \frac{1}{a^{2}}(2\dot{H} + 3H^{2})\delta S_{kk}\Big]\\
(13) & = \frac{\mu_{2}}{4}\delta_{ij}\Phi_{\alpha\beta}\Phi^{\alpha\beta} \rightarrow 
\frac{\mu_{2}}{2}\delta_{ij}(3\varphi_{1} - \varphi_{2})\frac{\delta\Phi_{kk}}{a^{2}}\\
(14) & = \frac{\mu_{2}}{4}\delta_{ij}R S_{ \ \alpha}^{\alpha} \rightarrow \frac{\mu_{2}}{4}
\delta_{ij}\Big[\left(\ddot{\mathcal{H}} + \frac{1}{a^{2}}(k^{2}\mathcal{H} - k_{k}k_{l}
\mathcal{H}_{kl}) + 3H\dot{\mathcal{H}}\right)(9\psi_{1} - 3\psi_{2})\\
 & + \frac{6}{a^{2}}(\dot{H} + 2H^{2})\delta S_{kk}\Big]\\
(15) & = - \frac{1}{2}\delta_{ij}\nabla^{\rho}S^{\alpha\beta}\nabla_{\rho}\Phi_{\alpha\beta}
\rightarrow \delta_{ij}\Big[- H(3\dot{\psi}_{1} - \dot{\psi}_{2})\frac{\delta\Phi_{kk}}{a^{2}}
- H(3\dot{\varphi}_{1} - \dot{\varphi}_{2})\frac{\delta S_{kk}}{a^{2}} + 2H^{2}\psi_{2}
\frac{\delta\Phi_{kk}}{a^{2}}\\
 & + 2H^{2}\varphi_{2}\frac{\delta S_{kk}}{a^{2}} + 4H\varphi_{2}\psi_{2}\dot{\mathcal{H}}\Big]\\
(16) & = - \frac{\mu_{2}}{2}\delta_{ij}\nabla_{\alpha}\nabla_{\beta}S^{\alpha\beta} \rightarrow
\frac{\mu_{2}}{2}\delta_{ij}\Big[\frac{k_{m}k_{n}}{a^{2}}\frac{\delta S_{mn}}{a^{2}} + (3\psi_{1}
- \psi_{2})\frac{k_{m}k_{n}}{a^{2}}\mathcal{H}_{mn} + 2H\dot{\mathcal{H}}(3\psi_{1} - \psi_{2})\\
 & + 9H\dot{\mathcal{H}}(\psi_{2} - \psi_{1}) + 2H^{2}\mathcal{H}(3\psi_{1} - \psi_{2}) + H
 \mathcal{H}(3\dot{\psi}_{1} - \dot{\psi}_{2}) + \dot{H}\frac{\delta S_{kk}}{a^{2}} + 2H\frac{
 \dot{\delta S}_{kk}}{a^{2}}\\
 & + \frac{1}{2}\ddot{\mathcal{H}}(3\psi_{1} - \psi_{2}) + \frac{1}{2}\dot{\mathcal{H}}(3\dot{\psi}
 _{1} - \dot{\psi}_{2})\Big]\\       
 (17) & = \frac{1}{2}\nabla^{i}\Phi_{\alpha\beta}\nabla_{j}S^{\alpha\beta} + (\Phi \leftrightarrow
S) \rightarrow - \Big[H^{2}\frac{\delta S_{ij}}{a^{2}}(3\varphi_{1} + \varphi_{2}) + H^{2}
\frac{\delta\Phi_{ij}}{a^{2}}(3\psi_{1} + \psi_{2})\\
 & + 2H\dot{\mathcal{H}}_{ij}(3\varphi_{1} - \varphi_{2})(3\psi_{1} - \psi_{2}) + 6H(\psi_{2}
 - \psi_{1})(3\varphi_{1} - \varphi_{2})\dot{\mathcal{H}}_{ij}\\
 & + 6H(\varphi_{2} - \varphi_{1})(3\psi_{1} - \psi_{2})\dot{\mathcal{H}}_{ij} + 18H(\varphi_{2}
 - \varphi_{1})(\psi_{2} - \psi_{1})\dot{\mathcal{H}}_{ij}\Big]\\
 (18) & = \nabla_{\beta}\Phi_{ \ \alpha}^{\beta}\nabla^{i}S_{ \ j}^{\alpha} + (\Phi\leftrightarrow 
S) \rightarrow - 3H(\dot{\varphi}_{2} - \dot{\varphi}_{1})\frac{\delta S_{ij}}{a^{2}} - 3H(
\dot{\psi}_{2} - \dot{\psi}_{1})\frac{\delta\Phi_{ij}}{a^{2}}\\
 & + 3(\dot{\varphi}_{2} - \dot{\varphi}_{1})(3\psi_{1} - \psi_{2})H\mathcal{H}_{ij} + 3(\dot{\psi}
 _{2} - \dot{\psi}_{1})(3\varphi_{1} - \varphi_{2})H\mathcal{H}_{ij} - H^{2}\delta_{ij}(3\psi_{1}
 - \psi_{2})\frac{\delta\Phi_{kk}}{a^{2}}\\
\end{split}
\end{equation}
\begin{equation}\label{S-F19}
\begin{split}
& - H^{2}\delta_{ij}(3\varphi_{1} - \varphi_{2})\frac{\delta S_{kk}}{a^{2}} + 2H^{2}\delta_{ij}
 (3\psi_{1} - \psi_{2})(3\varphi_{1} - \varphi_{2})\mathcal{H} - 3(\dot{\varphi}_{2} - \dot{\varphi}
 _{1})\psi_{2}\dot{\mathcal{H}}_{ij}\\
 & - 3(\dot{\psi}_{2} - \dot{\psi}_{1})\varphi_{2}\dot{\mathcal{H}}_{ij} - 9H(\varphi_{2} - 
 \varphi_{1})\psi_{2}\dot{\mathcal{H}}_{ij} - 9H(\psi_{2} - \psi_{1})\varphi_{2}\dot{\mathcal{H}}
 _{ij} - 3H(3\varphi_{1} - \varphi_{2})\psi_{2}\dot{\mathcal{H}}_{ij}\\
 & - 3H(3\psi_{1} - \psi_{2})\varphi_{2}\dot{\mathcal{H}}_{ij} + 6H^{2}(3\psi_{1} - \psi_{2})
 \varphi_{2}\mathcal{H}_{ij} + 6H^{2}(3\varphi_{1} - \varphi_{2})\psi_{2}\mathcal{H}_{ij} - 6H^{2}
 \varphi_{2}\frac{\delta S_{ij}}{a^{2}}\\
 & - 6H^{2}\psi_{2}\frac{\delta\Phi_{ij}}{a^{2}} - 2H\dot{\mathcal{H}}\delta_{ij}(9\varphi_{1}
 \psi_{1} + 5\varphi_{2}\psi_{2} - 6\varphi_{1}\psi_{2} - 6\psi_{1}\varphi_{2})\\
(19) & = - S_{ \ j}^{\alpha}\nabla_{\beta}\nabla^{i}\Phi_{ \ \alpha}^{\beta} + (\Phi\leftrightarrow
 S) \rightarrow - 2\dot{H}\varphi_{2}\frac{\delta S_{ij}}{a^{2}} - 2H\dot{\varphi}_{1}\frac{\delta
 S_{ij}}{a^{2}} - \dot{\mathcal{H}}_{ij}\dot{\varphi}_{2}(3\psi_{1} - \psi_{2})\\
 & - \ddot{\mathcal{H}}_{ij}\varphi_{2}(3\psi_{1} - \psi_{2}) + \frac{\delta\Phi_{ij}}{a^{2}}
 (3\psi_{1} - \psi_{2})(2H^{2} - \dot{H}) - \frac{\dot{\delta\Phi}_{ij}}{a^{2}}H(3\psi_{1} - 
 \psi_{2})\\
 & - 2\dot{H}\psi_{2}\frac{\delta\Phi_{ij}}{a^{2}} - 2H\dot{\psi}_{1}\frac{\delta\Phi_{ij}}{a^{2}}
 - \dot{\mathcal{H}}_{ij}\dot{\psi}_{2}(3\varphi_{1} - \varphi_{2}) - \ddot{\mathcal{H}}_{ij}
 \psi_{2}(3\varphi_{1} - \varphi_{2})\\
 & + \frac{\delta S_{ij}}{a^{2}}(3\varphi_{1} - \varphi_{2})(2H^{2} - \dot{H}) - \frac{\dot{\delta
 S}_{ij}}{a^{2}}(3\varphi_{1} - \varphi_{2})\\
(20) & = - \Phi_{ \ j}^{\alpha}\Box S_{ \ \alpha}^{i} + (\Phi \leftrightarrow S) = (3\varphi_{1}
- \varphi_{2})\frac{\ddot{\delta S_{ij}}}{a^{2}} + (3\psi_{1} - \psi_{2})\frac{\ddot{\delta\Phi}
_{ij}}{a^{2}} + (3\ddot{\psi}_{1} - \ddot{\psi}_{2})\frac{\delta\Phi_{ij}}{a^{2}} + (3\ddot{\varphi}
_{1} - \ddot{\varphi}_{2})\frac{\delta S_{ij}}{a^{2}}\\
 & + 3H(3\dot{\psi}_{1} - \dot{\psi}_{2})\frac{\delta\Phi_{ij}}{a^{2}} + 3H(3\dot{\varphi}_{1} - 
 \dot{\varphi}_{2})\frac{\delta S_{ij}}{a^{2}} - 4H^{2}\psi_{2}\frac{\delta\Phi_{ij}}{a^{2}} - 4H^{2}
 \varphi_{2}\frac{\delta S_{ij}}{a^{2}} - 2\dot{H}(3\varphi_{1} - \varphi_{2})\frac{\delta S_{ij}}
 {a^{2}}\\
 & - 2\dot{H}(3\psi_{1} - \psi_{2})\frac{\delta\Phi_{ij}}{a^{2}} - (3\varphi_{1} - \varphi_{2})H
 \frac{\dot{\delta S}_{ij}}{a^{2}} - (3\psi_{1} - \psi_{2})H\frac{\dot{\delta\Phi}_{ij}}{a^{2}}
 - 4H^{2}(3\varphi_{1} - \varphi_{2})\frac{\delta S_{ij}}{a^{2}}\\
 & - 4H^{2}(3\psi_{1} - \psi_{2})\frac{\delta\Phi_{ij}}{a^{2}} + \frac{k^{2}}{a^{2}}(3\varphi_{1}
 - \varphi_{2})\frac{\delta S_{ij}}{a^{2}} + \frac{k^{2}}{a^{2}}(3\psi_{1} - \psi_{2})\frac{\delta
 \Phi_{ij}}{a^{2}} - 2(3\varphi_{1} - \varphi_{2})(3\psi_{1} - \psi_{2})\frac{k^{2}}{a^{2}}
 \mathcal{H}_{ij}\\
 & + \frac{1}{2}\dot{H}(3\varphi_{1} - \varphi_{2})(3\dot{\psi}_{1} - \dot{\psi}_{2})\dot{
 \mathcal{H}}\delta_{ij} + \frac{1}{2}\dot{H}(3\psi_{1} - \psi_{2})(3\dot{\varphi}_{1} - \dot{
 \varphi}_{2})\dot{\mathcal{H}}\delta_{ij} - 6(\psi_{2} - \psi_{1})(3\varphi_{1} - \varphi_{2})
 H\dot{\mathcal{H}}_{ij}\\
 & - 6(\varphi_{2} - \varphi_{1})(3\psi_{1} - \psi_{2})H\dot{\mathcal{H}}_{ij}\\
 (21) & = \Phi^{\alpha\beta}\nabla_{\beta}\nabla^{i}S_{j\alpha} + (\Phi\leftrightarrow S)\rightarrow
3H\dot{\mathcal{H}}_{ij}(3\psi_{1} - \psi_{2})(\varphi_{2} - \varphi_{1}) + 3H\dot{H}_{ij}(3\varphi
_{1} - \varphi_{2})(2\psi_{1} + \psi_{2})\\
 & - 3(\varphi_{2} - \varphi_{1})\ddot{\mathcal{H}}_{ij}\psi_{2} + 12H^{2}\frac{\delta S_{ij}}{a^{2}}
 \varphi_{1} + 6H^{2}\frac{\delta\Phi_{ij}}{a^{2}} + H\dot{\mathcal{H}}\delta_{ij}(3\varphi_{1} - 
 \varphi_{2})\psi_{2} + 2\delta_{ij}H^{2}\frac{\delta\Phi_{kk}}{a^{2}}\psi_{2}\\
 & - 3(\varphi_{2} - \varphi_{1})\dot{\mathcal{H}}_{ij}\dot{\psi}_{2} - (3\varphi_{1} - \varphi_{2})
 \frac{k_{k}k_{i}}{a^{2}}\frac{\delta S_{jk}}{a^{2}} + (3\varphi_{1} - \varphi_{2})(3\psi_{1} - 
 \psi_{2})\frac{k_{l}k_{i}}{a^{2}}\mathcal{H}_{jl}\\
 & - 3(\varphi_{2} - \varphi_{1})\frac{1}{a^{2}}(\dot{H}\delta S_{ij} + H\dot{\delta S}_{ij})
 - 9(\varphi_{2} - \varphi_{1})[H\mathcal{H}_{ij}(\dot{\psi}_{2} - \dot{\psi}_{1}) + (\psi_{2} 
 - \psi_{1})(\dot{H}\mathcal{H}_{ij} + 2H^{2}\mathcal{H}_{ij})]\\
 & + 3H\dot{\mathcal{H}}_{ij}(3\varphi_{1} - \varphi_{2})(\psi_{2} - \psi_{1}) + 3H\dot{\mathcal{H}}
 _{ij}(3\psi_{1} - \psi_{2})(2\varphi_{1} + \varphi_{2}) - 3(\psi_{2} - \psi_{1})\varphi_{2}
 \ddot{\mathcal{H}}_{ij} + 12H^{2}\frac{\delta\Phi_{ij}}{a^{2}}\psi_{1}\\
 & + 6H^{2}\frac{\delta S_{ij}}{a^{2}}\varphi_{1} + H\dot{\mathcal{H}}\delta_{ij}(3\psi_{1} - 
 \psi_{2})\varphi_{2} + 2\delta_{ij}H^{2}\frac{\delta S_{kk}}{a^{2}}\varphi_{2} - 3(\psi_{2} - 
 \psi_{1})\dot{\mathcal{H}}_{ij}\dot{\varphi}_{2} - (3\psi_{1} - \psi_{2})\frac{k_{k}k_{i}}{a^{2}}
 \frac{\delta\Phi_{jk}}{a^{2}}\\
 & + (3\varphi_{1} - \varphi_{2})(3\psi_{1} - \psi_{2})\frac{k_{l}k_{i}}{a^{2}}\mathcal{H}_{jl}
 - 3(\psi_{2} - \psi_{1})\frac{1}{a^{2}}(\dot{H}\delta\Phi_{ij} + H\dot{\delta\Phi}_{ij}) - 9
 (\psi_{2} - \psi_{1})[H\mathcal{H}_{ij}(\dot{\varphi}_{2} - \dot{\varphi}_{1})\\
 & + (\varphi_{2} - \varphi_{1})(\dot{H}\mathcal{H}_{ij} + 2H^{2}\mathcal{H}_{ij})] \\
%
(22) & = - \nabla^{\beta}S^{\alpha i}\nabla_{j}\Phi_{\alpha\beta} + (\Phi \leftrightarrow S)
\rightarrow 2H^{2}\frac{\delta\Phi_{ij}}{a^{2}}\psi_{2} + 2H^{2}\frac{\delta S_{ij}}{a^{2}}
\varphi_{2} + 6H^{2}\frac{\delta S_{ij}}{a^{2}}\varphi_{2} + 6H^{2}\frac{\delta\Phi_{ij}}{a^{2}}
\psi_{2}\\
 & + 8H\dot{\mathcal{H}}_{ij}\varphi_{2}\psi_{2} - 2H\frac{\dot{\delta S}_{ij}}{a^{2}}\varphi_{2}
 - 2H\frac{\dot{\delta\Phi}_{ij}}{a^{2}}\psi_{2} - \dot{\mathcal{H}}_{ij}(3\dot{\psi}_{1} - 
 \dot{\psi}_{2})\varphi_{2} - \dot{\mathcal{H}}_{ij}(3\dot{\varphi}_{1} - \dot{\varphi}_{2})\psi_{2}\\
 & - H(3\dot{\psi}_{1} - \dot{\psi}_{2})\frac{\delta\Phi_{ij}}{a^{2}} - H(3\dot{\varphi}_{1} - 
 \dot{\varphi}_{2})\frac{\delta S_{ij}}{a^{2}}
\end{split}
\end{equation}
Combining all these terms, we finally obtain the following expression of the spatial components 
of field equations
{\small
\begin{equation}\label{S-F21}
\begin{split}
& (1 - 2\mu_{1}S - 4\mu_{2}(3\psi_{1} - \psi_{2}))\Big[\frac{1}{2}(\ddot{\mathcal{H}}_{ij}
- \delta_{ij}\ddot{\mathcal{H}}) + \frac{3H}{2}(\dot{\mathcal{H}}_{ij} - \delta_{ij}\dot{
\mathcal{H}}) + \frac{k^{2}}{2a^{2}}(\mathcal{H}_{ij} - \delta_{ij}\mathcal{H})\\
 & + \frac{1}{2}\frac{k_{i}k_{j}}{a^{2}}\mathcal{H} + \frac{1}{2}\delta_{ij}\frac{k_{k}k_{l}}
 {a^{2}}\mathcal{H}_{kl} - \frac{1}{2a^{2}}(k_{i}k_{k}\mathcal{H}_{kj} + k_{j}k_{k}\mathcal{H}
 _{ki})\Big] - \mu_{1}\dot{S}\dot{\mathcal{H}}_{ij} + 4\mu_{2}(3\psi_{1} - \psi_{2})(\dot{H}
 + 3H^{2})\mathcal{H}_{ij}\\
 & - 2\mu_{2}(\dot{H} + 2H^{2})\frac{\delta S_{ij}}{a^{2}} + \mu_{2}\frac{\delta\Phi_{ij}}{a^{2}}
 - \mu_{2}\mathcal{H}_{ij}(3\varphi_{1} - \varphi_{2}) + 2\mu_{2}H\frac{\dot{\delta S}_{ij}}{a^{2}}
 + \mu_{2}[3(\dot{\psi}_{2} - \dot{\psi}_{1}) + 8H\psi_{2}]\dot{\mathcal{H}}_{ij}\\
 & + 2\mu_{2}\psi_{2}\ddot{\mathcal{H}}_{ij} + \mu_{2}\delta_{ij}\Big[\frac{\ddot{\mathcal{H}}}
 {2}(5\psi_{2} - 9\psi_{1}) + \frac{3H\dot{\mathcal{H}}}{2}(13\psi_{2} - 15\psi_{1}) + 3\frac{
 k^{2}}{a^{2}}\mathcal{H}(\psi_{2} - 2\psi_{1}) + 3\frac{k_{k}k_{l}}{a^{2}}\mathcal{H}_{kl}(
 \psi_{1} - \psi_{2})\\
 & + \frac{\delta S_{kk}}{a^{2}}(2\dot{H} + 9H^{2}) + \frac{\delta\Phi_{kk}}{a^{2}}(3\varphi_{1}
 - \varphi_{2}) - 2H(3\dot{\psi}_{1} - \dot{\psi}_{2})\frac{\delta\Phi_{kk}}{a^{2}} - 2H(3\dot{
 \varphi}_{1} - \dot{\varphi}_{2})\frac{\delta S_{kk}}{a^{2}}\\
 & + 2H^{2}\frac{\delta\Phi_{kk}}{a^{2}}(5\psi_{2} - 3\psi_{1}) + \frac{k_{m}k_{n}}{a^{2}}
 \frac{\delta S_{mn}}{a^{2}} + 2H\dot{\mathcal{H}}(15\varphi_{1}\psi_{2} + 15\psi_{1}\varphi_{2}
 - 8\varphi_{2}\psi_{2} - 18\varphi_{1}\psi_{1})\\
 & + 2H^{2}(3\psi_{1} - \psi_{2})\mathcal{H} + (3\dot{\psi}_{1} - \dot{\psi}_{2})H\mathcal{H} + 
 2H\frac{\dot{\delta S}_{kk}}{a^{2}} + \frac{1}{2}(3\dot{\psi}_{1} - \dot{\psi}_{2})\dot{
 \mathcal{H}} + 2\mu_{2}\Big[H^{2}\frac{\delta S_{ij}}{a^{2}}(21\varphi_{1} - \varphi_{2})\\
 & + H^{2}\frac{\delta\Phi_{ij}}{a^{2}}(21\psi_{1} - \psi_{2}) + 3H\dot{\mathcal{H}}_{ij}
 (6\psi_{1}\varphi_{1} + 5\psi_{1}\varphi_{2} + 5\psi_{2}\varphi_{1} - 4\varphi_{2}\psi_{2})
 + 3H^{2}\mathcal{H}_{ij}(18\psi_{1}\varphi_{2} + 18\psi_{2}\varphi_{1} \\
%
 & - 12\psi_{1}\varphi_{1} - 16\psi_{2}\varphi_{2}) - 2H(\dot{\varphi}_{1} + \dot{\varphi}_{2})
 \frac{\delta S_{ij}}{a^{2}} - 2H(\dot{\psi}_{1} + \dot{\psi}_{2})\frac{\delta\Phi_{ij}}{a^{2}}
 - 2\dot{\mathcal{H}}_{ij}(\psi_{2}\dot{\varphi}_{2} + \dot{\psi}_{2}\varphi_{2}) + 6H(\dot{\psi}
 _{2} - \dot{\psi}_{1})\\
 & \times(3\varphi_{1} - 2\varphi_{2})\mathcal{H}_{ij} + 6H(\dot{\varphi}_{2} - \dot{\varphi}_{1})
 (3\psi_{1} - 2\psi_{2})\mathcal{H}_{ij} - 4H\psi_{2}\frac{\dot{\delta\Phi}_{ij}}{a^{2}} - 4H
 \varphi_{2}\frac{\dot{\delta S}_{ij}}{a^{2}} - 4\dot{H}\varphi_{2}\frac{\delta S_{ij}}{a^{2}}
 - 4\dot{H}\psi_{2}\frac{\delta\Phi_{ij}}{a^{2}} \\
 & - 4\varphi_{2}\psi_{2}\dot{\mathcal{H}}_{ij} - 2\dot{\mathcal{H}}_{ij}(\varphi_{2}\dot{\psi}
 _{2} + \dot{\varphi}_{2}\psi_{2}) + 2(3\varphi_{1} - \varphi_{2})(3\psi_{1} - \psi_{2})\frac{k
 _{l}k_{(i}}{a^{2}}\mathcal{H}_{j)l} - \frac{3\varphi_{1} - \varphi_{2}}{2}\frac{k_{k}k_{(i}}{
 a^{2}}\frac{\delta S_{j)k}}{a^{2}}\\
 & - \frac{3\psi_{1} - \psi_{2}}{2}\frac{k_{k}k_{(i}}{a^{2}}\frac{\delta\Phi_{j)k}}{a^{2}} - 18
 \dot{H}\mathcal{H}_{ij}(\psi_{2} - \psi_{1})(\varphi_{2} - \varphi_{1}) + 2H^{2}\mathcal{H}
 \delta_{ij}(3\psi_{1} - \psi_{2})(3\varphi_{1} - \varphi_{2})\Big]\\
 & + 2\mu_{2}\Big[- (3\varphi_{1} - \varphi_{2})\frac{\delta\Phi_{ij}}{a^{2}} + 3(\varphi_{1} - 
 \varphi_{2})^{2}\mathcal{H}_{ij} - (3\varphi_{1} - \varphi_{2})(3\ddot{\psi}_{1} - \ddot{\psi}
 _{2})\mathcal{H}_{ij} - 2(3\varphi_{1} - \varphi_{2})(3\psi_{1} - \psi_{2})\ddot{\mathcal{H}}
 _{ij}\\
 & + 3H(3\varphi_{1} - \varphi_{2})\frac{\dot{\delta S}_{ij}}{a^{2}} - 4H^{2}(3\varphi_{1} - 
 \varphi_{2})\frac{\delta S_{ij}}{a^{2}} - (3\varphi_{1} - \varphi_{2})\frac{k^{2}}{a^{2}}
 \frac{\delta S_{ij}}{a^{2}} + (3\psi_{1} - \psi_{2})(\dot{H} + 3H^{2})\mathcal{H}_{ij}\\
 & + \frac{1}{2}(3\psi_{1} - \psi_{2})\ddot{\mathcal{H}}_{ij} + \frac{1}{2}(3\psi_{1} - \psi_{2})
 \frac{k_{i}k_{j}}{a^{2}}\mathcal{H} + \frac{1}{2}(3\psi_{1} - \psi_{2})\frac{k^{2}}{a^{2}}
 \mathcal{H}_{ij}- (3\psi_{1} - \psi_{2})\frac{1}{2a^{2}}(k_{i}k_{k}\mathcal{H}_{kj}\\
 & + k_{j}k_{k}\mathcal{H}_{ki}) + \frac{3}{2}H(3\psi_{1} - \psi_{2})\dot{\mathcal{H}}_{ij} + 
 \frac{1}{2}(3\psi_{1} - \psi_{2})H\dot{\mathcal{H}}\delta_{ij} - (3\psi_{1} - \psi_{2})
 (3\ddot{\varphi}_{1} - \ddot{\varphi}_{2})\mathcal{H}_{ij} + 3H(\psi_{1} - \psi_{2})\frac{
 \dot{\delta\Phi}_{ij}}{a^{2}}\\
 & - 4H^{2}(3\psi_{1} - \psi_{2})\frac{\delta\Phi_{ij}}{a^{2}} - (3\psi_{1} - \psi_{2})\frac{
 k^{2}}{a^{2}}\frac{\delta\Phi_{ij}}{a^{2}} + 3H(3\dot{\varphi}_{1} - \dot{\varphi}_{2})\frac{
 \delta S_{ij}}{a^{2}} + (3\ddot{\psi}_{1} - \ddot{\psi}_{2})\frac{\delta\Phi_{ij}}{a^{2}} + 
 (3\ddot{\varphi}_{1} - \ddot{\varphi}_{2})\frac{\delta S_{ij}}{a^{2}}\\
 & + 3H(3\dot{\psi}_{1} - \dot{\psi}_{2})\frac{\delta\Phi_{ij}}{a^{2}} - 4H^{2}\psi_{2}\frac{
 \delta\Phi_{ij}}{a^{2}} - 4H^{2}\varphi_{2}\frac{\delta S_{ij}}{a^{2}} - 4H^{2}(3\varphi_{1}
 - \varphi_{2})\frac{\delta S_{ij}}{a^{2}} - 4H^{2}(3\psi_{1} - \psi_{2})\frac{\delta\Phi_{ij}}
 {a^{2}}\\
 & - 2(3\varphi_{1} - \varphi_{2})(3\psi_{1} - \psi_{2})\frac{k^{2}}{a^{2}}\mathcal{H}_{ij}
 + \frac{1}{2}\dot{H}(3\varphi_{1} - \varphi_{2})(3\dot{\psi}_{1} - \dot{\psi}_{2})\dot{
 \mathcal{H}}\delta_{ij} + \frac{1}{2}\dot{H}(3\psi_{1} - \psi_{2})(3\dot{\varphi}_{1} - 
 \dot{\varphi}_{2})\dot{\mathcal{H}}\delta_{ij}\\
 & - 6(\psi_{2} - \psi_{1})(3\varphi_{1} - \varphi_{2})H\dot{\mathcal{H}}_{ij} - 6(\varphi_{2}
 - \varphi_{1})(3\psi_{1} - \psi_{2})H\dot{\mathcal{H}}_{ij}\Big] = 0.
\end{split}
\end{equation}
}
Now we compute sequentially the $00$th component of the perturbed field equations which are the following
\begin{equation}\label{S-F22}
\begin{split}
(1) & \rightarrow - (1 - 2\mu_{1}S)\Big[2H\dot{\mathcal{H}} + \frac{1}{2}\ddot{\mathcal{H}}
+ \frac{1}{a^{2}}(k^{2}\mathcal{H} - k_{i}k_{j}\mathcal{H}_{ij})\Big]\\
(2) & = - 2S_{ \ \alpha}^{0}G_{ \ 0}^{\alpha} \rightarrow - 6(\psi_{2} - \psi_{1})\Big[2H
\dot{\mathcal{H}} + \frac{1}{2}\ddot{\mathcal{H}} + \frac{1}{a^{2}}(k^{2}\mathcal{H} - k_{i}
k_{j}\mathcal{H}_{ij})\Big]\\
(3) & = - RS_{ \ 0}^{0} \rightarrow 3(\psi_{2} - \psi_{1})\Big[\ddot{\mathcal{H}} + \frac{1}
{a^{2}}(k^{2}\mathcal{H} - k_{i}k_{j}\mathcal{H}_{ij}) + 3H\dot{\mathcal{H}}\Big]\\
(4) & , \ (5) \rightarrow 0\\
(6) & = \nabla_{\alpha}\nabla^{0}S_{ \ 0}^{\alpha} \rightarrow \dot{\mathcal{H}}\dot{\psi}_{2}
+ 3H(\dot{\psi}_{2} - \dot{\psi}_{1})\mathcal{H} - 2\psi_{2}H\dot{\mathcal{H}} - 3H^{2}
\frac{\delta S_{kk}}{a^{2}}\\
(7) & - (11) \ \text{are the same as for ij components}\\
(12) & = \frac{1}{2}\nabla^{0}\Phi_{\alpha\beta}\nabla_{0}S^{\alpha\beta} + (\Phi\leftrightarrow
S) \rightarrow 2\Big[ - 2H^{2}(3\varphi_{1} - \varphi_{2})\frac{\delta S_{kk}}{a^{2}} - 2H
(3\varphi_{1} - \varphi_{2})(3\psi_{1} - \psi_{2})\dot{\mathcal{H}}\\
 & - 2H^{2}(3\psi_{1} - \psi_{2})\frac{\delta\Phi_{kk}}{a^{2}}\Big]\\
%
(13) & = \nabla_{\beta}\Phi_{ \ \alpha}^{\beta}\nabla^{0}S_{ \ 0}^{\alpha} + (\Phi\leftrightarrow
S) \rightarrow - 3(\dot{\psi}_{2} - \dot{\psi}_{1})H\frac{\delta\Phi_{kk}}{a^{2}} - 3(\dot{\psi}
_{2} - \dot{\psi}_{1})\varphi_{2}\dot{\mathcal{H}}\\
 & - 3(\dot{\varphi}_{2} - \dot{\varphi}_{1})H\frac{\delta S_{kk}}{a^{2}} - 3(\dot{\varphi}_{2}
 - \dot{\varphi}_{1})\psi_{2}\dot{\mathcal{H}}\\
(14) & = - S_{ \ 0}^{\alpha}\nabla_{\beta}\nabla^{0}\Phi_{ \ \alpha}^{\beta} + (\Phi\leftrightarrow
 S) \rightarrow 3(\psi_{2} - \psi_{1})H^{2}\frac{\delta\Phi_{kk}}{a^{2}} + 3(\varphi_{2} - \varphi
 _{1})H^{2}\frac{\delta S_{kk}}{a^{2}} - 3(\psi_{2} - \psi_{1})\dot{\varphi}_{2}\dot{\mathcal{H}}\\
 & - 3(\varphi_{2} - \varphi_{1})\dot{\psi}_{2}\dot{\mathcal{H}} - 18(\psi_{2} - \psi_{1})\varphi
 _{1}H\dot{\mathcal{H}} - 18(\varphi_{2} - \varphi_{1})\psi_{1}H\dot{\mathcal{H}}\\
(15) & = - \Phi_{ \ 0}^{\alpha}\Box S_{ \ \alpha}^{0} + (\Phi\leftrightarrow S) \rightarrow
\frac{9}{2}(\varphi_{2} - \varphi_{1})(\dot{\psi}_{2} - \dot{\psi}_{1})\dot{\mathcal{H}} - 12
(\varphi_{2} - \varphi_{1})\psi_{2}H\dot{\mathcal{H}} - 6(\varphi_{2} - \varphi_{1})H^{2}
\frac{\delta S_{kk}}{a^{2}}\\
 & + \frac{9}{2}(\psi_{2} - \psi_{1})(\dot{\varphi}_{2} - \dot{\varphi}_{1})\dot{\mathcal{H}} - 
 12(\psi_{2} - \psi_{1})\varphi_{2}H\dot{\mathcal{H}} - 6(\psi_{2} - \psi_{1})H^{2}\frac{\delta
 \Phi_{kk}}{a^{2}}\\
(16) & = \Phi^{\alpha\beta}\nabla_{\beta}\nabla^{0}S_{0\alpha} + (\Phi\leftrightarrow S) 
\rightarrow - (9\varphi_{1} - \varphi_{2})H^{2}\frac{\delta S_{kk}}{a^{2}} - (9\psi_{1} - \psi_{2})
H^{2}\frac{\delta\Phi_{kk}}{a^{2}} - 2H\psi_{2}(3\varphi_{1} - \varphi_{2})\dot{\mathcal{H}}\\
 & - 2H\varphi_{2}(3\psi_{1} - \psi_{2})\dot{\mathcal{H}} + \dot{\psi}_{2}(3\varphi_{1} - \varphi
 _{2})\dot{\mathcal{H}} + \dot{\varphi}_{2}(3\psi_{1} - \psi_{2})\dot{\mathcal{H}} + H(3\varphi_{1}
 - \varphi_{2})\frac{\dot{\delta S}_{kk}}{a^{2}}\\
 & + H(3\psi_{1} - \psi_{2})\frac{\dot{\delta\Phi}_{kk}}{a^{2}} + 2H\dot{\psi}_{2}\frac{\delta
 \Phi_{kk}}{a^{2}} + 2H\dot{\varphi}_{2}\frac{\delta S_{kk}}{a^{2}}\\
(17) & = \nabla^{\beta}S^{\alpha 0}\nabla_{0}\Phi_{\alpha\beta} + (\Phi\leftrightarrow S)
\rightarrow - (3\dot{\varphi}_{1} - \dot{\varphi}_{2})\psi_{2}\dot{\mathcal{H}} - (3\dot{\psi}
 _{1} - \dot{\psi}_{2})\varphi_{2}\dot{\mathcal{H}} - H(3\dot{\varphi}_{1} - \dot{\varphi}_{2})
 \frac{\delta S_{kk}}{a^{2}}\\
 & - H(3\dot{\psi}_{1} - \dot{\psi}_{2})\frac{\delta\Phi_{kk}}{a^{2}} + 4H^{2}\psi_{2}\frac{
 \delta\Phi_{kk}}{a^{2}} + 4H^{2}\varphi_{2}\frac{\delta S_{kk}}{a^{2}}.  
\end{split}
\end{equation}
Combining all the above information, we finally obtain the following expression of the $00$th
component of the perturbed field equations in linear order of tensor perturbations
\begin{equation}\label{S-F23}
\begin{split}
 &- (1 - 2\mu_{1}S)\Big[2H\dot{\mathcal{H}} + \frac{1}{2}\ddot{\mathcal{H}} + \frac{1}{a^{2}}
(k^{2}\mathcal{H} - k_{i}k_{j}\mathcal{H}_{ij})\Big] + \mu_{2}\Big[- 3(\psi_{2} - \psi_{1})\\
 & \times\left(H\dot{\mathcal{H}} + \frac{1}{a^{2}}(k^{2}\mathcal{H} - k_{i}k_{j}\mathcal{H}
 _{ij})\right) + \dot{\mathcal{H}}\dot{\psi}_{2} + H\mathcal{H}\left(\frac{5\dot{\psi}_{2} - 3
 \dot{\psi}_{1}}{2}\right) + \frac{3}{4}H\dot{\mathcal{H}}(5\psi_{2} - 3\psi_{1})\\
 & + \frac{1}{4}\ddot{\mathcal{H}}(3\psi_{1} - \psi_{2}) + \frac{\delta S_{kk}}{a^{2}}\left(
 \dot{H} - \frac{3H^{2}}{2}\right) + \frac{1}{2}\frac{k^{2}}{a^{2}}\psi_{2}\mathcal{H} - 
 \frac{1}{2}\frac{k_{k}k_{l}}{a^{2}}\mathcal{H}_{kl}(3\psi_{1} + \psi_{2}) + \frac{3\varphi_{1}
 - \varphi_{2}}{2}\frac{\delta\Phi_{kk}}{a^{2}}\\
 & - \frac{\delta\Phi_{kk}}{a^{2}}H(3\dot{\psi}_{1} - \dot{\psi}_{2}) - \frac{\delta S_{kk}}
 {a^{2}}H(3\dot{\varphi}_{1} - \dot{\varphi}_{2}) + 2H^{2}\frac{\delta S_{kk}}{a^{2}}(4\varphi_{2}
 - 9\varphi_{1}) + 2H^{2}\frac{\delta\Phi_{kk}}{a^{2}}(4\psi_{2} - 9\psi_{1})\\
 & + H(3\varphi_{1} - \varphi_{2})\frac{\dot{\delta S}_{kk}}{a^{2}} + H(3\psi_{1} - \psi_{2})
 \frac{\dot{\delta\Phi}_{kk}}{a^{2}} + \frac{1}{2}\frac{k_{m}k_{n}}{a^{2}}\frac{\delta S_{mn}}
 {a^{2}} - 20H\psi_{2}\varphi_{2}\dot{\mathcal{H}} + H^{2}\mathcal{H}(3\psi_{1} - \psi_{2})\\
 & + \frac{1}{4}\dot{\mathcal{H}}(3\dot{\psi}_{1} - \dot{\psi}_{2}) + \dot{\mathcal{H}}\left(
 \frac{9}{2}(\varphi_{2} - \varphi_{1})(\dot{\psi}_{2} - \dot{\psi}_{1}) + \frac{9}{2}(\psi_{2}
 - \psi_{1})(\dot{\varphi}_{2} - \dot{\varphi}_{1}) - 2\varphi_{2}\dot{\psi}_{2} - 2\dot{\varphi}
 _{2}\psi_{2}\right)\\
 & + 2\dot{\psi}_{2}\dot{\mathcal{H}}(3\varphi_{1} - 2\varphi_{2}) + 2\dot{\varphi}_{2}\dot{
 \mathcal{H}}(3\psi_{1} - 2\psi_{2})\Big] = 0
\end{split}
\end{equation}
Now we look at the perturbed field equations for auxiliary field in spatial components
\begin{equation}\label{S-F24}
\begin{split}
\Box\Phi_{ij} & = - R_{ij}\\ 
\implies & \frac{\ddot{\delta\Phi}_{ij}}{a^{2}} + (3\ddot{\varphi}_{1} - \ddot{\varphi}_{2})
\mathcal{H}_{ij} + (3\varphi_{1} - \varphi_{2})\ddot{\mathcal{H}}_{ij} - 2\dot{H}\frac{\delta
\Phi_{ij}}{a^{2}} - 4H\frac{\dot{\delta\Phi}_{ij}}{a^{2}} + 4H^{2}\frac{\delta\Phi_{ij}}{a^{2}}
+ \frac{k^{2}}{a^{2}}\frac{\delta\Phi_{ij}}{a^{2}}\\
 & = \mathcal{H}_{ij}(\dot{H} + 3H^{2}) + \frac{1}{2}\ddot{\mathcal{H}}_{ij} + \frac{1}{2}
 \frac{k_{i}k_{j}}{a^{2}}\mathcal{H} + \frac{1}{2}\frac{k^{2}}{a^{2}}\mathcal{H}_{ij} - \frac{1}
 {2a^{2}}(k_{i}k_{k}\mathcal{H}_{kj} + k_{j}k_{k}\mathcal{H}_{ki})\\
 & + \frac{3}{2}H\dot{H}_{ij} + \frac{H}{2}\dot{\mathcal{H}}\delta_{ij},
\end{split}
\end{equation}
and
\begin{equation}\label{S-F25}
\begin{split}
\Box S_{ij} & = - \Phi_{ij}\\
\implies & \frac{\ddot{\delta S}_{ij}}{a^{2}} + (3\ddot{\psi}_{1} - \ddot{\psi}_{2})\mathcal{H}
_{ij} + (3\psi_{1} - \psi_{2})\ddot{\mathcal{H}}_{ij} - 2\dot{H}\frac{\delta S_{ij}}{a^{2}} - 
4H\frac{\dot{\delta S}_{ij}}{a^{2}}\\
 & + 4H^{2}\frac{\delta S_{ij}}{a^{2}} + \frac{k^{2}}{a^{2}}\frac{\delta S_{ij}}{a^{2}} = 
 \frac{\delta\Phi_{ij}}{a^{2}} + (3\varphi_{1} - \varphi_{2})\mathcal{H}_{ij}.
\end{split}
\end{equation}
We may now note that the equation \ref{S-F25}) can also be expressed as
\begin{equation}\label{S-F26}
\begin{split}
\frac{d^{2}}{dt^{2}} & \left(\frac{\delta S_{ij}}{a^{2}}\right) + \frac{k^{2}}{a^{2}}\left(
\frac{\delta S_{ij}}{a^{2}}\right) = - (3\ddot{\psi}_{1} - \ddot{\psi}_{2})\mathcal{H}_{ij} 
- (3\psi_{1} - \psi_{2})\ddot{\mathcal{H}}_{ij} + \frac{\delta\Phi_{ij}}{a^{2}} + (3\varphi_{1}
- \varphi_{2})\mathcal{H}_{ij},
\end{split}
\end{equation}
where the terms on the \textit{r.h.s} are source terms. In the similar manner, we may write 
the equation in \ref{S-F24}) as
\begin{equation}\label{S-F27}
\begin{split}
\left(\frac{d^{2}}{dt^{2}} + \frac{k^{2}}{a^{2}}\right) & \frac{\delta\Phi_{ij}}{a^{2}} = 
- (3\ddot{\varphi}_{1} - \ddot{\varphi}_{2})\mathcal{H}_{ij} - (3\varphi_{1} - \varphi_{2})
\ddot{\mathcal{H}}_{ij} + \mathcal{H}_{ij}(\dot{H} + 3H^{2}) + \frac{1}{2}\ddot{\mathcal{H}}_{ij}
\\
 & + \frac{1}{2}\frac{k_{i}k_{j}}{a^{2}}\mathcal{H} + \frac{1}{2}\frac{k^{2}}{a^{2}}\mathcal{H}_{ij}
 - \frac{1}{2a^{2}}(k_{i}k_{k}\mathcal{H}_{kj} + k_{j}k_{k}\mathcal{H}_{ki}) + \frac{3}{2}H
 \dot{\mathcal{H}}_{ij} + \frac{1}{2}\delta_{ij}H\dot{\mathcal{H}},
\end{split}
\end{equation}
where the terms on the \textit{r.h.s} are as earlier the source terms. Thus, we may write the
formal solutions of \ref{S-F26}) and \ref{S-F27}) as
\begin{equation}\label{S-F28}
\begin{split}
\frac{\delta\Phi_{ij}}{a^{2}} & = \int_{-\infty}^{\infty}dt' \ \mathcal{G}_{k}(t ; t')
\mathcal{S}_{1}(t')\\
\frac{\delta S_{ij}}{a^{2}} & = \int_{-\infty}^{\infty}dt' \ \mathcal{G}_{k}(t ; t')
\mathcal{S}_{2}(t'),
\end{split}
\end{equation}
where $\mathcal{S}_{1}(t)$, and $\mathcal{S}_{2}(t)$ terms are the source terms on the \textit{r.h.s} 
of the equations \ref{S-F27}) and \ref{S-F26}), respectively, and $\mathcal{G}_{k}(t ; t')$ is the Green's function satisfying the following differential equation
\begin{equation}\label{S-F29}
\left(\frac{d^{2}}{dt^{2}} + \frac{k^{2}}{a^{2}}\right)\mathcal{G}_{k}(t ; t') = \delta(t - t').
\end{equation}
In writing the equation \ref{S-F28}), we suppressed the spatial indices on the \textit{r.h.s}.
If we plug the formal solutions in equation \ref{S-F28}) in the metric perturbed field equations
derived earlier, we will essentially obtain a set of non-local field equations in $\mathcal{H}_{ij}$.
This also shows the effect of de-gravitation in this class of non-local gravity theories.

\bibliographystyle{utphys}
\bibliography{Ver-Fin-CQGStyle}

\providecommand{\href}[2]{#2}\begingroup\raggedright\begin{thebibliography}{10}

\bibitem{Deser:2007jk}
S.~Deser and R.~P. Woodard, ``{Nonlocal Cosmology},''
  \href{http://dx.doi.org/10.1103/PhysRevLett.99.111301}{{\em Phys. Rev. Lett.}
  {\bfseries 99} (2007) 111301},
  \href{http://arxiv.org/abs/0706.2151}{{\ttfamily arXiv:0706.2151 [gr-qc]}}.

\bibitem{Deser:2013uya}
S.~Deser and R.~P. Woodard, ``{Observational Viability and Stability of
  Nonlocal Cosmology},''
  \href{http://dx.doi.org/10.1088/1475-7516/2013/11/036}{{\em JCAP} {\bfseries
  11} (2013) 036}, \href{http://arxiv.org/abs/1307.6639}{{\ttfamily
  arXiv:1307.6639 [astro-ph.CO]}}.

\bibitem{Dodelson:2013sma}
S.~Dodelson and S.~Park, ``{Nonlocal Gravity and Structure in the Universe},''
  \href{http://dx.doi.org/10.1103/PhysRevD.90.043535}{{\em Phys. Rev. D}
  {\bfseries 90} (2014) 043535},
  \href{http://arxiv.org/abs/1310.4329}{{\ttfamily arXiv:1310.4329
  [astro-ph.CO]}}. [Erratum: Phys.Rev.D 98, 029904 (2018)].

\bibitem{Maggiore:2014sia}
M.~Maggiore and M.~Mancarella, ``{Nonlocal gravity and dark energy},''
  \href{http://dx.doi.org/10.1103/PhysRevD.90.023005}{{\em Phys. Rev. D}
  {\bfseries 90} no.~2, (2014) 023005},
  \href{http://arxiv.org/abs/1402.0448}{{\ttfamily arXiv:1402.0448 [hep-th]}}.

\bibitem{Capozziello:2021krv}
S.~Capozziello and F.~Bajardi, ``{Nonlocal gravity cosmology: An overview},''
  \href{http://dx.doi.org/10.1142/S0218271822300099}{{\em Int. J. Mod. Phys. D}
  {\bfseries 31} no.~06, (2022) 2230009},
  \href{http://arxiv.org/abs/2201.04512}{{\ttfamily arXiv:2201.04512 [gr-qc]}}.

\bibitem{Modesto:2013jea}
L.~Modesto and S.~Tsujikawa, ``{Non-local massive gravity},''
  \href{http://dx.doi.org/10.1016/j.physletb.2013.10.037}{{\em Phys. Lett. B}
  {\bfseries 727} (2013) 48--56},
  \href{http://arxiv.org/abs/1307.6968}{{\ttfamily arXiv:1307.6968 [hep-th]}}.

\bibitem{Nojiri:2019dio}
S.~Nojiri, S.~D. Odintsov, and V.~K. Oikonomou, ``{Ghost-free non-local $F(R)$
  Gravity Cosmology},''
  \href{http://dx.doi.org/10.1016/j.dark.2020.100541}{{\em Phys. Dark Univ.}
  {\bfseries 28} (2020) 100541},
  \href{http://arxiv.org/abs/1911.07329}{{\ttfamily arXiv:1911.07329 [gr-qc]}}.

\bibitem{Nojiri:2007uq}
S.~Nojiri and S.~D. Odintsov, ``{Modified non-local-F(R) gravity as the key for
  the inflation and dark energy},''
  \href{http://dx.doi.org/10.1016/j.physletb.2007.12.001}{{\em Phys. Lett. B}
  {\bfseries 659} (2008) 821--826},
  \href{http://arxiv.org/abs/0708.0924}{{\ttfamily arXiv:0708.0924 [hep-th]}}.

\bibitem{Elizalde:2018qbm}
E.~Elizalde, S.~D. Odintsov, E.~O. Pozdeeva, and S.~Y. Vernov, ``{De Sitter and
  power-law solutions in non-local Gauss\textendash{}Bonnet gravity},''
  \href{http://dx.doi.org/10.1142/S0219887818501888}{{\em Int. J. Geom. Meth.
  Mod. Phys.} {\bfseries 15} no.~11, (2018) 1850188},
  \href{http://arxiv.org/abs/1805.10810}{{\ttfamily arXiv:1805.10810 [gr-qc]}}.

\bibitem{Jhingan:2008ym}
S.~Jhingan, S.~Nojiri, S.~D. Odintsov, M.~Sami, I.~Thongkool, and S.~Zerbini,
  ``{Phantom and non-phantom dark energy: The Cosmological relevance of
  non-locally corrected gravity},''
  \href{http://dx.doi.org/10.1016/j.physletb.2008.04.054}{{\em Phys. Lett. B}
  {\bfseries 663} (2008) 424--428},
  \href{http://arxiv.org/abs/0803.2613}{{\ttfamily arXiv:0803.2613 [hep-th]}}.

\bibitem{DiValentino:2021izs}
E.~Di~Valentino, O.~Mena, S.~Pan, L.~Visinelli, W.~Yang, A.~Melchiorri, D.~F.
  Mota, A.~G. Riess, and J.~Silk, ``{In the realm of the Hubble
  tension\textemdash{}a review of solutions},''
  \href{http://dx.doi.org/10.1088/1361-6382/ac086d}{{\em Class. Quant. Grav.}
  {\bfseries 38} no.~15, (2021) 153001},
  \href{http://arxiv.org/abs/2103.01183}{{\ttfamily arXiv:2103.01183}}.

\bibitem{Kamionkowski:2022pkx}
M.~Kamionkowski and A.~G. Riess, ``{The Hubble Tension and Early Dark
  Energy},'' \href{http://dx.doi.org/10.1146/annurev-nucl-111422-024107}{{\em
  Ann. Rev. Nucl. Part. Sci.} {\bfseries 73} (2023) 153--180},
  \href{http://arxiv.org/abs/2211.04492}{{\ttfamily arXiv:2211.04492}}.

\bibitem{2016-Joyce.etal-ARN}
A.~Joyce, L.~Lombriser, and F.~Schmidt, ``{Dark Energy Versus Modified
  Gravity},'' \href{http://dx.doi.org/10.1146/annurev-nucl-102115-044553}{{\em
  Ann. Rev. Nucl. Part. Sci.} {\bfseries 66} (2016) 95--122},
  \href{http://arxiv.org/abs/1601.06133}{{\ttfamily arXiv:1601.06133}}.

\bibitem{Shankaranarayanan:2022wbx}
S.~Shankaranarayanan and J.~P. Johnson, ``{Modified theories of gravity: Why,
  how and what?},'' \href{http://dx.doi.org/10.1007/s10714-022-02927-2}{{\em
  Gen. Rel. Grav.} {\bfseries 54} no.~5, (2022) 44},
  \href{http://arxiv.org/abs/2204.06533}{{\ttfamily arXiv:2204.06533 [gr-qc]}}.

\bibitem{Mandal:2025xuc}
S.~Mandal and S.~Shankaranarayanan, ``{Modified theories of gravity at
  different curvature scales},''
  \href{http://arxiv.org/abs/2502.07437}{{\ttfamily arXiv:2502.07437 [gr-qc]}}.

\bibitem{Boulware:1972yco}
D.~G. Boulware and S.~Deser, ``{Can gravitation have a finite range?},''
  \href{http://dx.doi.org/10.1103/PhysRevD.6.3368}{{\em Phys. Rev. D}
  {\bfseries 6} (1972) 3368--3382}.

\bibitem{Dvali:2007kt}
G.~Dvali, S.~Hofmann, and J.~Khoury, ``{Degravitation of the cosmological
  constant and graviton width},''
  \href{http://dx.doi.org/10.1103/PhysRevD.76.084006}{{\em Phys. Rev. D}
  {\bfseries 76} (2007) 084006},
  \href{http://arxiv.org/abs/hep-th/0703027}{{\ttfamily arXiv:hep-th/0703027}}.

\bibitem{deRham:2007rw}
C.~de~Rham, S.~Hofmann, J.~Khoury, and A.~J. Tolley, ``{Cascading Gravity and
  Degravitation},'' \href{http://dx.doi.org/10.1088/1475-7516/2008/02/011}{{\em
  JCAP} {\bfseries 02} (2008) 011},
  \href{http://arxiv.org/abs/0712.2821}{{\ttfamily arXiv:0712.2821 [hep-th]}}.

\bibitem{Jaccard:2013gla}
M.~Jaccard, M.~Maggiore, and E.~Mitsou, ``{Nonlocal theory of massive
  gravity},'' \href{http://dx.doi.org/10.1103/PhysRevD.88.044033}{{\em Phys.
  Rev. D} {\bfseries 88} (2013) 044033},
  \href{http://arxiv.org/abs/1305.3034}{{\ttfamily arXiv:1305.3034 [hep-th]}}.

\bibitem{Nojiri:2010pw}
S.~Nojiri, S.~D. Odintsov, M.~Sasaki, and Y.-l. Zhang, ``{Screening of
  cosmological constant in non-local gravity},''
  \href{http://dx.doi.org/10.1016/j.physletb.2010.12.035}{{\em Phys. Lett. B}
  {\bfseries 696} (2011) 278--282},
  \href{http://arxiv.org/abs/1010.5375}{{\ttfamily arXiv:1010.5375 [gr-qc]}}.

\bibitem{Capozziello:2008gu}
S.~Capozziello, E.~Elizalde, S.~Nojiri, and S.~D. Odintsov, ``{Accelerating
  cosmologies from non-local higher-derivative gravity},''
  \href{http://dx.doi.org/10.1016/j.physletb.2008.11.060}{{\em Phys. Lett. B}
  {\bfseries 671} (2009) 193--198},
  \href{http://arxiv.org/abs/0809.1535}{{\ttfamily arXiv:0809.1535 [hep-th]}}.

\bibitem{Gambuti:2021meo}
G.~Gambuti and N.~Maggiore, ``{Fierz\textendash{}Pauli theory reloaded: from a
  theory of a symmetric tensor field to linearized massive gravity},''
  \href{http://dx.doi.org/10.1140/epjc/s10052-021-08962-8}{{\em Eur. Phys. J.
  C} {\bfseries 81} no.~2, (2021) 171},
  \href{http://arxiv.org/abs/2102.10813}{{\ttfamily arXiv:2102.10813 [gr-qc]}}.

\bibitem{Bautista:2017enk}
T.~Bautista, A.~Benevides, and A.~Dabholkar, ``{Nonlocal Quantum Effective
  Actions in Weyl-Flat Spacetimes},''
  \href{http://dx.doi.org/10.1007/JHEP06(2018)055}{{\em JHEP} {\bfseries 06}
  (2018) 055}, \href{http://arxiv.org/abs/1711.00135}{{\ttfamily
  arXiv:1711.00135 [hep-th]}}.

\bibitem{Barvinsky:2023exr}
A.~O. Barvinsky and W.~Wachowski, ``{Notes on conformal anomaly, nonlocal
  effective action, and the metamorphosis of the running scale},''
  \href{http://dx.doi.org/10.1103/PhysRevD.108.045014}{{\em Phys. Rev. D}
  {\bfseries 108} no.~4, (2023) 045014},
  \href{http://arxiv.org/abs/2306.03780}{{\ttfamily arXiv:2306.03780
  [hep-th]}}.

\bibitem{Barvinsky:2002uf}
A.~O. Barvinsky and V.~F. Mukhanov, ``{New nonlocal effective action},''
  \href{http://dx.doi.org/10.1103/PhysRevD.66.065007}{{\em Phys. Rev. D}
  {\bfseries 66} (2002) 065007},
  \href{http://arxiv.org/abs/hep-th/0203132}{{\ttfamily arXiv:hep-th/0203132}}.

\bibitem{Elias:2017wkr}
M.~El\'\i{}as, F.~D. Mazzitelli, and L.~G. Trombetta, ``{Nonlocal effective
  actions in semiclassical gravity: thermal effects in stationary
  geometries},'' \href{http://dx.doi.org/10.1103/PhysRevD.96.105007}{{\em Phys.
  Rev. D} {\bfseries 96} no.~10, (2017) 105007},
  \href{http://arxiv.org/abs/1709.10435}{{\ttfamily arXiv:1709.10435
  [hep-th]}}.

\bibitem{Donoghue:2015nba}
J.~F. Donoghue and B.~K. El-Menoufi, ``{Covariant non-local action for massless
  QED and the curvature expansion},''
  \href{http://dx.doi.org/10.1007/JHEP10(2015)044}{{\em JHEP} {\bfseries 10}
  (2015) 044}, \href{http://arxiv.org/abs/1507.06321}{{\ttfamily
  arXiv:1507.06321 [hep-th]}}.

\bibitem{Barvinsky:2003kg}
A.~O. Barvinsky, ``{Nonlocal action for long distance modifications of gravity
  theory},'' \href{http://dx.doi.org/10.1016/j.physletb.2003.08.055}{{\em Phys.
  Lett. B} {\bfseries 572} (2003) 109--116},
  \href{http://arxiv.org/abs/hep-th/0304229}{{\ttfamily arXiv:hep-th/0304229}}.

\bibitem{Barvinsky:1994cg}
A.~O. Barvinsky, Y.~V. Gusev, G.~A. Vilkovisky, and V.~V. Zhytnikov, ``{The One
  loop effective action and trace anomaly in four-dimensions},''
  \href{http://dx.doi.org/10.1016/0550-3213(94)00585-3}{{\em Nucl. Phys. B}
  {\bfseries 439} (1995) 561--582},
  \href{http://arxiv.org/abs/hep-th/9404187}{{\ttfamily arXiv:hep-th/9404187}}.

\bibitem{Cusin:2015rex}
G.~Cusin, S.~Foffa, M.~Maggiore, and M.~Mancarella, ``{Nonlocal gravity with a
  Weyl-square term},'' \href{http://dx.doi.org/10.1103/PhysRevD.93.043006}{{\em
  Phys. Rev. D} {\bfseries 93} no.~4, (2016) 043006},
  \href{http://arxiv.org/abs/1512.06373}{{\ttfamily arXiv:1512.06373
  [hep-th]}}.

\bibitem{Bernard:2014bfa}
L.~Bernard, C.~Deffayet, and M.~von Strauss, ``{Consistent massive graviton on
  arbitrary backgrounds},''
  \href{http://dx.doi.org/10.1103/PhysRevD.91.104013}{{\em Phys. Rev. D}
  {\bfseries 91} no.~10, (2015) 104013},
  \href{http://arxiv.org/abs/1410.8302}{{\ttfamily arXiv:1410.8302 [hep-th]}}.

\bibitem{deRham:2014zqa}
C.~de~Rham, ``{Massive Gravity},''
  \href{http://dx.doi.org/10.12942/lrr-2014-7}{{\em Living Rev. Rel.}
  {\bfseries 17} (2014) 7}, \href{http://arxiv.org/abs/1401.4173}{{\ttfamily
  arXiv:1401.4173 [hep-th]}}.

\bibitem{Mazuet_2018}
C.~Mazuet and M.~S. Volkov, ``Massive spin-2 field in arbitrary
  spacetimes---the detailed derivation,''
  \href{http://dx.doi.org/10.1088/1475-7516/2018/07/012}{{\em Journal of
  Cosmology and Astroparticle Physics} {\bfseries 2018} no.~07, (Jul, 2018)
  012}. \url{https://dx.doi.org/10.1088/1475-7516/2018/07/012}.

\bibitem{Gumrukcuoglu:2021gua}
A.~E. Gumrukcuoglu, R.~Kimura, M.~Kenna-Allison, and K.~Koyama, ``{Vainshtein
  mechanism in Generalised Massive Gravity},''
  \href{http://dx.doi.org/10.1088/1475-7516/2021/09/023}{{\em JCAP} {\bfseries
  09} (2021) 023}, \href{http://arxiv.org/abs/2107.01423}{{\ttfamily
  arXiv:2107.01423 [hep-th]}}.

\bibitem{Mandal:2024nhv}
S.~Mandal and S.~Shankaranarayanan, ``{Generation of effective massive Spin-2
  fields through spontaneous symmetry breaking of scalar field},''
  \href{http://dx.doi.org/10.1007/s10714-025-03367-4}{{\em Gen. Rel. Grav.}
  {\bfseries 57} no.~2, (2025) 36},
  \href{http://arxiv.org/abs/2407.06572}{{\ttfamily arXiv:2407.06572
  [hep-th]}}.

\bibitem{Fierz:1939ix}
M.~Fierz and W.~Pauli, ``{On relativistic wave equations for particles of
  arbitrary spin in an electromagnetic field},''
  \href{http://dx.doi.org/10.1098/rspa.1939.0140}{{\em Proc. Roy. Soc. Lond. A}
  {\bfseries 173} (1939) 211--232}.

\bibitem{vanDam:1970vg}
H.~van Dam and M.~J.~G. Veltman, ``{Massive and massless Yang-Mills and
  gravitational fields},''
  \href{http://dx.doi.org/10.1016/0550-3213(70)90416-5}{{\em Nucl. Phys. B}
  {\bfseries 22} (1970) 397--411}.

\bibitem{Zakharov:1970cc}
V.~I. Zakharov, ``{Linearized gravitation theory and the graviton mass},'' {\em
  JETP Lett.} {\bfseries 12} (1970) 312.

\bibitem{Hell:2022wci}
A.~Hell, \href{http://dx.doi.org/10.5282/edoc.30317}{{\em {The massless limit
  of massive gauge theories}}}.
\newblock PhD thesis, Munich U., 2022.

\bibitem{Banerjee:2025fph}
N.~Banerjee and J.~Singh, ``{Massless limit of massive self-interacting vector
  fields},'' \href{http://arxiv.org/abs/2505.22119}{{\ttfamily arXiv:2505.22119
  [hep-th]}}.

\bibitem{Wang:2009yz}
A.~Wang and R.~Maartens, ``{Linear perturbations of cosmological models in the
  Horava-Lifshitz theory of gravity without detailed balance},''
  \href{http://dx.doi.org/10.1103/PhysRevD.81.024009}{{\em Phys. Rev. D}
  {\bfseries 81} (2010) 024009},
  \href{http://arxiv.org/abs/0907.1748}{{\ttfamily arXiv:0907.1748 [hep-th]}}.

\bibitem{Charmousis:2009tc}
C.~Charmousis, G.~Niz, A.~Padilla, and P.~M. Saffin, ``{Strong coupling in
  Horava gravity},''
  \href{http://dx.doi.org/10.1088/1126-6708/2009/08/070}{{\em JHEP} {\bfseries
  08} (2009) 070}, \href{http://arxiv.org/abs/0905.2579}{{\ttfamily
  arXiv:0905.2579 [hep-th]}}.

\bibitem{2018-Planck-AA}
{\bfseries Planck} Collaboration, N.~Aghanim {\em et~al.}, ``{Planck 2018
  results. VI. Cosmological parameters},''
  \href{http://dx.doi.org/10.1051/0004-6361/201833910}{{\em Astron. Astrophys.}
  {\bfseries 641} (2020) A6}, \href{http://arxiv.org/abs/1807.06209}{{\ttfamily
  arXiv:1807.06209 [astro-ph.CO]}}. [Erratum: Astron.Astrophys. 652, C4
  (2021)].

\bibitem{2016-Riess.others-Astrophys.J.}
A.~G. Riess {\em et~al.}, ``A 2.4\% determination of the local value of the
  hubble constant,'' \href{http://dx.doi.org/10.3847/0004-637X/826/1/56}{{\em
  Astrophys. J.} {\bfseries 826} no.~1, (2016) 56},
  \href{http://arxiv.org/abs/1604.01424}{{\ttfamily arXiv:1604.01424
  [astro-ph.CO]}}.

\bibitem{Carron:2022eyg}
J.~Carron, M.~Mirmelstein, and A.~Lewis, ``{CMB lensing from Planck
  PR4~maps},'' \href{http://dx.doi.org/10.1088/1475-7516/2022/09/039}{{\em
  JCAP} {\bfseries 09} (2022) 039},
  \href{http://arxiv.org/abs/2206.07773}{{\ttfamily arXiv:2206.07773
  [astro-ph.CO]}}.

\bibitem{DESI:2025fii}
{\bfseries DESI} Collaboration, K.~Lodha {\em et~al.}, ``{Extended dark energy
  analysis using DESI DR2 BAO measurements},''
  \href{http://dx.doi.org/10.1103/w4c6-1r5j}{{\em Phys. Rev. D} {\bfseries 112}
  no.~8, (2025) 083511}, \href{http://arxiv.org/abs/2503.14743}{{\ttfamily
  arXiv:2503.14743 [astro-ph.CO]}}.

\bibitem{DESI:2025zgx}
{\bfseries DESI} Collaboration, M.~Abdul~Karim {\em et~al.}, ``{DESI DR2
  results. II. Measurements of baryon acoustic oscillations and cosmological
  constraints},'' \href{http://dx.doi.org/10.1103/tr6y-kpc6}{{\em Phys. Rev. D}
  {\bfseries 112} no.~8, (2025) 083515},
  \href{http://arxiv.org/abs/2503.14738}{{\ttfamily arXiv:2503.14738
  [astro-ph.CO]}}.

\bibitem{2025-Gu.others-}
{\bfseries DESI} Collaboration, G.~Gu {\em et~al.}, ``Dynamical dark energy in
  light of the desi dr2 baryonic acoustic oscillations measurements,''
  \href{http://arxiv.org/abs/2504.06118}{{\ttfamily arXiv:2504.06118
  [astro-ph.CO]}}.

\bibitem{2025-KiDS}
A.~H. Wright {\em et~al.}, ``{KiDS-Legacy: Cosmological constraints from cosmic
  shear with the complete Kilo-Degree Survey},''
  \href{http://arxiv.org/abs/2503.19441}{{\ttfamily arXiv:2503.19441
  [astro-ph.CO]}}.

\bibitem{Allen:2023kib}
B.~Allen, S.~Dhurandhar, Y.~Gupta, M.~McLaughlin, P.~Natarajan, R.~M. Shannon,
  E.~Thrane, and A.~Vecchio, ``{The International Pulsar Timing Array checklist
  for the detection of nanohertz gravitational waves},''
  \href{http://arxiv.org/abs/2304.04767}{{\ttfamily arXiv:2304.04767
  [astro-ph.IM]}}.

\bibitem{NANOGrav:2020spf}
{\bfseries NANOGrav} Collaboration, N.~S. Pol {\em et~al.}, ``{Astrophysics
  Milestones for Pulsar Timing Array Gravitational-wave Detection},''
  \href{http://dx.doi.org/10.3847/2041-8213/abf2c9}{{\em Astrophys. J. Lett.}
  {\bfseries 911} no.~2, (2021) L34},
  \href{http://arxiv.org/abs/2010.11950}{{\ttfamily arXiv:2010.11950
  [astro-ph.HE]}}.

\bibitem{NANOGrav:2023gor}
{\bfseries NANOGrav} Collaboration, G.~Agazie {\em et~al.}, ``{The NANOGrav 15
  yr Data Set: Evidence for a Gravitational-wave Background},''
  \href{http://dx.doi.org/10.3847/2041-8213/acdac6}{{\em Astrophys. J. Lett.}
  {\bfseries 951} no.~1, (2023) L8},
  \href{http://arxiv.org/abs/2306.16213}{{\ttfamily arXiv:2306.16213
  [astro-ph.HE]}}.

\bibitem{LIGOScientific:2017adf}
{\bfseries LIGO Scientific, Virgo, 1M2H, Dark Energy Camera GW-E, DES, DLT40,
  Las Cumbres Observatory, VINROUGE, MASTER} Collaboration, B.~P. Abbott {\em
  et~al.}, ``{A gravitational-wave standard siren measurement of the Hubble
  constant},'' \href{http://dx.doi.org/10.1038/nature24471}{{\em Nature}
  {\bfseries 551} no.~7678, (2017) 85--88},
  \href{http://arxiv.org/abs/1710.05835}{{\ttfamily arXiv:1710.05835
  [astro-ph.CO]}}.

\bibitem{Ezquiaga:2017ekz}
J.~M. Ezquiaga and M.~Zumalac{\'a}rregui, ``{Dark Energy After GW170817: Dead
  Ends and the Road Ahead},''
  \href{http://dx.doi.org/10.1103/PhysRevLett.119.251304}{{\em Phys. Rev.
  Lett.} {\bfseries 119} no.~25, (2017) 251304},
  \href{http://arxiv.org/abs/1710.05901}{{\ttfamily arXiv:1710.05901
  [astro-ph.CO]}}.

\bibitem{LISA:2022yao}
{\bfseries LISA} Collaboration, P.~A. Seoane {\em et~al.}, ``{Astrophysics with
  the Laser Interferometer Space Antenna},''
  \href{http://dx.doi.org/10.1007/s41114-022-00041-y}{{\em Living Rev. Rel.}
  {\bfseries 26} no.~1, (2023) 2},
  \href{http://arxiv.org/abs/2203.06016}{{\ttfamily arXiv:2203.06016 [gr-qc]}}.

\bibitem{Fasiello:2012rw}
M.~Fasiello and A.~J. Tolley, ``{Cosmological perturbations in Massive Gravity
  and the Higuchi bound},''
  \href{http://dx.doi.org/10.1088/1475-7516/2012/11/035}{{\em JCAP} {\bfseries
  11} (2012) 035}, \href{http://arxiv.org/abs/1206.3852}{{\ttfamily
  arXiv:1206.3852 [hep-th]}}.

\bibitem{Mazuet:2015pea}
C.~Mazuet and M.~S. Volkov, ``{De Sitter vacua in ghost-free massive gravity
  theory},'' \href{http://dx.doi.org/10.1016/j.physletb.2015.10.012}{{\em Phys.
  Lett. B} {\bfseries 751} (2015) 19--24},
  \href{http://arxiv.org/abs/1503.03042}{{\ttfamily arXiv:1503.03042
  [hep-th]}}.

\bibitem{Alberte:2011ah}
L.~Alberte, ``{Massive Gravity on Curved Background},''
  \href{http://dx.doi.org/10.1142/S0218271812500587}{{\em Int. J. Mod. Phys. D}
  {\bfseries 21} (2012) 1250058},
  \href{http://arxiv.org/abs/1110.3818}{{\ttfamily arXiv:1110.3818 [hep-th]}}.

\bibitem{Starobinsky:2007hu}
A.~A. Starobinsky, ``{Disappearing cosmological constant in f(R) gravity},''
  \href{http://dx.doi.org/10.1134/S0021364007150027}{{\em JETP Lett.}
  {\bfseries 86} (2007) 157--163},
  \href{http://arxiv.org/abs/0706.2041}{{\ttfamily arXiv:0706.2041
  [astro-ph]}}.

\bibitem{Hu:2007nk}
W.~Hu and I.~Sawicki, ``{Models of f(R) Cosmic Acceleration that Evade
  Solar-System Tests},''
  \href{http://dx.doi.org/10.1103/PhysRevD.76.064004}{{\em Phys. Rev. D}
  {\bfseries 76} (2007) 064004},
  \href{http://arxiv.org/abs/0705.1158}{{\ttfamily arXiv:0705.1158
  [astro-ph]}}.

\bibitem{Johnson:2019vwi}
J.~P. Johnson and S.~Shankaranarayanan, ``{Low-energy modified gravity
  signatures on the large-scale structures},''
  \href{http://dx.doi.org/10.1103/PhysRevD.100.083526}{{\em Phys. Rev. D}
  {\bfseries 100} no.~8, (2019) 083526},
  \href{http://arxiv.org/abs/1904.07608}{{\ttfamily arXiv:1904.07608
  [astro-ph.CO]}}.

\bibitem{Bansal:2024bbb}
P.~Bansal, J.~P. Johnson, and S.~Shankaranarayanan, ``{Interacting dark sector
  from Horndeski theories and beyond: Mapping fields and fluids},''
  \href{http://dx.doi.org/10.1103/PhysRevD.111.024071}{{\em Phys. Rev. D}
  {\bfseries 111} no.~2, (2025) 024071},
  \href{http://arxiv.org/abs/2408.12341}{{\ttfamily arXiv:2408.12341
  [astro-ph.CO]}}.

\bibitem{Bansal:2025usn}
P.~Bansal, J.~P. Johnson, and S.~Shankaranarayanan, ``{Disformal interactions
  in the Dark Sector: From driving Early Dark Energy to confronting
  cosmological tensions},'' \href{http://arxiv.org/abs/2508.17003}{{\ttfamily
  arXiv:2508.17003 [astro-ph.CO]}}.

\bibitem{Dolgov:2003px}
A.~D. Dolgov and M.~Kawasaki, ``{Can modified gravity explain accelerated
  cosmic expansion?},''
  \href{http://dx.doi.org/10.1016/j.physletb.2003.08.039}{{\em Phys. Lett. B}
  {\bfseries 573} (2003) 1--4},
  \href{http://arxiv.org/abs/astro-ph/0307285}{{\ttfamily
  arXiv:astro-ph/0307285}}.

\bibitem{Vainshtein:1972sx}
A.~I. Vainshtein, ``{To the problem of nonvanishing gravitation mass},''
  \href{http://dx.doi.org/10.1016/0370-2693(72)90147-5}{{\em Phys. Lett. B}
  {\bfseries 39} (1972) 393--394}.

\bibitem{Halder:2024uao}
S.~Halder, J.~de~Haro, T.~Saha, and S.~Pan, ``{Phase space analysis of
  sign-shifting interacting dark energy models},''
  \href{http://dx.doi.org/10.1103/PhysRevD.109.083522}{{\em Phys. Rev. D}
  {\bfseries 109} no.~8, (2024) 083522},
  \href{http://arxiv.org/abs/2403.01397}{{\ttfamily arXiv:2403.01397 [gr-qc]}}.

\bibitem{Xu:2012jf}
C.~Xu, E.~N. Saridakis, and G.~Leon, ``{Phase-Space analysis of Teleparallel
  Dark Energy},'' \href{http://dx.doi.org/10.1088/1475-7516/2012/07/005}{{\em
  JCAP} {\bfseries 07} (2012) 005},
  \href{http://arxiv.org/abs/1202.3781}{{\ttfamily arXiv:1202.3781 [gr-qc]}}.

\bibitem{Boehmer:2023knj}
C.~G. Boehmer, E.~Jensko, and R.~Lazkoz, ``{Dynamical Systems Analysis of f(Q)
  Gravity},'' \href{http://dx.doi.org/10.3390/universe9040166}{{\em Universe}
  {\bfseries 9} no.~4, (2023) 166},
  \href{http://arxiv.org/abs/2303.04463}{{\ttfamily arXiv:2303.04463 [gr-qc]}}.

\bibitem{Ghosh:2023amt}
S.~Ghosh, R.~Solanki, and P.~K. Sahoo, ``{Dynamical system analysis of scalar
  field cosmology in coincident f(Q) gravity},''
  \href{http://dx.doi.org/10.1088/1402-4896/ad39b5}{{\em Phys. Scripta}
  {\bfseries 99} no.~5, (2024) 055021},
  \href{http://arxiv.org/abs/2309.11198}{{\ttfamily arXiv:2309.11198 [gr-qc]}}.

\end{thebibliography}\endgroup

\end{document}